\documentclass[utf8]{frontiersSCNS} % for Science, Engineering and Humanities and Social Sciences articles
\usepackage{url,hyperref,lineno,microtype,subcaption}
\usepackage[onehalfspacing]{setspace}
\usepackage{multirow}
\def\apss{Astroph.   Space Sci.}

\def\apj{Astroph. J.}
\def\aj{Astron. J.}
\def\apjl{Astroph. J. Lett.}\def\apjs{Astroph. J. Supp.}
\def\mnras{Mon. Not. R. Astron. Soc.}
\def\aap{Astron. Astroph. }

\def\nat{Nature}\def\araa{Ann. Rev. Astron. Astroph.}
\defcitealias{marzianietal16}{M16}
\defcitealias{marzianietal16a}{M16a}
\def\mbh{$M_\mathrm{BH}$}

\def\kms{km s$^{-1}$\/}

\def\lledd{$L/L_\mathrm{Edd}$}

\newcolumntype{C}[1]{>{\centering\let\newline\\\arraybackslash\hspace{0pt}}m{#1}}
\def\keyFont{\fontsize{8}{11}\helveticabold }
\def\firstAuthorLast{D'Onofrio \& Marziani  } %use et al only if is more than 1 author
\def\Authors{
M. D'Onofrio\,$^{1,*}$, P. Marziani\,$^{2,*}$}
\def\Address{$^{1}$ Department of Physics \& Astronomy, University of Padova, Padova, Italia\\
$^{2}$ National Institute for Astrophysics (INAF), Padua Astronomical Observatory, Italy}
\def\corrAuthor{Mauro D'Onofrio}
\def\corrEmail{mauro.donofrio@unipd.it}
\begin{document}
\onecolumn
\firstpage{1}
\title[]{A multimessenger view of galaxies and quasars from now to mid-century} 
\author[\firstAuthorLast ]{\Authors} %This field will be automatically populated
\address{} %This field will be automatically populated
\correspondance{} %This field will be automatically populated
\extraAuth{}% If there are more than 1 corresponding author, comment this line and uncomment the next one.
%\extraAuth{corresponding Author2 \\ Laboratory X2, Institute X2, Department X2, Organization X2, Street X2, City X2 , State XX2 (only USA, Canada and Australia), Zip Code2, X2 Country X2, email2@uni2.edu}

\maketitle

\begin{abstract}
%\section{} 
In the next 30 years, a new generation of space and ground-based telescopes will  permit to obtain multi-frequency observations of  faint sources and, for the first time in human history, to achieve a deep, almost synoptical monitoring of the whole sky. Gravitational wave observatories  will detect a Universe of unseen black holes in the merging process over a broad spectrum of mass. Computing facilities will permit new high-resolution simulations with a deeper physical analysis of the main phenomena occurring at different scales. 
 
Given these development lines,  we first sketch a panorama of the main instrumental developments expected in the next thirty years, dealing not only with electromagnetic radiation, but also from a multi-messenger perspective that includes gravitational waves, neutrinos, and cosmic rays. We  then present how the new instrumentation will make it possible to foster advances in our present understanding of galaxies and quasars. We  focus on selected scientific themes that are hotly debated today, in some cases advancing conjectures on the solution of major problems that may become solved in the next 30 years.

\tiny
 \keyFont{ \section{Keywords:}  galaxy evolution -- quasars -- cosmology  --  supermassive black holes -- black hole physics} %All article types: you may provide up to 8 keywords; at least 5 are mandatory.
\end{abstract}

%\tableofcontents
\section{Introduction: toward multimessenger astronomy}
\label{intro}

The development of  astronomy in the second half of the XX$^\mathrm{th}$ century   followed two major lines of improvement: the increase in light gathering power (i.e., the ability to detect fainter objects), and the extension of the frequency domain in the electromagnetic spectrum beyond the traditional optical domain.  Around 40 optical, ground based, reflecting telescopes of diameter larger than 3m are operational at the time of writing: only eleven of them became operational before 1990, which means that 3/4 of the largest telescopes have been built in the last 25 years. The 14 telescopes of the 8-10m class (counting 2 for the Large Binary Telescope, LBT, and 4 for the Very Large Telescope, VLT)  all became operational around the year 2000 or afterwards, with the exception of the first Keck telescope (at Keck science observations began in 1993). Telescopes of the 6-10m class in space and telescopes of the 30-40m class on the ground belong to the near future: the James Webb Space Telecope (JWST) will be soon launched and the ground-based telescopes that are under construction are the ESO Extremely Large Telescope (E-ELT) and the Giant Magellan Telescope (GMT). The light gathering power is steadily increasing and will benefit spectroscopic studies of distant quasars, galaxies and supernov\ae. 

The second line  of development involved the extension over unexplored frequency domains of the electromagnetic radiation: from the tiny optical range from 3700 to 8000 \AA, ground and space based instrumentation now yields the ability to cover the electromagnetic spectrum from meter wavelengths to the $\gamma$-ray domain. If there is a safe prediction is that the progress will follow along the line of a deeper, wider and faster coverage of cosmic sources at all accessible wavelengths of the electromagnetic spectrum.  

A third line -- the increase in resolving power (i.e., the ability to resolve finer details of distant objects\footnote{The angular resolutions is defined by the angular radius of the Airy disk which is given as $\theta \approx 0.025 \lambda_{1\mu m}/D_\mathrm{1m}$ seconds of arc, where the radiation wavelength is in units of microns and $D$ (either the disk diameter of the interferometer baseline) is in meters.}) -- has somehow lagged behind due to the formidable technical challenges at the typical wavelengths of the electromagnetic radiation in the visual and UV bands. This third line of  instrumental development is based on reaching diffraction limit performances for single dish telescopes and on  interferometry, from ground and space.  In April 2016 Physics Today challenged its readers to write a fictional news story about a discovery made in 2116. Robert Austin's winning entry describes a huge space telescope fashioned from laser-machined asteroids \citep{austin16}, the Asteroid Belt Astronomical Telescope (ABAT). Even if this idea is perhaps too far-fetched to become true even  one century from now, we can expect substantial developments in interferometry from ground and space. For instance, right at the time of writing, the new instrument MATISSE, an interferometric imagining spectrograph working in the L, M, and N bands in the mid-infrared region (3-13 $\mu$m), is becoming operational at the Very Large Telescope Interferometer (VLTI) \citep{lopezetal14} and is expected to reach a spatial resolution of 0.01 arcsec, comparable to the Hubble Space Telescope (HST), the 1.54m telescope orbiting our Earth since 1990.  
 
%\subsection{Multimessenger: gravitational waves, neutrinos, and cosmic rays}

There is however a most exciting development that  will break the almost complete monopoly of electromagnetic radiation as carrier of information from extragalactic sources.  The detection of {\it gravitational waves} is probably the achievement of the century. Much of frontiers astrophysics will be devoted not only to expand and create new gravitational observatories, but also at  conventional electromagnetic studies of counterparts of gravitational wave sources, following the first positive identification just a few months ago \citep{abbottetal17}.  

At the same time new types of telescopes and instruments will be soon in operation for the detection of neutrinos and cosmic rays.All these endeavors contribute to {\it multi-messenger astronomy}: the new magic word under which all future researchers will operate. We would try below to summarize the main aspects of such future instrumental developments. %\textcolor{magenta}{The intended readership of this review are  mainly students, from  senior high-school to advanced undergraduates.}
%which one can disguise conventional observations if they are just minimally related to non-electromagnetic detections.  This is not to downplay the role of multimessenger astronomy, on the contrary. The point is that we can expect much in the next decades. Much of our data are still {\it single-messenger}, as data from gravitational waves, neutrinos, and cosmic rays are limited, at least as far as their relevance for extragalactic astronomy and quasar studies is concerned.  

% (and excludes the possibility of paradigm shifts that might be necessary for a deeper understanding of the physical world)

\section{The main instrumental developments}
\label{sec:instr} 

An eminent physicist foresaw that we will be able to have a complete account of the  reality, from the Big Bang to humans, in physical and chemical terms, within this century. Although this idea is perhaps too optimistic, it is reasonable to presume that we will have a reasonably complete view of the constituents of the visible Universe and of its evolution from the dark ages (beyond $z \approx 6$, currently sampled up to $z \approx 10 - 12$) until the present cosmic epoch. 

Once an overall physical understanding will be reached, we expect that science will progress toward the explanation and modeling  of finer physical details. This has been the case for the study of stars in the XX$^\mathrm{th}$\ century. The stellar evolution theory provides a detailed general physical framework with predictive power, although there are still many challenging aspects in the physics of stellar atmospheres and stellar structure (magnetic reconnection, flares, internal oscillation, internal turbulence) that are at the frontiers of present-day research. In this sense, we may hope to reach a global understanding of the nature of the active galactic nuclei\footnote{In the following we will use the term AGN to indicate all accretion-powered black holes, and ``quasars'' to indicate mainly high luminosity AGN. However, it is important to stress that there isn't any critical luminosity divide. } (AGN), and of their connection with host galaxy across cosmic times, but this will open a new order of very complex problems concerning the spatial and temporal modeling of the nuclear activity and its interplay with the host galaxies involving turbulence, chemical evolution, dust physics and magnetic fields.  All areas of investigations where spatial resolution is crucial (such as the study of nuclear feedback in galaxies, which are now in their infancy, as outlined below) will benefit from the advancements expected in active optics, and ground- and space-based interferometry.

Several developments  expected in the next 30--40 years are not so difficult to foresee,  not last because  the  new astronomical instruments require  careful planning that may imply at least a 10-year lapse between early proposals and first light at the facility. We will first review the major observational projects that are ongoing or planned and are expected to have frontier capability,\footnote{Even if some of the proposed projects may not be completed, or developed as assumed here,  we are confident that instrumentation with similar technical specifications will become operational over the next decades.} and then discuss their impact on several   present-day  scientific themes of extragalactic astronomy, in particular for galaxies and quasars (Sect. \ref{sec:prog}).
%Here below we discuss some of the main possible instrumental developments of the next 30 years.

 \subsection{Full-sky coverage across the electromagnetic spectrum} 
 
Generally speaking, the most modern ground-based telescopes and even more so the forthcoming ones are best suited for spectroscopy to deep limits (currently  25.6 AB mag in the Z-band in 3 hours exposure at LBT) or high spectral resolution ($R \sim 10,000  - 100,000$). They can also provide  wide-field coverage (up to $\approx$ 1 deg) and spectroscopic multiplexing ($\sim 10^{3}$ sources per fields, \citealt{dalcantonetal15}).  These capabilities make it possible to observe large samples of sources. For example, the Dark Energy Spectroscopic Instrument (DESI) is an optical spectrograph capable of performing highly-multiplexed observations to measure the BAOs (5000 robotized fibers can be distributed in an 8.0 square degree field, \citealt{levietal13}).

Ground-based telescopes equipped with adaptive optics systems have also made great advances in high-resolution near-IR imaging.  For example, GRAVITY, a VLT-based imaging IR interferometer is expected to reach resolving power of   three milliarcsec \citep{eisenhaueretal11}.  GRAVITY will yield an unprecedented view of the nuclei of most nearby galaxies. Adaptive optics (AO) systems are expected to improve, although  not beyond $\approx$ 1 arcmin fields and not at the shortest wavelengths.  However, multi-Conjugate Adaptive Optics systems can in principle yield diffraction-limited images of a relatively wide field (30 arcsec across), meaning a resolution of 20 milli arcsec in the V band at one of the VLT telescopes  \citep{espositoetal16}.  The  Multi-conjugate Adaptive Optics RelaY for ELT (MAORY) \citep{Fiorentinoetal2017}  may reach a 5$\sigma$detection limit of a point source in $J_\mathrm{AB} \approx 29$ with an exposure time of 5 hours \citep{bellazzini16}. 

Space-based telescopes provide complementary capabilities. Since backgrounds  are 10--100 times lower than at the ground,  space offers  a stable operating environment.  Space-based  observations reach the same broadband imaging depths of ground-based observations up to 100 times faster, achieve fainter magnitude limits with the next generation instruments ($AB \sim 32 - 34$ mag), and extend stable diffraction-limited PSFs to larger  fields of view (3--4 arcmin) and visual and UV wavelengths \citep{dalcantonetal15}.  {Foreseeable space astrometric missions (for example  the future ESA mission ``Theia" \citep{Theia2017}, on the path of the hugely successful Hipparcos and Gaia missions) may eventually achieve sub-$\mu$arcsec precision in astrometric measurement \citep{vallenari18}.   The return would be, for the first time in history, the ability to detect proper motions in extragalactic sources through angular displacements and not only by radial velocity differences with respect to a reference frame.  }

%These advantages favor science programs requiring high-resolution imaging of faint objects, high photometric precision, low-resolution and/or slitless spectroscopy, imaging in crowded  fields, and long-term variability and proper motion studies.  
Space telescopes also uniquely cover the information-rich UV, FIR, and sub-mm spectra that are  blocked by Earth's atmosphere.  Planned projects such as LUVOIR  \citep{franceetal17} can provide $ \sim 10 $ milliarsec resolution in the visual and UV bands that  may be unobtainable for the ELT if it reaches diffraction limited capabilities only at wavelength longer than 0.5 $\mu $m.

\begin{figure}[h!]
\vspace{0cm}
	\begin{center}
		\includegraphics[width=16cm, angle=0]{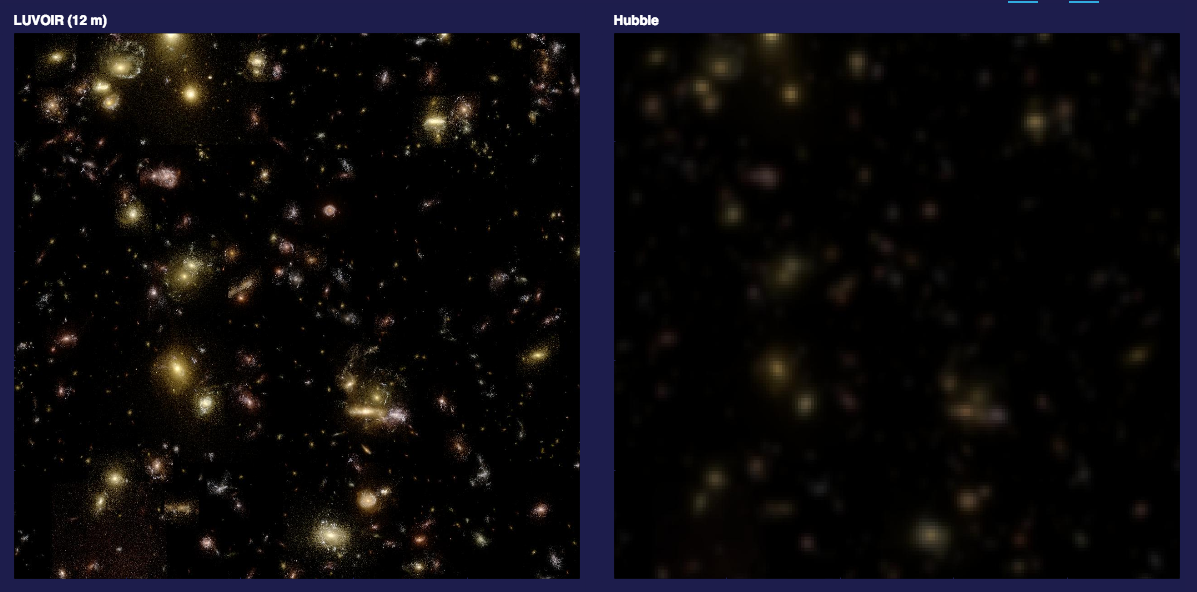}
				\includegraphics[width=16cm, angle=0]{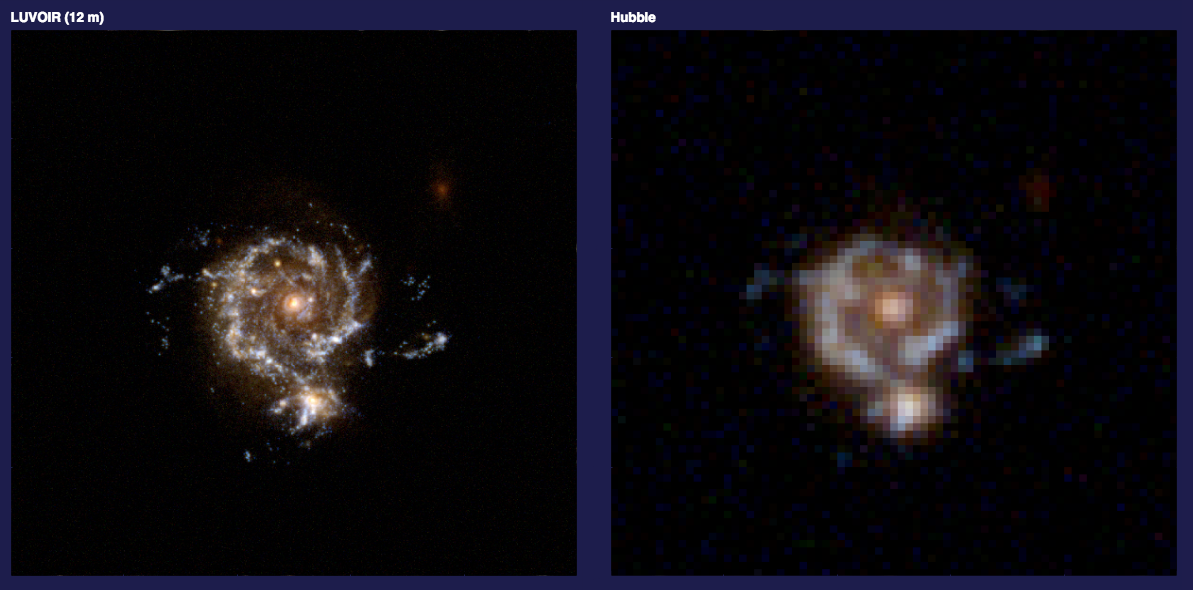}
	 \end{center}\vspace{0cm}
	\caption{{Comparison between expected performance of LUVOIR, an optical space based telescope of 12m aperture (left panels), and HST (right panels), for a deep field (top) and for the morphology of a galaxy at redshift $\approx 2$\ (bottom). Credit: LUVOIR, \href{http://luvoir.stsci.edu/image_comparison}{image comparison}  page, at  http://luvoir.stsci.edu. } \label{fig:luvoir} }
\end{figure}

\subsubsection{NIR-optical-UV}
 
%The space missions and grand-based observatories that are in development now and are expected to have a major impact in the next decades are listed in Table 1. 

A most notable advancement should be the building of fully steerable 40m class telescopes in the optical and near infrared: the GMT and the ESO ELT. The first  will consist of seven 8.4 m diameter segments with the collecting area equivalent to a 22.0 m single-mirror telescope \citep{mccarthy06}.  The E-ELT is a 40-m class with included adaptive optics \citep{hooketal09}. Both will be equipped with integral field units for spatially-resolved spectroscopy \citep{esoelt}. 

The James Webb Space Telescope (JWST) is expected to  be another milestone, providing a ten fold improvement in sensitivity and   in spatial resolution with respect to the Hubble Space Telescope \citep[HST, ][]{gardneretal06}. The JWST being optimized for optical and infrared, the UV domain (not accessible from ground) needs to be covered by a new generation of space based observatory. At present   WSO-UV and LUVOIR may have some chances to become operational in the 2020s, \citep{wsouv,franceetal17}. It is difficult to undermine the importance of these new developments. The present danger is that the UV range may not be any more accessible after the decommissioning of HST, a loss without precedent, since the UV has been accessible since the late 1978 after the launch of the International Ultraviolet explorer \citep{macchettopenston78}. The perspective is on the converse to have  instruments with unprecedented resolving power and sensitivity, able to yield a full map star cluster and dwarf galaxies in the local Universe (apart from the heavily-extinguished regions in the Galactic plane).    
 
The { Euclid}  space mission, designed to study the baryonic acoustic oscillations (BAOs) from the large scale distribution of galaxies and quasars (see \S \ref{obscosmo} for a brief discussion) is expected to yield millions of moderate-resolution quasar spectra over the visual wavelength range \citep{euclid}. The Euclid spatial resolution of 0.2 arcsec  is comparable to the Hubble Space Telescope and should provide an accurate  redshift   for  the majority of the new sources identified by future imaging observatories, from radio to X-rays, at  $z \lesssim 2$.  The Euclid output is especially welcome as  ground-based survey telescopes such as the Large Synoptic Survey Telescope (LSST) will make it possible to discover millions of quasar candidates. LSST telescope will have an 8.4 m (6.5 m effective) primary mirror, a 9.6 square-degree field of view, and a 3.2 Gigapixel camera. This system can image about 10,000 square degrees of sky in three clear nights using pairs of 15-second exposures twice per night, with typical depth for point sources of $r\approx  24.5$\ (AB) \citep{ivezic17}.  About 90\%\ of the LSST time will be devoted to a deep-wide-fast survey mode which will uniformly observe a 18,000 deg$^{2}$\ region about 800 times (summed over six photometric bands). For 10 years of operations, the stacked observations will yield a coadded map up a limiting magnitude  $r \lesssim  27.5$. These data will result in databases including 20 billion galaxies and a similar number of stars \citep{ivezic17}. Most of the candidate quasars identified by LSST will be too faint for spectroscopic studies with 4m class telescope, yielding only a handful of telescopes from which spectroscopy could be obtained.    The SDSS V should  offer the equivalent to LSST in terms of spectroscopic monitoring:  all-sky, multi-epoch spectroscopic survey of over six million objects \citep{kollmeieretal17}, able to reveal spectral changes on timescales from 20 minutes to 20 years. It will be based on two 2.5m telescopes that should yield a sky-survey rate of $\approx 40{^\circ}^{2}$ per hour, and  complement high-energy surveys in the soft- and hard-X domains (eROSITA and Athena, introduced below), at least for the brightest sources, $i \lesssim 20$. 

The SDSS V should  still leave  the faint end of the quasar luminosity function sampled by LSST. It may  however serve   as a testing experiment for dedicated, larger aperture telescopes. The need to cover spectroscopically  faint sources  ($i \gtrsim 21$) is already  felt \citep[see \S\ \ref{redfront}\ and e.g., the discussion by][]{sulenticetal14},  and will be even more felt in the next decades. It may become a major drive toward the development of more 30m class and even larger aperture telescopes. Apart from costs, a 100m telescope is already technologically feasible, and a phase I study for an ``Overwhelmingly Large Telescope'' was completed at ESO  \citep{esoowl}. A 100m telescope working  at   diffraction limit  of 1 milli-arcsec can reach $V \approx 37$. This means  the  ability to sample and study objects in the Universe at the epoch of reionization with high efficiency. We are confident that one or more telescopes of the $100$ m class will  be eventually built around the 2050s.

\subsubsection{X-ray/$\gamma$-ray} 
 
The obscured Universe  is an important part of extra-galactic studies. The word ``obscured'' refers here to the optical/UV  domain that is subject to extinction by interstellar dust.\footnote{{Dark is used to indicate an early cosmic age at $z \gtrsim 6$ when the Universe was still at least in part opaque to the Lyman continuum radiation i.e., before reionization epoch.} }  At present, astronomers looking at an all sky survey  in the soft-X-ray domain   have  to relay on the ROSAT All Sky Survey (RASS) carried out over 6 months between 1990 and 1991. The RASS  was tremendously successful   and detected more than 60,000 X-ray sources over the whole celestial sphere. There is no all-sky survey in the hard X-ray domain above 2keV presently available, even if archival observations with XMM and Chandra now include several hundred thousands of exposures.   New-generation instruments should be able to probe obscured sources (among them most quasars) in the next decades.  eROSITA (extended ROentgen Survey with an Imaging Telescope Array) is expected to perform a deep survey of the entire X-ray sky. In the soft X-ray band (0.5-2 keV) eROSITA is expected to be about 20 times more sensitive than the ROSAT all sky survey with a resolution of 15-30 arcsec with a field of view of 0.8 deg$^{2}$, while in the hard band (2-10 keV) it would provide the first ever true imaging survey of the sky at those energies  \citep{predheletal10}. Athena (over two orders of magnitude more powerful than current facilities)  should  also provide imaging capabilities in the 0.2-15 keV energy band over a  40 arcmin$^{2}$ field of view with angular resolution of 5 arcsec on-axis, simultaneously with spectrally and time-resolved photon counting \citep{rauetal17,guainazzi17}.  The NuSTAR (Nuclear Spectroscopic Telescope Array)  focuses light in the hard X-ray domain (3 - 79 keV) with an angular resolution of $\approx 1'$ (HPD) \citep{harrisonetal13}. It is conceivable that angular resolution will be improved in forthcoming missions.  

 { A powerful (albeit costly)  strategy is to detect the track of $\gamma $-rays (as well as cosmic rays) through dedicated arrays of telescopes sensitive to the  Cerenkov radiation emitted in the earth atmosphere. The  Cerenkov Telescope Array (CTA), based on this approach,  is expected to detect Cerenkov radiation from very  high-energy (VHE) cosmic rays and $\gamma$-rays. } The CTA, an array of $\sim 100$\ built-for-purpose optical telescopes, should be able to cover a huge range in photon energy from 20 GeV to 300 TeV, and be a factor $\sim 10^{3}$\ more sensitive on hour timescales than  the space-based  Fermi Large Area Telescope (LAT) at 30 GeV. The angular resolution of CTA will approach 1 arc-minute  allowing detailed imaging of a large number of  $\gamma$-ray sources.  CTA will be the first VHE observatory that will reach the angular resolution needed for easy  cross-identification of optical counterparts \citep{CTAconsortium17}. 

%The Nu Star telescope with unprecedented imaging capabilities ($\lesssim 10$ arcsec) has a   small field 

\subsubsection{FIR/mm/radio}

The Atacama Large Millimeter/submillimeter Array (ALMA)  is an aperture synthesis array with 66 radio telescopes for sub-/millimetre  astronomy. ALMA bands cover from 30 to 1000 GHz (300 $\mu$m), with typical resolutions of a few hundredths of arcsec \citep[][]{testiwalsh13}. The telescope has been operational since  2011, and it will remain operational in the next decades, providing an unprecedented view of the cold Universe. Some of the ongoing and potential research  will be  described in the next sections.  After the huge success of the FIR/mm space observatories (ISO, WISE, Spitzer, Herschel), the future might involve new telescopes for surveys such as WFIRST   or mainly dedicated for spectroscopy, for example SPICA (Space Infrared Telescope for Cosmology and Astrophysics) is targeted for launch in the late 2020s. With SPICA, the goals are to reveal   ``metal and dust enrichment through galaxy evolution" \citep{spinoglio16}  through low-to-high resolution capabilities in the MIR/FIR spectral range. Planned FIR observatories are single dish  although it is conceivable that an ALMA-like heterodyne interferometer could eventually become in part space-based. 

Currently, the Very Large Array-based FIRST \citep{beckeretal95} is the main radio survey of the Northern hemisphere carried out with a resolution of about $\approx 5$\ arcsec. Its catalog is overwhelmingly dominated by AGNs (at a detection limit of $\approx 1 $ mJy), and only a small fraction of  sources are star-forming galaxies. At radio wavelengths, advances in digital signal processing and cost reductions  are making it possible to build arrays composed of a large number of relatively small antenna elements, giving large fields of view with large collecting areas.  Most sources detectable by the next-generation surveys (which will reach sensitivities of the order of   $10 \mu$Jy) are expected to be star-forming galaxies, so that these surveys are dominated by the same galaxies that are studied by optical and infrared (IR) surveys \citep{padovani17}. The Square Kilometer Array (SKA) with its unprecedented collecting area and baseline, will offer improvements over the JVLA, LOFAR, for resolving power (5 milliarcsec at 6cm for the widest array configuration, 3000 \kms), sensitivity and survey speed due to its relatively large field of view.   

The RadioAstron telescope is an array of ground based telescopes with an    antenna of 10m diameter  in space (with a baseline of 350,000 km) that   achieved the unprecedented spatial resolution of $\approx 35 \mu$arcsec at 6cm. RadioAstron is  a successful effort that demonstrated the feasibility of VLBI with space antenn\ae. It is reasonable to think that by 2050 more sensitive arrays with more antenn\ae\ in space will be operational.  Space based radio-antenn\ae\ obviously yields an extension of the baseline, but can also take advantage of a much lower background noise that allows for a much wider dynamic range (especially helpful when trying to map a boosted jet and much fainter unboosted, extended features, a typical condition in extragalactic radio sources). 

{A poorly sampled radio frequency domain corresponds to the meter wavelength domain. There are several radio telescopes operating at low frequency, among them the   Giant Metrewave Radio Telescope \citep{anathakrishnan05} that covers  several bands between 50 and 1420 MHz, and the  Murchison Widefield Array that covers the frequency range 80-300 MHz. The GMRT completed a survey at 150 MHz with resolution $\sim$ 25 arcsec and noise level $\lesssim 5$mJy \citep{intemaetal17}.  The VLA Low-Frequency Sky Survey Redux (VLSSr) provides a Northern view at 74 MHz   with resolution about 80" and sensitivity 0.1 Jy rms \citep{cohenetal07,laneetal12}. With LOFAR and  SKA   low-frequency arrays of unprecedented    sensitivity and resolution will  become operational ($\approx 0.1 $ arcsec at 100 MHz with SKA2 and $\approx$ 10 arcsec for LOFAR). LOFAR exploits an array of simple omni-directional antennas in place of a dish antenna \citep{vanhaarlemetal13}. A space interferometer would allow to extend the frequency range down to about 0.3 MHz, a frequency domain where earth's ionosphere opacity makes grand-based observations unfeasible \citep{basartetal97}.

%These arrays enable comprehensive sky surveys with implications for cosmology, for the study of transients, and for charting the evolution of galaxies and clusters over cosmic timescales.  SKA -- Low and high Frequency precursors.     and its pathfinders (most notably LOFAR operating from 15 to 240 MHz at resolution 30 to 2 arcsec for a 100 km aperture)

\subsection{Multimessenger}
 
\subsubsection{Gravitational waves}
\label{gravwaves}

%\textcolor{magenta}{The attempts a detecting gravitational waves through massive, resonant bars have a long history, but even the second generation  cryogenic Weber bars are just sensitive enough to detect only extremely powerful gravitational waves. At the time of writing, no detection of gravitational waves by  Weber bars has been  reported. Gravitational waves induce changes in the distance between  objects induced perturbing the metric of the spacetime.  }

Since spatial changes induced by gravitational waves occur with opposing sign for orthogonal directions, a Michelson  interferometer has a well-suited geometry to maximize the tiny effect on the detector. Laser interferometry is employed for motion sensing \citep{houghrowan05}. The next generation of ground-based gravitational observatories   will be more than ten times more sensitive than Advanced LIGO that made possible the first detection of gravitational waves ever. Several laser-interferometric telescopes are under development or planned. Of them, the third-generation Einstein telescope in Germany   has the potential to dramatically increase the detection rate of gravitational wave sources. The current design posits two independent interferometers located underground with vertices separates by 10 km in a triangular configuration. With this baseline the interferometer may detect merging of intermediate mass black holes, below the smallest masses that are found in the nearest AGN, $\sim 10^6$ M$_\odot$. Some of them may however be detected in the nuclei of non-active galaxies and be precursors of more massive BHs detected as AGN.  LISA is expected to adopt a triangular configuration in space but the separation is planned to be 2.5 million km, making the instrument unique to cover a low-frequency domain not accessible from the ground (Figure \ref{fig:bh}). It will be necessary to wait even beyond the first generation of space-borne observatories such as LISA  to detect  merging super massive  black holes (SMBH, $\gtrsim 10^{9}$ M$_{\odot}$, see   Figure \ref{fig:bh}) . 

{A complementary technique to detect low-frequency gravitational waves is to consider an array of millisecond pulsars and measure the pulse arrival times. Differences in arrival times (the timing residuals) should be correlated if produced by gravitational waves \citep{hobbsetal10}. Pulsar timing observations are used to place constraints on the rate of coalescence of supermassive black-hole (SMBH) binaries as a function of mass and redshift  from the GW background in the nHz domain \citep{wenetal11}. At present the very existence of SMBH binaries is established in only a handful of cases. Larger pulsar timing arrays (i.e., involving a larger number of pulsar and a long temporal baseline will allow for the detection of a SMBH binary in a nearby galaxy, provided that the gravitational wave background can be adequately subtracted \citep{mingarellietal17}.} 

\begin{figure}[h!]
\vspace{0cm}
	\begin{center}
		\includegraphics[width=16cm, angle=0]{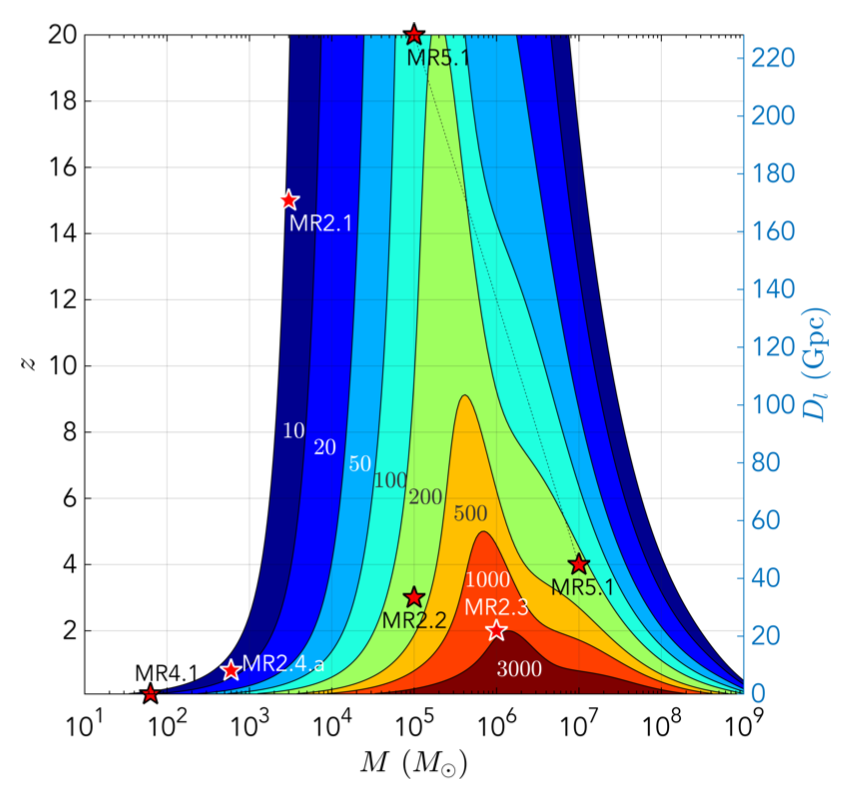}
			 
	 \end{center}\vspace{0cm}
	\caption{{S/N levels as a function or redshift (left scale) and luminosity distance (right scale) and of total source frame mass for the baseline configuration of LISA, for a fixed mass ratio of $q=0.2$. The stars identify  threshold cases to define mission requirements. From \citet{danzmann17}. \label{fig:bh}} }
\end{figure}

\subsubsection{Cosmic rays}

The CTA will detect Cerenkov  photons emitted by the cosmic rays before their first interaction with the atmosphere.  While the CTA will remain state-of-the-art in the next decades for atmospheric Cerenkov CR detection, it is expected that other type of CR detectors may undergo significant developments. For instance, a large collecting area can be obtained exploiting  Cerenkov radiation emitted in dense media. In this case, the detection of  Cerenkov flashes is achieved  by using photomultipliers submerged in water tanks over a large surface area, as in the case of ``HAWC" \citep{carraminanaetal17}. Extensive air shower (EAS) arrays presently  employ a variety of detecting techniques, according to the energies of the air shower particles that may include,  drift chambers, scintillators and Geiger tubes, and other devices. The Auger Observatory  employs a hybrid technique monitoring the sky  with a fast UV-sensitive camera able to record the track of fluorescent emission (a near-UV line of sodium) associated with the interaction between the atmosphere and  particle showers induced by CRs, in addition to water tanks for monitoring Cerenkov flashes. 

Cosmic ray detection from extra-galactic sources has been a serious problems in the past decades. Low energy ($\lesssim 1$ GeV)\ cosmic rays are frequent (1 event/s/m$^2$) and their detection   straightforward with fog chambers. The  cosmic ray energy spectrum \citep{beringeretal12} is a power law that shows a steepening  (a ``knee") at energy $\sim 1 $PeV, and then a flattening (an ``ankle) at  $\sim 1 $EeV (in the domain of ultra-high energy cosmic rays, UHECRs).  Acceleration from supernov\ae\ to energies $\sim 1 $EeV is ruled out. Around and above this energy  the UHECR energy distribution is believed to be dominated by  extra-galactic CRs. Since the spectral energy distribution of CRs is a steep power law,   events due to extra-galactic CRs are exceedingly rare (at energies of the order of  $\sim 1 $PeV, the expected flux is just 1 event/yr/m$^2$).  A major driver in the development of detectors with large collecting areas is therefore the  nature of the CR with energies $\gtrsim$ 1 PeV whose acceleration mechanism  is unclear but believed to be at the heart of the inner working of radio jet acceleration. 
}  
 %Only the highest energy CRs are probably of extra-galactic origin \citep{}.
%Plans have been made at carrying the telescopes monitoring the UV flourescent track into space.  

\subsubsection{Neutrinos}

{Neutrinos offer a unique diagnostic of extremely high energy processes, and they can, unlike cosmic rays, travel unimpeded across the magnetic field of the Galaxy. Given the weak interaction of neutrinos with matter,  large masses are needed to reveal neutrinos. Mechanisms for the production of high energy neutrinos will also produce $\gamma$\ rays of similar energies. $\gamma$-ray telescopes such as CTA are expected to achieve   precision pointing and sensitivity to identify populations of accelerators. 

A common type of design  of a neutrino observatory involves an array of photomultiplier tubes housed in transparent containers which are suspended within a large tank of pure water or ice (or other suitable materials) and aimed at the detecting the Cerenkov radiation due to leptons (typically muons) or to other  decay products induced by the interaction with the neutrinos. Over the years, neutrino observatories have employed larger and larger volumes placed underground, to improve the detection rate and lower the background. } The IceCube Neutrino Observatory is a  cubic-kilometer detector that uses ice as a medium   which  detects Cerenkov radiation through an array of photomultipliers. At present, IceCube has detected   $\lesssim$ 100 PeV neutrinos of astrophysical origin.  The PeV  detections of IceCube might be substantially increased by a second-generation observatory, IceCube-Gen2 which should be based on  a 10 km$^{3}$ volume of  ice at the South Pole \citep{icecubegen214}. The large collecting volumes are needed in order to detect extra-galactic neutrinos in the PeV energy domain. They are extremely rare and  likely the ones produced in active nuclei. The neutrinos detected using reconstructed muon tracks, are   unambiguous tracers of hadronic acceleration, up to high redshifts and beyond PeV energies.  This means that at present, data rest  on several tens of neutrino detections, and that larger volumes are needed to have a good statistics of UHE neutrinos to make them  of relevance for the study of relativistic jet formation physics (Sect. \ref{sec:vhe}). A high-energy neutrino background has been revealed by IceCube since 2013 \citep{icecubeetal13}, and blazars have been suggested as a likely source \citep{padovanietal16}. Very recently,  IceCube detected an $\approx$ 290 TeV neutrino for which follow up $\gamma$-ray observations have made possible, for the first time, the identification with an electromagnetic counterpart,  the flaring $\gamma$-ray blazar TXS 0506+056 \citep{icecubeetal18}.

\subsection{Computing power and software development}

 {Increase in computing power cannot follow the Moore’s law forever. The Moore’s law posits that the surface density of transistors on integrated circuits doubles approximately every two years. However, the limiting size of the technology is about 2-3 nm (down from about 14 of today, which may still implying a more than tenfold increase in computing speed), and may be reached around 2025. At present it is not clear what may follow. The clock speed  saturated at about 3GHz because faster speeds produce too much heating \citep{evans14}. The increase in speed  in the last years has been achieved by building  multicore processors.  200 petaflops can be obtained in supercomputers that house a large number of cores, and computers reaching the 1000 petaflops will be build toward the 2030s. These computers are the best hopes for modeling of chaotic astrophysical environments. 
Quantum computing is still in infancy, and it is not yet clear whether an all-purpose mainframe can be built. However, the  potential for contribution  to computational astrophysics in the next decades is enormous, even epoch-making \citep{cross16}.}

\subsection{Artificial intelligence for data-driven science} 

{Large surveys and simulations will provide data in the order of the petabyte; SKA is expected to store more than 1 petabyte of data per day. These ``big data'' will require ad hoc solutions for storage and data processing and retrieval.  The radio interferometric data processing should become especially data intensive, as instrumental configurations such as the one of LOFAR demand special approaches to compensate for the individual antennae lack of directionality. Numerical models of increasing complexity may involve the accretion disk structure as well as the interaction between the accretion disk and other compact objects orbiting the same black hole. Magneto-hydrodynamics simulations in general relativity are needed to provide realistic model of the multi-phase quasar winds at the origin of feedback effects. Large scale cosmological simulations like the Illustris \citep[][]{springeletal05} still fail to reproduce the diversity in galaxy properties, and treat central BH in a primitive way. There is a large latitude of improvement for SPH codes that exploit enhanced computing power. }

Major surveys are expected to  produce huge amounts of reduced data as well as public-domain Virtual Observatory (VO) compliant catalogues of measurements.
The analysis of large amounts of data provided by instruments from the new generation of telescopes and numerical experiments is expected to become increasingly cumbersome for human researchers. Ultimately, neural networks and other forms of artificial intelligence may become a necessity to manage the sheer amount of data. Deep machine learning has been considered for applications to astronomy since the late 1980s. For instance,  a convolutional neural network   is a class of deep, feed-forward artificial neural networks that has successfully been applied to analyzing visual imagery. Deep neural networks are being exploited for a host of problems associated with visual morphological classification, a frequent necessity in astronomy.  For instance, they proved to be very effective in evaluating galaxy morphology, and extracting morphological parameters such as Sersic  index and isophotal magnitudes  \citep{tuccilloetal18}, as well as in the classification of radio morphology \citep{aniyanthorat17}.  

\section{Expected progresses on selected, present-day scientific themes}
\label{sec:prog}

Summing up the previous discussion of instrumental capabilities, we can say that we can expect the ability to probe much deeper than today with planned instrumentation. Observations with active optics may become commonplace with the largest telescopes, yielding an overall improvement of a factor $\sim$10\ in resolution for ``every night" observations in the optical and NIR domains. The largest telescopes are expected to reach magnitudes $\sim$30, with a more-than-tenfold improvement with respect of today. In ranges where spatial resolution has been poor (low-frequency radio, hard X-ray, $\gamma$-ray)   improvements are expected to bring the resolution $\sim 1'$. This means that, for example,   bright optical extra-galactic sources could be  unambiguously identified with their $\gamma$\ counterparts   in the wide majority of cases.  We can now focus in more detail on the possibilities that the new instrumental capabilities will offer in the study of galaxies and quasars, and nuclear activity in general.

\subsection{Galaxies: studies in the next 30 years}

Imagining the future of our understanding of galaxies is not an easy task, in particular for   researchers  who formed their background of astrophysics during the epoch of transition from photographic plates to CCDs, when the only big telescope was the 5m Hale in Palomar and few space missions were already lunched (e.g. IUE, Uhuru, Ariel 5) and ROSAT as well as HST were still to come. At that time radio observations already revealed the spiral structure of our  Galaxy and the first radio sources were identified with optical counterparts. This is the epoch when computers started the first data reduction of astronomical images and spectra and the most-widely used compiler was Fortran 77. Galaxies were singularly studied through deep CCD images or photographic plates (micro-densitometers were still largely used) with optical filters (B and V Johnson) or long slit spectra, and the first numerical models started to appear in the astronomical literature.

The  progress in all fields of astronomy has been so far-reaching during these 30 years that a  single researcher could not be up-to-date of the whole literature with the exception of his/her specific interests. When we think that only 100 years ago the humankind was not aware of the existence of galaxies, we are legitimate to feel very happy of being part of such a fantastic development of our field. 

Coming back to the theme of this review, the first step to imagining the future of the study of galaxies is to  keep in mind first why we study galaxies. The second step is to make a list of the hottest research topics in this field. The first item is necessary because it acts as the helm of a ship. Remembering why we study galaxies it is important to stay on course, to follow the aims of our projects. The second item is fundamental because we plan today our future researches and this implies to know many things, last but not least how much they cost in terms of economical resources.

Why do we study galaxies? Galaxies are the largest gravitational bound systems where stars are ``organized" to trace the baryonic matter in the Universe. If our aim is to reconstruct the history of the Universe, we must  understand how such organization of stars in galaxies changed during the cosmic epochs. This means to examine how stars are distributed at all spatial scales, how and when galaxies and stars formed and in which way the population of stars evolved. With such motivation, we can easily predict that the focus of our future researches will be to understand the origin of galaxies. This implies that most of the efforts will be dedicated to the studies of high redshift objects.

When dis galaxies emerge from the dark era? Which kind of stars formed first? How long was the epoch of re-ionization of the Universe and what kind of sources contributed to it? Were the first galaxies similar in structure and shape to those we see at low $z$? What was the contribution of merging in shaping the galaxies observed at recent cosmic epochs? How did the chemical elements produced by stars contaminate the galactic medium that formed  new stars enriching it of metals? What does trigger the star formation (SF) and in which way this phenomenon evolved in the various types of galaxies? These are only a few examples of  questions that still have only an approximate answer. They acts as a beacon in the night and they will likely remain fundamental questions for many years to come.

{Up to now coordinated efforts exploiting multi-wavelength observations have permitted to trace a preliminary picture of the evolution of the star formation in galaxies. In particular we have measured the star formation  rate density (SFRD). This is the rate at which stars formed within galaxies in comoving volumes of the Universe. 
Most of the merits of such achievement can be attributed to space missions like HST, Spitzer, Herschel and Galex. The surveys carried out with these telescopes permitted to acquire a large database of galaxies observed at different cosmic epochs. We should not forget however the important contribution of the optical surveys at smaller redshifts, like e.g. the SDSS. The SDSS sample contain $\sim 300,000$ galaxies brighter than $L^*$, the turning point of the galaxy luminosity function, within a volume of $\approx 0.05 h^{-3}$ Gpc$^3$\ \citep{belletal09}.} 

{\cite{madaudickinson14} provided a comprehensive review of  the history of star formation across  cosmic epochs. Their most famous figure is reproduced here in Fig. \ref{figMadau}. The data indicate that the SFRD peaked at $z\sim2$, i.e. approximately 3.5 Gyrs after the Big Bang and that nearly half of the stellar mass observed today was already in place at $z\sim1.3$. The Universe was very active in forming stars in the past, with a rate much larger than what we measure today, while at the present epoch the SFR is very low.}

{The observed trend and the well known $SFR-M^*$ relation between SFR and stellar mass, suggest that such behavior is due to a balance between the accretion of gas and the feedback effects from SNe and active galactic nuclei. It seems that the stochastic events provided by mergers do not play a significant role. This might appear in contrast with the widely accepted view that we live in a hierarchical Universe, where structures form by subsequent merging events, but it is clearly a symptom of our still poor knowledge of the processes that occurred in galaxies at different cosmic epochs. The most recent numerical simulations running in the cosmological framework are in fact able to reproduce many observed features of present day galaxies (see below).}

{In this respect the recent work by \cite{Choisetal2017} provides one possible theoretical explanation of the trend seen in Fig.\ref{figMadau} which is due to a combination of effects: the number of galaxies of a given mass at each redshift (regulated by the hierarchical structure of our Universe), and the time-delayed star formation of galaxies. The star formation in galaxies starts at a low rate, grows to a maximum and then declines. Within the formalism of the infall models, the gas is gradually converted into stars ($\dot{M_s} \propto (t/\tau) exp(-t/\tau)$). This kind of behavior is different for galaxies of different masses. While massive objects have a rapid increase to a maximum SFR followed by a prolonged decline, smaller systems have a slow progressive increase of the SFR followed by episodic events. In their models the large feedback effects from AGN (\S \ref{feedhost}) do not seem to be relevant. This is in contrast with the conclusions of \cite{Tescarietal2014} and \cite{Katsianisetal2017} who found that the key factor for reproducing the SFRD is the combination of feedback effects.} {Interestingly, a very similar trend is observed when we look at the growing history of the super-massive black holes (Fig. \ref{figQSOgrowth}), suggesting a close connection between the two physical phenomena \citep[see e.g][]{Heckman&Best2014}.

\begin{figure}[h!]
\vspace{0cm}
	\begin{center}
		\includegraphics[width=16cm, angle=0]{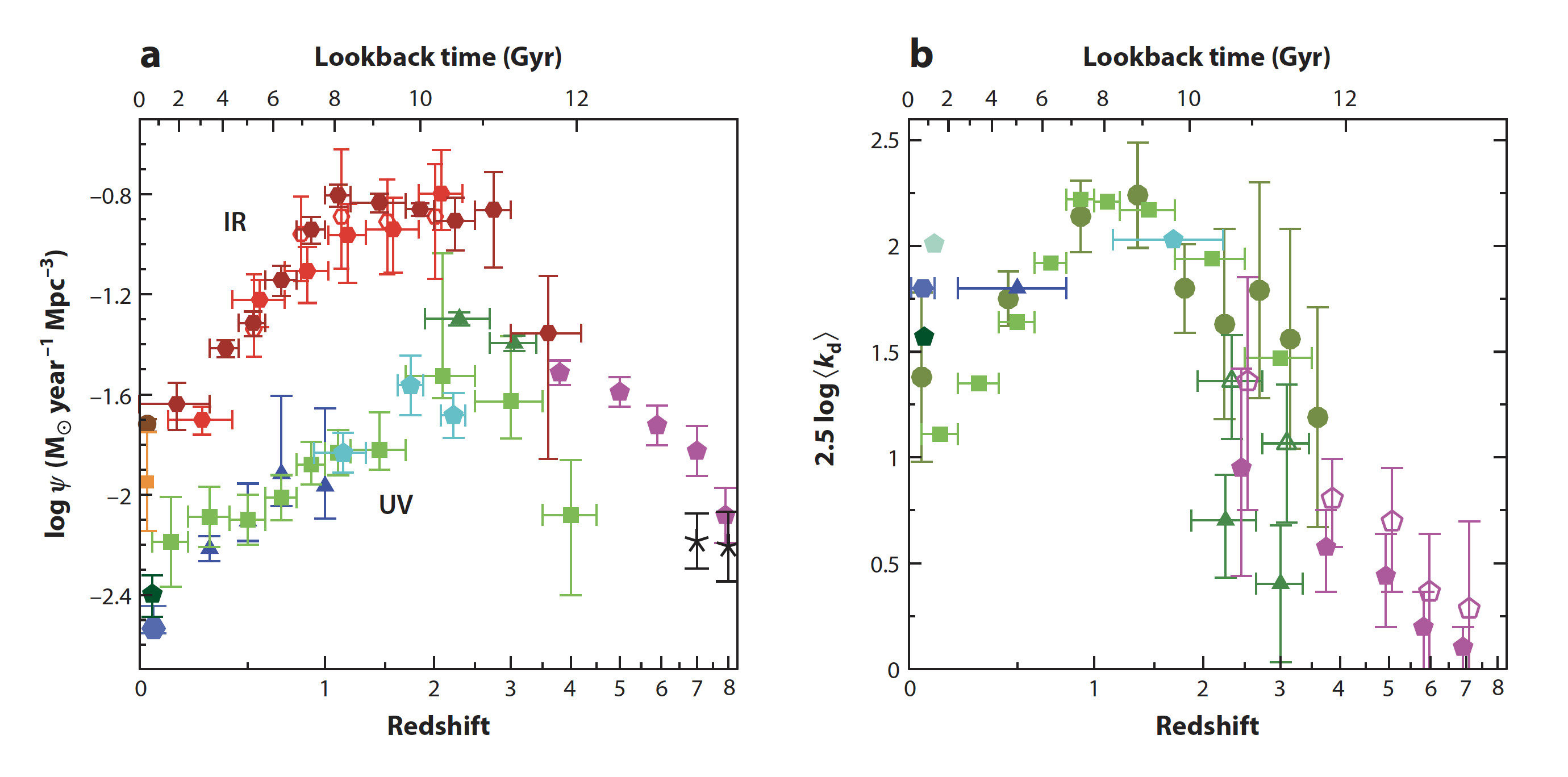}							\end{center}\vspace{0cm}
	\caption{The observed star formation rate density in galaxies at different redshifts (Credit to \cite{madaudickinson14}, their Fig.8). \label{figMadau} }
\end{figure}

\begin{figure}[h!]
\vspace{0cm}
	\begin{center}
		\includegraphics[width=16cm, angle=0]{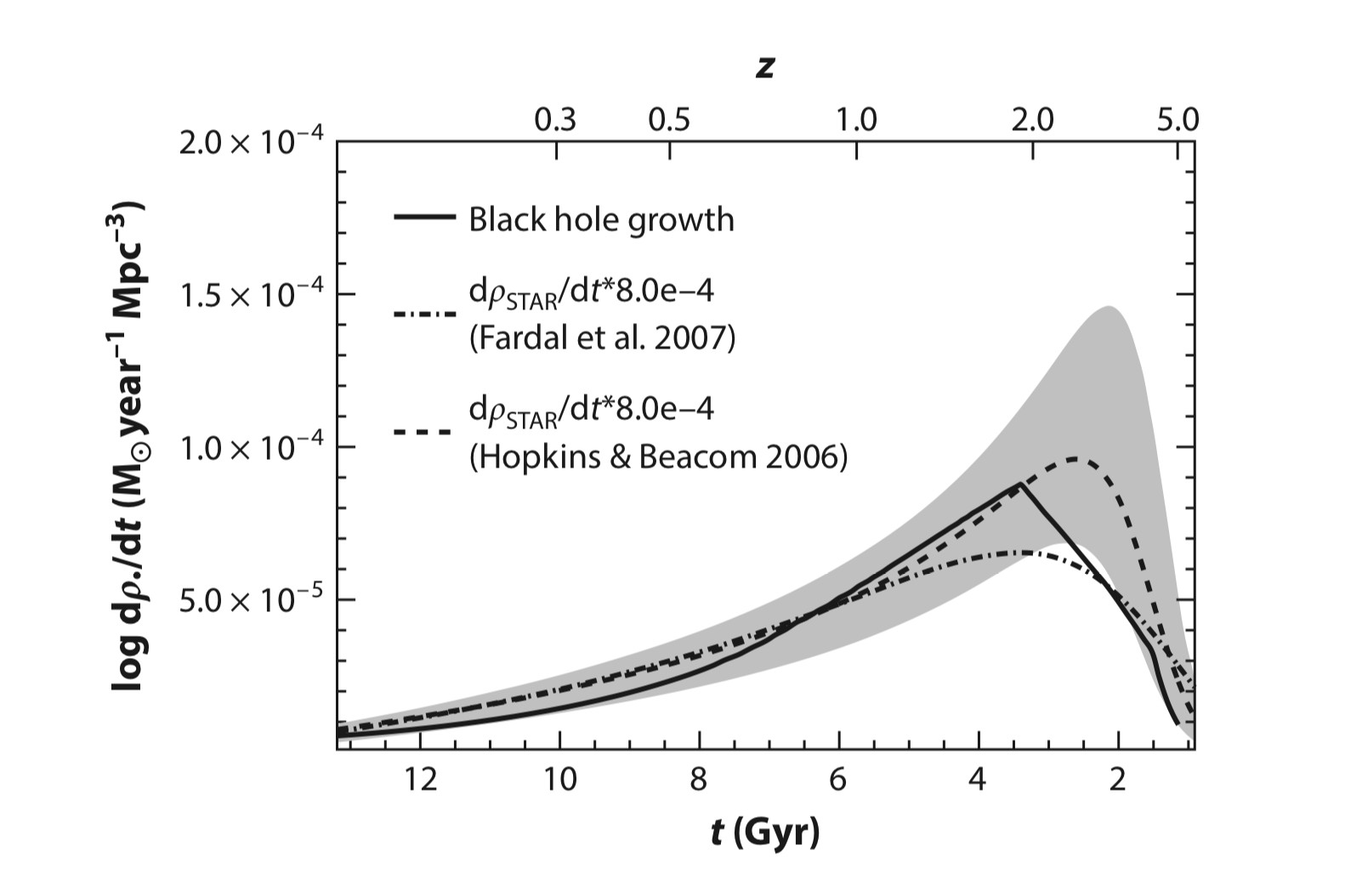}							\end{center}\vspace{0cm}
	\caption{The history of the BH growth compared with that of stellar mass growth (Credit to \citet{Heckman&Best2014}, their Fig.1). \label{figQSOgrowth} }
\end{figure}

{The idea about the formation and evolution of galaxies is that stellar systems grow primarily for the accretion of gas from the cosmic web. Major mergers of gas-rich systems happen and provide strong bursts of star formation, but do not seem to contribute to the bulk of star formation (they might do so for $\sim10\%$). This seems consistent with numerical simulations and with observations of the star-forming galaxies 
\citep[see e.g.][]{Dekeletal2009,Wuytsetal2011,Lillyetal2013}. In an analogous way, the accretion of SMBHs must be related to the growth of the inner region of galaxies. The efficiency of the two processes are different, but both rates reflect the way in which gas is accreted and transformed.}

Clearly the possibility of making significant progresses in this area is closely linked to our ability of planning new powerful telescopes for the ground and space. 
Among the various projects that will have a big impact on the next 20 years we should mention the JWST and the ELT whose construction has started in the Chilean desert. The groud-based telescopes will in fact largely benefit of the progresses made in adaptive optic systems, like e.g. MAORY.  

%We put the quotation marks around telescopes to mean all systems able to gather photons from the whole EM spectrum, but today after the discovery of gravitational waves we should to include a new class of telescopes, i.e. those sensitive to the waves emitted by accelerated masses. The realization of this largely depends on many factors such as the technological advance and the economic budget available, since the cost of such big enterprises cannot be afforded by single nations, as it has been demonstrated by the LIGO and Virgo consortium.

In the next decade high redshift observations would likely permit to formulate a coherent picture of galaxy evolution linking the data available for the different cosmic epochs. What is important is to establish which physical processes play the major effects and are responsible of the major transformation observed in size, morphology and stellar population content. 

By studying the earliest galaxies JWST and ELT will contribute to understand how galaxies grow and evolve. These telescopes will gather data on the types of stars that populated the very early objects. The spectroscopic follow-up observations will help to clarify how elements heavier than hydrogen formed and built galaxies through the cosmic ages. These studies would also contribute to understand the role of merging among galaxies and to have a much better knowledge of the mechanisms of feedback from supernovae (SNe) and  AGN.

\begin{figure}[h!]
\vspace{0cm}
	\begin{center}
		\includegraphics[width=16cm, angle=0]{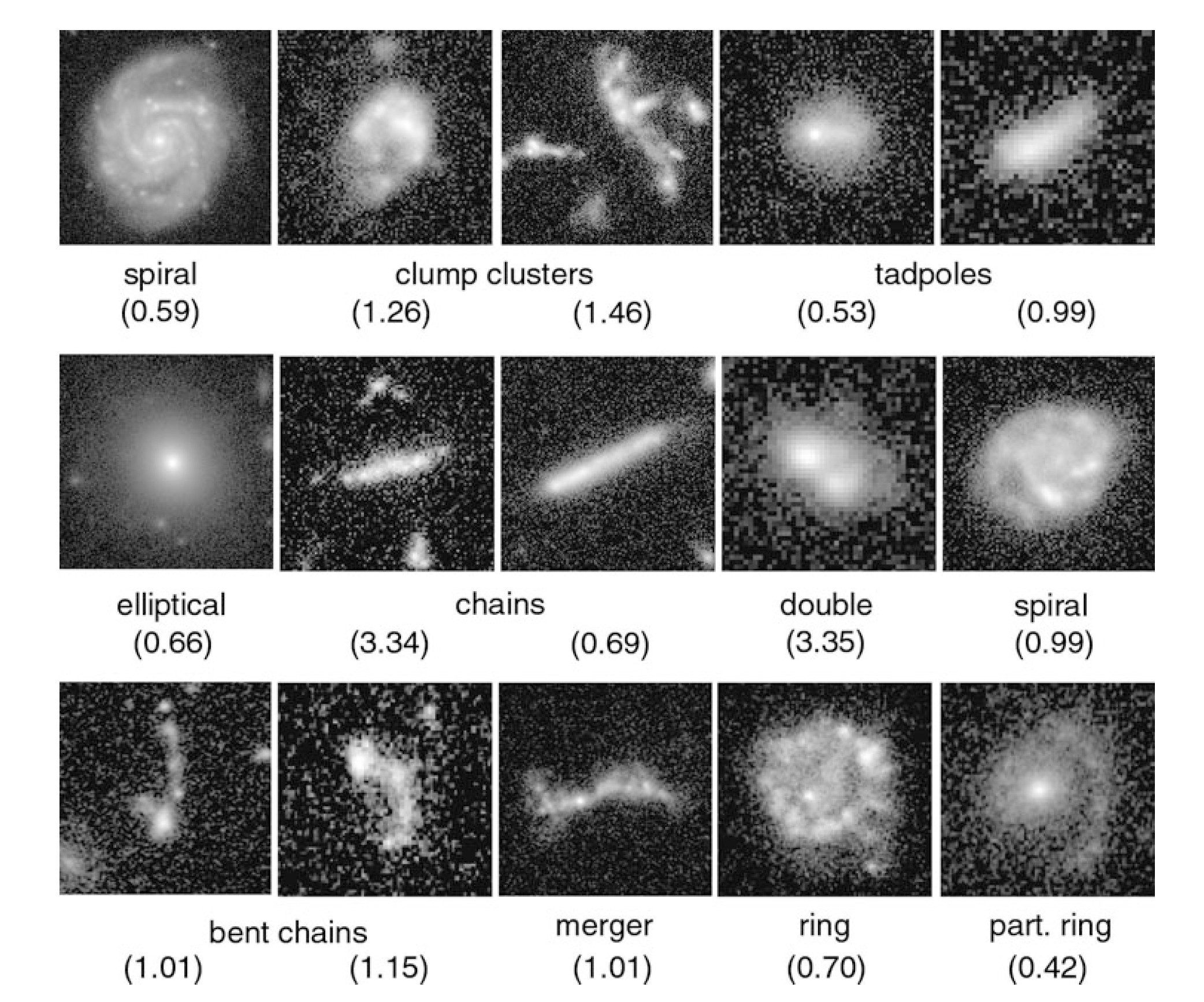}							\end{center}\vspace{0cm}
	\caption{Morphology of galaxies at intermediate and high redshift. Redshift is given by numbers in parentheses. (Credit to \cite{Buta2013}, his Fig.1-46, from Planet and Stellar Systems). \label{FigHighzobj} }
\end{figure}

The AGN feedback has characterized the evolution of galaxies in particular during the epoch of major interest for the stellar population evolution in galaxies at $z \sim 2 - 3$. The galaxies we see today around us have already experienced the turbulent period that characterize the first structures. Their SF is almost null or very small and the enormous feedback effects are no more present. So, the Universe we are going to discover in the new decades contains galaxies at their infancy and we know that this is a turbulent period not only for galaxies. The new generation of astronomers will face and will be called to explain all the processes that galaxies have experienced during their lifetime.

A further step forward requires the analysis of data coming from large optical and spectroscopic near infrared surveys. In particular it will be extremely important to address the nature of galaxies fainter than $L^*$ at the redshift were these objects assembled, i.e. around $z=2-3$ (where we observe the peak of star formation density and black hole accretion activity) \citep{madaudickinson14}.

The first galaxies with redshift larger than 1 were discovered with a color selection technique (see \citealt{steideletal99}) and called Lyman Break Galaxies (LBGs). These are galaxies with moderate masses ($10^9-10^{10}$ M$_{\odot}$) and metallicities ($\sim0.3$ solar), characterized by a rapid SF and significant outflows from SN winds. Later on a class of distant red galaxies (DRGs) was discovered (see \citealt{franxetal03}). These objects cannot be isolated through the color selection technique. They are
generally more massive and red than LBGs. Some of them suffer a conspicuous obscuration by dust. This the class of high redshift galaxies increased in number since new telescopes were able to detect more objects at $z=2-3$. We have now a zoo of galaxy types (see e.g. Fig.\ref{FigHighzobj}). This clearly reflects that we are at the beginning of a new era, very similar to that of 100 years ago, when galaxies were discovered as isolated ``island universes". A lot of work is therefore needed to assemble a coherent picture, to classify the galaxy types and to understand the transformations that have produced the objects we see today.

{The morphological transformation of galaxies is generally accompanied by an evolution in size of stellar systems. Galaxy size is typically measured through the effective radius $R_e$, the radius of the circle enclosing half the total luminosity. Observations have shown that both local early-type and late-type galaxies follow a direct relation between size and stellar mass $M^*$ 
\citep[see e.g.][]{Shenetal2003,Valentinuzzietal2010,Cappellarietal2013,Lietal2018}. The distribution of galaxies in the mass-size plane appreciably vary with velocity dispersion, age, metallicity and stellar mass-to-light ratio \citep{Cappellarietal2013,Langeetal2015}. This behavior confirms that there is a fine-tuning between the stellar population that each galaxy forms and its structural shape and dynamics. The same mechanism is likely at the origin of the Fundamental Plane, the scaling relation involving size, effective surface brightness and velocity dispersion 
\citep[see e.g.][]{Donofrioetal2017}. The existence of a fine-tuning implies that the dynamical condition of a galaxy have an active role in determining the amount of star formation of each system and the epoch of its occurrence. This is surprising being the SF a local phenomenon triggered by the local density of gas and dust.}

{At high redshift the observed quiescent early-type galaxies have more compact sizes (by a factor of 3–5)  than local objects
\citep{Daddietal2005,Trujilloetal2006,vanDokkumetal2008,Straatmanetal2015,Bellietal2017}. The kinematical studies measured increasingly larger stellar velocity dispersions and smaller sizes up to redshifts $z\sim2$ 
\citep{Newmanetal2015,Bellietal2017}.}

{The questions posed by these observations are therefore: how massive and compact systems could be already in place at early cosmic epochs in a hierarchical Universe where the large structures are the last to form, and what are their progenitors? What processes drive the evolution of these systems and the quenching of star formation? All these questions will find in the future planned sky surveys from ground and space, with billion of new data, the possibility of being answered.}

Before the consolidation of the so-called "precision cosmology" that is today represented by the $\Lambda$-CDM model, according to which the Universe consists of 70\% of dark energy (DE), 25\%\ of dark matter (DM) and 5\%\ of ordinary matter, galaxies were the main objects that enclosed the information on the cosmological parameters. Constraints on $\Omega$ and $H_0$ were often derived by studying galaxies at different redshifts, determining their distances, metal abundances, masses and ages. Today the works on the cosmic microwave background (CMB), the SNe as distance indicators, the Sunayen-Zeldovic effect on galaxy clusters, the gravitational lensing, the quasars and the large scale structure of the Universe have largely eclipsed the importance of the stellar population studies at high redshifts for the measurement of the cosmlogical parameters. However, once galaxies will fill the gap due to the lack of a coherent picture of their stellar populations at redshift $z=2-3$, they could contribute very important information. We should remember that stellar evolution is a natural clock that we can use to measure the age of systems in the Universe. So when we will have a much clear view of the stellar populations that dominate at large redshifts, we could be in the position of linking descendants and progenitors simply on the base of the prediction of stellar evolution.

A still open problem is that we do not know the nature of DM, although we can constrain its distribution through dynamical studies of clusters and satellite systems, working on the intergalactic absorption visible in the spectra of high redshift objects, and studying gravitational lenses. 

{These are particularly promising since future large scale imaging surveys will likely increase the number of strong lensing candidates. These objects are difficult to find, but great results are expected from the automated search methods in rapid development 
\citep[see e.g.][]{Alard2007,Seidel&Bartelmann2007,Bometal2017,Gavazzietal2014,Josephetal2014}. A strong gravitational lens occurs when a distant galaxy or quasar is aligned with a foreground galaxy or cluster of galaxies whose gravitational field might create multiple, highly distorted images of the background object. Strong lensing could also magnify the brightness of a source. 
General relativity has permitted a profound understanding of the lensing phenomenon enabling their use for the study of the dark matter distribution around galaxies and clusters 
\citep[see e.g.][]{Koopmans&Treu2003,Waythetal2005,Vegetti&Koopmans2009,Tessoreetal2016}. Lenses have also been used to measure the Hubble constant and the cosmological parameters 
\citep[e.g.][]{Blandford&Narayan1992,Wittetal2000,Suyuetal2013,Treu&Marshall2016}. This can be better accomplished when the constraints from strong lenses are coupled with those coming from weak lenses 
\citep[see e.g.][]{Bartelmannetal2002}. This definitely more important at larger scales. The weak lens from galaxy clusters might distort the shape of background galaxies and affect their apparent brightness. This phenomenon may then permit the determination of the primordial power spectrum of the DM distribution through measurements of the shear on large angular scales \citep[see e.g.][]{Maturietal2011}.} 

{The SKA,  LSST and the Euclid space telescope will likely increase the number of lenses by orders of magnitude 
\citep[see e.g.][]{Oguri&Marshall2010,Collett2015}. Estimates give $\sim200,000$ observable galaxy-galaxy lenses from Euclid. A big jump considering that up to now less than a thousand lenses have been discovered across many heterogeneous surveys.} 

{ Today, despite the past efforts, there is still a large uncertainty latitude concerning the mass of the DM halos \citep[see the contribution of P. Kroupa in ][]{donofrioetal16a}, especially for the smallest galaxies whose mass merges with the mass distribution of globular clusters; in such cases, DM halos might be completely absent \citep{vandokkumetal18}.}

The situation is promising in particular because astronomers have learned how to model the behavior of the DM component with N-body simulations on large computers. Unfortunately, the behavior of the baryonic component is complex. This will be a key question for the future. The problem is the large dynamic range of the baryon interactions, from the scale of stars to that of galaxies. What we know is that baryons collapse in the DM halos forming the first stars and that the gas often feed large super-massive BH at the center of galaxies, originating enormous feedback effects in terms of energy and matter moved all across the galaxy body. 

The future astrometric mission Theia is aimed at probing the dark matter distribution in galaxies and the power spectrum of density perturbations. Theia could permit a detailed study of the shape of the dark matter profiles (core or cuspy) that are known to depend on different processes
induced by the baryon physics, such as star formation, self-interaction, BH growth, etc. \citep{Readetal2016}. At the same time the sub-micro-arcsec astrometry of Theia should permit for the first time high-precision proper motion measurements (even of nearby galaxies) that could remove the degeneracy between radial dark matter profile and orbital anisotropy, clarifying the nature of DM particles.

The standard hierarchical model of galaxy formation and evolution has permitted up to now to follow the evolution of the cosmic structures, to observe the creation of the first galaxies up to the appearance of the galaxies we see today. The ``Illustris" project is one of these large-scale cosmological simulations \citep[see e.g.][]{springeletal05,springeletal05a,Vogelsbergeretal2014,Sijackietal2015}. The model tracks the expansion of the Universe, the effects of gravity and the hydrodynamics of the gas, as well as the formation of stars and black holes. 
The simulation starts from very initial cosmological conditions and maps the evolution of a big volume of the Universe up the present epoch. A wide range of masses, rates of star formation, shapes, sizes of the galaxies we see today are reproduced.
 
Despite the success of these simulations a number of severe problems still affect   galaxies and the structures that
are reproduced. One is that it is difficult to form realistic disk structures (the so-called angular momentum catastrophe). Another one is that the amount of stars that can be predicted with a simple physical receipt is largely overabundant with respect to what observations tell us. There are today a number of tensions between theory and observation that will likely characterize the future epoch. Probably many of these are due to the fact that we do not know so well the complex physics of baryons.

In any case the enormous growth of numerical simulations will likely characterize the years to come. At present we are still testing the power of simulations in representing reality. The complexity of the problem does not permit to numerical calculations to fully capture the correct answer across all scales of space and time. The finite resolution, i.e. the size of the smallest details that can be reproduced implies that some processes, such as the birth of individual stars, cannot be followed by cosmological simulations. As a consequence many physical approximations are necessary to accomplish the whole simulation.

\begin{figure}[h!]
\vspace{0cm}
	\begin{center}
		\includegraphics[width=8cm, angle=0]{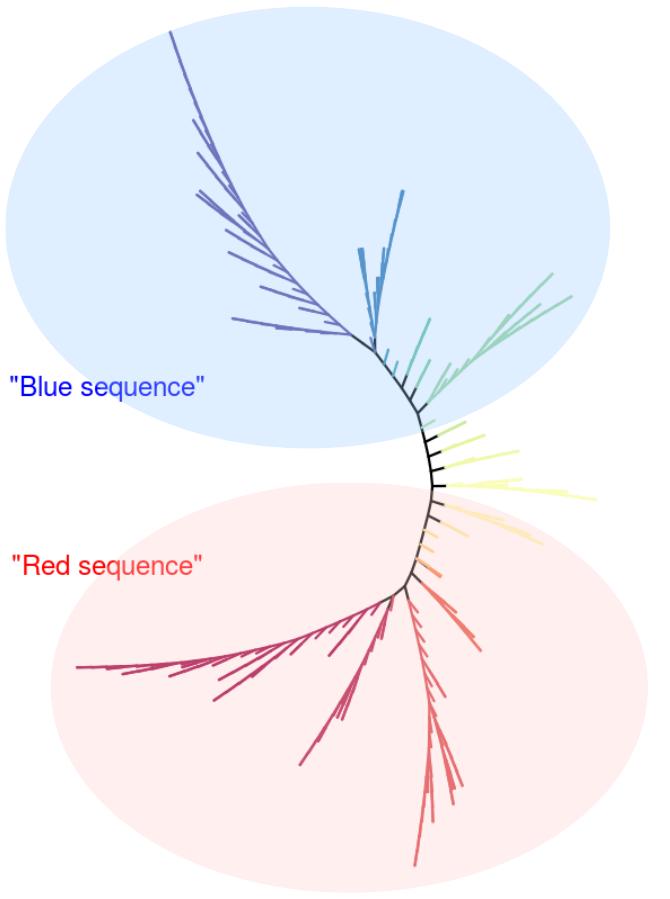}							\end{center}\vspace{0cm}
	\caption{The phylogenetic tree for the galaxies of the WINGS survey. Colors mark the different groups defined by the cladistic analysis on a set of  parameters which includes color index, absolute magnitude, effective radius, sersic index, among others.  \label{FigPhylo} From \citet{fraix-burnetetal17b}. }
\end{figure}

The expected increase in computing power will certainly help numerical simulations and will be
particularly useful, when managing the enormous databases of galaxies at different redshift that the various projects have in program. New statistical approaches to the data should be adopted to extract the driving processes of galaxy evolution. It is well known in fact that galaxy properties are mutually correlated: e.g. mass correlates with color, morphology, metallicity, SF rate, gas content, etc. and that galaxy environment also plays a role. It is therefore far from being simple to recognize the paths of evolution. In this respect the cladistic approach (see \citealt{fraix-burnetetal17b}) seems promising. The phylogenetic approach, also adopted in biology, try to establish a relationship among different species by minimizing the total evolutionary cost. The tree that results from cladistic analysis is not a genealogical tree, in the sense that it does not show ancestors and descendants (i.e., the ontogeny of individual galaxies), but the phylogenetic tool can take correctly into account the evolution of populations of objects. This technique requires an enormous amount of computing time \citep{Fraix-Burnet2015}). 

The phylogenetic analysis provides information equivalent to that of "scaling relations", but in a larger space defined by the number of parameters.  The classical 2D or 3D scaling relations, that identify some paths of evolution. For example the mass-metallicity relation constrains the amount of gas inflow and outflow during the cosmic epochs \citep[see e.g.][]{Truongetal2018,Lianetal2018,Hartwick2017,Torreyetal2017} and the mass-size relation  constrains the epoch and location of star formation \citep[see e.g.][]{Lietal2018,Sweetetal2017,Zanellaetal2016}.  The phylogenetic analysis offers  the possibility of testing some ancestral hypothesis by rooting the phylogenetic tree on a physical or observational parameter. We are confident that the method has a high scientific potential also in the field of galaxies and quasars. 

Fig. \ref{FigPhylo} shows an example of such analysis applied to the data of the WINGS survey (\citealt[see e.g.][]{fraix-burnetetal17b}). The colored lines mark the families of objects sharing similar properties according to a pre-clustering analysis (computing power was not sufficient to consider individual galaxies).  Fig. \ref{FigPhylo} shows the cladistic tree of 300 pre-clusters of the WINGS sample of 1494 galaxies.  Lower masses are at the top and each leave (ending branch) of the tree is  one pre-cluster.  Each group of galaxies thus corresponds either to a single branch or to a bunch of branches on the tree. The color progression from blue to red grossly matches the increase in mass of galaxies. Interestingly, the morphological type decreases along the tree downward. Even if the tree of Fig. \ref{FigPhylo} is unrooted, the cladistic  analysis is able to reconstruct a phylogenic sequence that separates the blue and red sequence of galaxies.

A further step forward for our understanding of galaxies will be obtained when we will be able to trace the behavior of the cold neutral and molecular gas up to the first cosmic epochs. The drivers of LOFAR and SKA are the capability to probe deep into the redshift range of the reionization epoch from 6 to 20, mapping the formation of massive galaxies, clusters and black holes using $z \le $ 6 radio and starburst galaxies as probes. 

The maps provided by e.g. ALMA and SKA will be important to understand the relation between SF, gas density and kinematics and could contribute to clarify the mechanisms of starburst and AGN activity with the associated feedback processes. The ALMA data for our own Galaxy will also provide the opportunity of resolving the gas transformation in the vicinity of the central BH. The environment around a BH is very poorly known. What we know is that the central region of galaxies are very dense and are dominated by gas cloud collisions and strong magnetic fields. The knowledge of the chemical enrichment in this region is crucial to understand the origin of BH itself. At the time of writing this contribution, strong molecular outflows have already be measured by ALMA using e.g. the CO lines and other molecular transitions. All these data will provide a significant advance for our understanding of the feedback process and the inter-stellar medium (ISM) enrichment. 

For  neutral Hydrogen,  SKA  with its unprecedented sensitivity will certainly contribute to create a better defined picture of the formation and evolution of the first stars and of the galaxies after the Big Bang,  and will provide important information on the role of the cosmic magnetism, as well as on the nature of gravity, and possibly even on the existence of life beyond Earth. Hydrogen is the most diffuse element in the Universe and we can exploit its distribution to afford one of the mysteries of the current cosmology: the nature and the role of dark energy. DE is responsible of the observed acceleration of the Universe, but its nature is unknown. The next 50 years will likely be dedicated to solve the puzzle posed by the current cosmological model. SKA should be able to detect the young forming galaxies at very high redshifts, so that HI  maps might include million of galaxies. The origin and evolution of cosmic magnetisms will be one the key researches of the new astrophysics that can change the future of our understanding of galaxies.

The nature of the DE can in principle be constrained by reconstructing the cosmic expansion history and the linear growth of cosmic structures. 
In this context the future ESA mission Euclid, by mapping billion of galaxies, will be able to provide the geometry of the dark Universe and classical spectroscopy is still mandatory to check for systematics effects in all measurements. Cluster velocity dispersions also require precise spectroscopy to reconstruct their evolution and  spectra will be fundamental to test AGN activity and systematic variations in the progenitor properties of SNe (a method that requires a good knowledge of metallicities, SFRs, and dust contents).

The large-scale photometric surveys used for example by BAOs or by the lensing statistics also require a precise spectroscopic calibration. 
{Baryonic Acoustic Oscillations (BAOs) are regular density fluctuations of a fluid of baryonic matter and photons present in the primordial Universe during the clustering of structures. Pressure generated expanding sound waves were imprinted on this fluid. With the expansion of the Universe the expansion of the pressure wave stopped and photons streamed away while BM and DM locked together for  gravitational attraction. This gave rise to the acoustic peak visible in the data of the SDSS and 2dFGRS as a characteristic scale bump of galaxy clustering in the power spectrum \citep{Coleetal2005,Eisensteinetal2005}. The position of this bump is a powerful cosmological probe that will be studied in detail by the Euclid mission.} {The BAOs provide what in cosmology is called a ``standard ruler": }  Tighter and tighter constraints may be obtained in a not-so-distant future by high resolution spectroscopy on the next generation of large-aperture telescopes. DESI aims at constructing a 3-dimensional map spanning the nearby universe to 3Gpc  \citep{levietal13}. First light is expected in 2019.

% \citep{nas16}               
\subsection{The next 30 years (and beyond) in quasar research}               
\subsubsection{Evolving quasars, the obscured Universe and the dawn of the present-day Universe} 

\begin{figure}[h!]
\vspace{0cm}
	\begin{center}
		\includegraphics[width=16cm, angle=0]{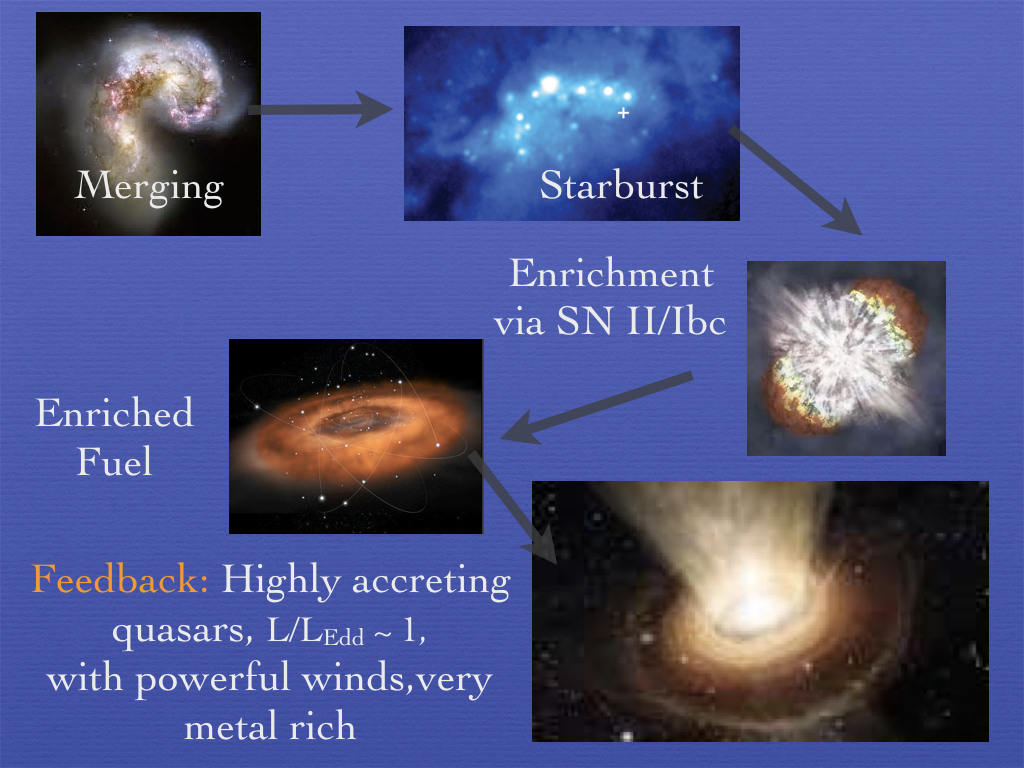}							\end{center}\vspace{0cm}
	\caption{The early stages in the evolution of AGN and quasars: merging and strong interaction lead to accumulation of gas in the galaxy central regions, inducing a burst of star formation (top panels). Mass loss due to stellar winds and supernova explosions eventually provides accretion fuel for the massive black hole at the galaxy center. Radiation force and mechanical energy can then sweep  the  dust surrounding the  black hole, at least within a cone coaxial with the accretion disk axis from where the radiative and mechanical output is free to escape into the host galaxy ISM (bottom panel). \label{fig:xa} }
\end{figure}

Fig. \ref{fig:xa} depicts the early stages in the evolution of quasars from low to high redshift ($0 \lesssim z \lesssim 6$): merging and strong interaction  lead to accumulation of gas in the galaxy central regions, inducing a burst of star formation. In this phase, the quasar may remain shrouded by a dense cocoon of gas and dust, opaque to UV and optical radiation. In the most extreme cases, the cocoon may be  Compton thick and therefore so opaque that soft-X-ray emission   can blanketed (if the column density of absorbing gas is $N_\mathrm{c} \sim 10^{25}$ cm$^{-2}$, a source is fully absorbed up to $\sim 10$ keV, \citealt{comastri04}), and even emission at $\sim 10$ KeV is significantly absorbed. These sources are expected to be visible only in the hard X-ray domain, by instrument such as Nustar. The ``obscured'' phase  may well occur after the collapse of the first seed black holes at redshift $\gtrsim 10$.   Mass loss due to stellar winds and supernova explosions  provide a large amount of enriched gas that is accretion fuel for the  black hole. The accretion rate is initially expected to be very high, yielding maximum radiative output per unit mass. Radiation force and mechanical energy can then sweep away the dust surrounding the accreting black hole, at least within a cone coaxial with the accretion disk axis. An unobscured (if seen not to far from the cone's axis) quasar is born. This scheme may apply to seed quasars of masses as low as 10$^{3 - 4}$ M$_{\odot}$ in the primordial Universe as well as to low-mass local quasars (local quasars that may be quasi-analogous of highly-accreting seed black holes have been identified since long, \citealt{sulenticetal00a,mathur00}).  In the case of primordial seeds, direct black hole collapse and enormous star formation rates may be associated with the build-up of the protogalaxy (instead of the merging phase depicted in Fig. \ref{fig:xa}). 

The sketch of Fig. \ref{fig:xa} identifies several key elements that are at the root of the complex physical processes in quasars: 
\begin{enumerate}
    \item    the connection between  black hole growth and the  build-up and evolution of galaxies, which involves the interplay between accretion and star formation. The first side of the issue  is what  the physical conditions (e.g. fueling mode, triggering mechanism) that initiate major black hole accretion events should be. The second side involves the mechanical and radiative output of the quasar (understood as an accreting black hole). What is the nature of AGN feedback? 
    \item   the accretion process itself. The basic process of accretion is self-similar although it may take different forms as a function of accretion rate, black hole mass, and spin, and these parameters are expected to be not only a function of cosmic epoch, but of environment as well (for example, merging leading to a sizable population of massive black hole binaries).    
\end{enumerate}
   
These issues are addressed by detailed studies of both nearby and distant SMBHs, and will of course benefit from the wide array of instruments providing very high spatial resolution from ground (active optics; interferometers) and space.

%They therefore provide a  method for compiling samples of   obscured black holes  that can lead to unbiased  distributions of several intrinsic properties (intrinsic luminosity, dust and gas column density, etc.). 

ALMA can locate   star-formation activity hidden by dust, and   identify spectroscopically the cooling of molecular clouds with primordial chemical composition. ALMA is a powerful tool with the potential of clarifying the inter-relation between star formation, metal enrichment and SMBH accretion-induced activity.  As an example of an application of the ALMA data, we can consider the [CII] 158 $\mu$m  line that is strong in star-forming galaxies, and is the dominant cooling mechanism for cold interstellar gas.  \citet{kimballetal15}  presented an analysis of an unusual [CII] emission line observed with ALMA of a very luminous quasi-stellar object (QSO) SDSS J155426.16+193703.0 at $z \approx 4.6$. The line is extremely broad, with FWHM $\approx$ 700 \kms  and a flat-topped or double-peaked line profile. These results   suggest the presence of a massive rotating disc  that may be the ultimate accretion fuel for the SMBH. Findings like these are hinting at the potential of sensitive, high resolution sub-mm spectroscopy  for relating gas dynamics and metal content to a quasar dark matter halo also at high redshift. X-rat observatories such as Athena  should maximize the synergies with ALMA and reveal hidden nuclear activity in star forming galaxies. {The issue of the relation between nuclear activity and circumnuclear star formation may be solved by the sinergy of radio, mm, and X observations which circumvent the effect of dust and provide a less biased view of the ``obscured" Universe  than the one we have now over a broad range of cosmic epochs. Gravitational wave observations from LISA (\S \ref{gravwaves}) may reveals BHs that are invisible even in the X-ray domain because deeply enshrouded by a Compton thick cocoon. The Einstein telescope may reveal intermediate mass BHs ($\sim 10^3$ M$_\odot$) that may be the progenitor of more massive BHs, perhaps filling the gap between the stellar mass domain and the massive black hole domin.}

The  radio surveys with SKA (\citealt{diamond08}) and the next-generation VLA will play an important role of multi-wavelength studies of galactic evolution, as they will detect  sources whose radio emission  is associated with stellar evolution processes up to the epoch of reionization ($z \gtrsim 6$; \citealt{nylandetal18}).  The 21 cm line will becomes detectable up to high redshift: an all-sky survey detecting $\sim 10^{9}$ galaxies up to redshift $\approx 2$ will become feasible with SKA (for comparison, at present there are only few detections beyond $z \approx 0.2$,  \citealt{abdallaetal15}).

\subsubsection{Feedback and reionization at the redshift frontiers}
\label{redfront}

 The existence of a SMBH (J1342+0928, \mbh\ $\approx 8 \cdot 10^{8}$  M$_{\odot}$)  at redshift 7.5  (when the Universe age was just 690 Myr) reinforces  black hole growth models that assume seed black holes with significant initial masses  ($\lesssim 10^{4}$ M$_{\odot}$)\ and super-Eddington accretion \citep{lodatonatarajan07}. There is time to build up an \mbh\ that large without challenge the $\Lambda$CDM cosmology?  Even if there has been a large delay between the previous record-holder and the last one and only $\sim 100$ quasars are known above redshift 6 \citep[][and references therein]{matsuokaetal18}, it is legitimate to surmise that more quasars at $z>6$\ will be discovered in the next decades through dedicated surveys with ad-hoc photometric bandpasses. Very high \mbh\ values may turn out to be more challenging for $\Lambda$CDM cosmology if they are found at higher redshift. { It is not only the high mass, but also the relatively high metal content (especially iron) of the line emitting gas that needs to be accounted for \citep{juarezetal09}. Clearly, metal enrichment in the proximity of a black hole may be decoupled from the host galaxy via circumnuclear star formation. The details of the enrichment process are however still not known. }
 
 {In the standard big bang cosmology, the Universe baryonic matter, following the hot phase after the big bang, should have undergone a rapid cooling, and have become mostly neutral. It may have remained so until the first accreting black holes and the first shining massive stars may have produced enough radiation to reionize it \citep[][ and references therein]{jiangetal16,yoshiuraetal17}. The evidence  of reionization (or better, of the existence of a significant fraction of neutral gas able to absorb Ly$\alpha$ photons from quasars) was associated with a deep, contiguous absorption through leading to zero flux on the blue side of Ly$\alpha$ \citep{gunnpeterson65}.  } The first detection  of the Gunn-Peterson effect \citep{beckeretal01}  started the exploration of the dark ages when the Universe was still partly neutral.  We see strong evidence of the quasar Ly$\alpha$ emission line being absorbed by a Gunn-Peterson damping wing from the intergalactic medium, as would be expected if the intergalactic hydrogen surrounding  is significantly neutral, indicating that the quasar allows to probe well within the reionization epoch \citep{banadosetal16}. The redshift frontier is now at  $z = 9-12$, within 500 million years of the Big Bang. Lyman-break  galaxies are detected in increasing number from $z = 10$ to $z = 6$, consistently with the predicted mass growth of their parent dark matter  halos \citep{kashikawaetal11}. On the converse, Hydrogen Ly$\alpha$ emission from these early galaxies appears to decline rapidly at $z > 6$, which suggests that gas at early epochs is becoming more and more opaque to the Ly$\alpha$ photons \citep{starketal11}.

Determining the relation  of star formation (a Population III of stars) and accretion (onto a direct collapse black hole?) and  reionization  during early cosmic epochs will connect the first light sources to the processes that assembled galaxies after reionization. Quasars are apparently not enough for the re-ionization of hydrogen:  the  number of ionizing photons from  the luminosity function of z $\approx$ 6 is apparently insufficient  to keep the Universe ionized, given also that the soft X-ray background sets limits on accretion power at high redshift \citep{mcquinn16}. In a cosmological context, X-rays are essential for addressing the issue, as they uniquely probe AGN at both the early heavily obscured stage and the later blow-out phase. X-rays can identify the ``buried'' evidence of heavily obscured black hole growth (e.g., the iron K$\alpha$\ line at 6.4 keV). Planned large-aperture X-ray observatories   will trace the cosmic history back to the time when the first luminous sources ignited,  and the subsequent evolution of galaxies and their supermassive black holes \citep{georgakakisetal13}.   
 
 Near- and mid IR spectroscopy in addition to X-ray observations  (discussed in mode detail below) are  crucial to understand the quasar accretion properties. JWST and E-ELT will be  suitable instruments to characterize the first luminous sources, in order  to reconstruct the ionization history of the early universe, and  to analyze how AGN and star formation evolved from the epoch of reionization to the present day \citep{gardneretal06}.

 {The redshift frontiers depend on luminosity. Relatively low mass black holes radiating at modest Eddingon ratio remain undetectable at high $z$, with the important consequence that our view of nuclear activity evolution suffers from a bias: at intermediate-to-high redshifts ($z \gtrsim 1$), we almost fully miss a population of low-luminosity quasars \citep{sulenticetal14}. The ability to collect moderate dispersion visual spectra down to  $V \sim 30 $ mag would allow to cover black holes down to $10^6$\ M$_\odot$\ radiating at low Eddington ratio up to $z \approx 3$. The LSST and the 40m-class telescopes are needed to unveil a much more comprehensive view of nuclear activity than the one available to us until now.}

\subsubsection{Feedback and the interplay  black hole / host galaxy} 
\label{feedhost}

%The  survey and spectral capabilities of Athena+/WFI (effective area, angular resolution,  field of view) will complete the census of black hole growth by yielding samples of up to 100 times larger than is currently possible of the most heavily obscured, including Compton thick, quasars to  $z\approx4$. Data available till now have yielded only very limited knowledge on the most deeply enshrouded (Compton-thick)  AGN population. 

The most important epoch for investigating the relation between accreting black holes and galaxies is the redshift  range $  ~ 1 - 4$, when most black holes gained most of their  masses and when most accretion power was released. X-rays are well-suited for studying in detail black hole feedback, although they are only one of the many spectral ranges that need to be covered to get a complete view of the phenomenon. Feedback  is a process that ultimately originates in the innermost regions close to the supermassive black hole and is dominated, in terms of energy and mass flow, by material over a wide range of ionization stages.  Current studies of the incidence, nature and energetics of AGN feedback are mainly restricted to the local Universe  (with only very limited knowledge on the most deeply enshrouded (Compton-thick)  black hole population), but systematic studies of AGN feedback to $z \sim 4$ via the identification and measurement of blue-shifted  absorption and emission  lines in the X-ray, optical and UV domains should become possible. We still have no clear global  view of the prevalence and evolution of outflows and their  relation to the growth of black holes as well as of their effects on galaxy evolution. Most galaxies host a SMBH at their center, with their mass  correlated with that of their host galaxy   \citep{ferraresemerritt00,gebhardtetal00}.  This correlation may suggest that the evolution of AGN and their galaxy hosts follow a parallel track (at least on a broad temporal average and for massive galaxies), although it says nothing of how this relation is built in physical term.  Some self-regulating process has to connect the growth of the SMBH (accretion-powered) to the  growth of the host galaxy (due to star  formation) to the point of suggesting coevolution. The observational evidences of such coevolution has been reviewed by \cite{Kormendy&Ho2013} while the theoretical motivation for the link between AGN feedback and galaxy evolution was reviewed by \cite{Fabian2012}. 

A crucial process  is the phase in which quasar winds ``invade'' the host galaxy (mechanical feedback).  According to models, quasar outflow rates may reach thousands of solar masses per year at high $z$\ \citep{baraietal18}. A key physical question is: how are the energy and metals accelerated in winds/outflows transferred and deposited into the circum-nuclear medium? The energy of such powerful AGN-driven winds is deposited into the host galaxy ISM, but it is as yet unclear, for instance, under which conditions quasar winds quench or trigger star formation. { Fast galactic-scale molecular outflows (e.g.  \citealt{ciconeetal14,zschaechneretal16}) are believed to be ultimately driven by nuclear activity   (\citealt{morganti17} and refences therein).}   In a  spectacular case,  a  galactic-scale molecular outflow \citep{feruglioetal15}  sweeps away the molecular gas ultimately suppressing  star formation, but there are counterexamples in which the quasar outflows trigger star formation \citep[e.g.,][]{ishibashifabian12}.  The most compelling evidence is now limited to a few case studies,  but it is reasonable to expect that consensus on a global view may be reached by the mid of the century.   A more general assessment will come from the next generation of imagining spectrographs  operating in the optical and near-IR  from ground (with active optics) and space, that may reach sub-0.1 arcsec resolution under ordinary observing conditions. 
                
%Within this context, the high speed ionized winds recently discovered by XMM-Newton, Chandra and Suzaku, from the radiatively efficient phase of the AGN, are regarded as the most effective way of transporting energy from the nuclear scale to host galaxy.
%, and may have its origin in a swept-up ISM or thermally driven wind from the molecular torus (Blustin et al 2005).   

A second key question is:  how do accretion disks around black holes launch winds/outflows, and how much energy do these carry?  The answer to this question suffers because of the poor understanding of the structure and dynamics of the broad line emitting regions, within 10$^{4}$ gravitational radii from the central black hole.  Certainly, not all quasars show powerful winds able to influence the global evolution of their hosts. Some authors distinguish between wind- and disk-dominated quasars \citep{richardsetal11}, a separation that is consistent with the one between high- and low- Eddington ratio quasars (Pop. A and B respectively, \citealt{sulenticetal00a}). The observational results point toward an origin very close to the SMBH for X and UV outflows alike. Measurements in the optical and UV rest frame of the quasars are important tracers of the outflow, but they are only part of the story.  In the X-ray bands we observe narrow absorption lines outflowing with moderate velocity of hundreds to few thousands km/s \citep{halpern84,longinottietal15}.  The {\em warm absorber}  is detected in $\approx$ 50\%\ of AGN \citep{reynolds97,piconcellietal05}.  In the UV band, narrow-absorption lines may be associated with the warm absorber and  broad absorption lines are seen in $\sim20-40$\% of AGN, and may be present with extreme properties (terminal velocity $0.1- 0.2 c$) in most Population A AGN  \citep{sulenticetal06a,trumpetal06,scaringietal09}, but detected only if the outflowing material intercepts the line of sight.  They probably arise in a  radiation driven wind from the accretion disk  \citep{elvis00,progakallman04} that seems to be a widespread and powerful phenomenon at least among Population A sources \citep[e.g.,][]{sulenticetal07}. However, most impressive outflows may appear   from gas  so highly ionized that the only bound transitions  are for Hydrogen- and Helium-like iron.  These X-ray winds are relatively frequent at low-$z$, with a prevalence $\approx$ 30-40\% of local AGN, and outflow velocities reaching $~0.3c$ \citep[Ultra-Fast Outflows (UFOs); ][]{tombesietal10}. The evidence supporting UFOs has been  growing  \citep[e.g.,][]{longinottietal15}. Originally it  was based on faint absorption features of unclear identification,  and UFO phenomenology is as yet poorly known, but the point here is that the mechanical feedback effect from the SMBH involve gas in widely different physical conditions.  A multifrequency analysis involving data from sub-mm to the soft and hard X-ray domain is needed to gain a full understanding of a multi-phase, likely highly-turbulent medium.

Even if there is agreement about the existence of an accretion disk and convincing evidence of outflows,  the launching mechanism and the physical processes involved are only crudely understood today.   Significant progress should come not only from multi-frequency simultaneous observations (optical, UV and X) but also  from SPH hydrodynamics simulations that are expected to improve in numerical sophistication and in the treatment of physical processes (see e.g., \citealt{sadowskietal14,liskaetal18}).

\subsubsection{Monitoring and Reverberation mapping on ``industrial scales''} 

Variability at all wavelengths is one of the defining properties of AGN. The most rapid variations in $\gamma$-rays are on the scale of only a few minutes. The very rapid variability of flares puts strong constraints on the size of the emitting region and its bulk velocity due to light crossing-time arguments. However a fundamental question such as: ``what causes the observed variability in AGN from time scales of a few years down to a few minutes?"  remains  without convincing answers at present.  As for the analysis of   single-epoch spectra of individual quasars, the potential of spectral variability to constrain quasar models has not been  sufficiently explored. The planned ``panoptic'' SDSS-V is intended to exploit this potential on wide scales. The SDSS-V plans to do spectroscopic reverberation mapping sampling hundreds of epochs for  $\sim 10^{3}$\ quasars ($0.1<z<4.5$ ) and $L_\mathrm{bol} \sim 10^{45} - 10^{47}$ erg/s. This is a tenfold increase in the number of  sources monitored until now.  With a more modest number of epochs (a few to a dozen per target), the SDSS-V will also characterize the optical spectral variability of approximately 25,000 quasars. Implications on the  SMBH accretion disks, dynamical structure in the broad line region (BLR) including the optical UV outflows, and especially on signatures of binary BHs might be far reaching, perhaps yielding predictive ability on spectral variations of AGN, from broad band luminosity to emission line profiles.  Optical spectrographs planned at major telescopes  should bring similar scales to other spectrographic surveys. 

Changing-look AGN, in which the broad lines in the AGN spectra either appear or disappear (i.e., passing from type-1 to type-2, or viceversa), an extreme case of line profile variability \citep{yangetal17,lamassaetal15}) may pose that challenges to standard accretion disk theory \citep{lawrence18}. In this respect, the most interesting sources are the ones  hinting at a periodic behavior that in turn may suggest the presence of a sub-pc SMBH binary \citep{bonetal12,grahametal15}. We are forced the word ``hint'' for both the observational data and inferences, as the temporal baseline is still not long enough to confirm the periods excluding the ``red noise'' typical of nearby AGN light curves \citep[][note that the ``red noise'' behavior may not be stochastic but due to processes occurring on timescales comparable to the monitoring temporal baseline]{vaughanetal16}. The inference of  a SMBH binary is bound by several assumptions at present rather ad hoc \citep{lietal16}. 

  The new capabilities offered by multiplexing spectrographs as well as  a significantly longer temporal baseline for the monitoring (periods -- often corresponding to the dynamical timescale of the BLR -- are of the order of tens of years) will likely lead to an assessment of the prevalence of supermassive binary black holes in the local and in the remote Universe, as well as of the interplay magnetic fields/viscosity/turbulence in accretion disk. Astrometric measurements with resolution $\sim 1 \mu$arcsec have the potential to resolve sub-pc binary black hole systems up to   high  redshifts.

%The spectrum of the reprocessed emission is strongly modified by both General and Special Relativistic effects, and can be fitted with models to derive the main disk parameters, including the inner radius of the disk. Under the assumption that it coincides with the ISCO, and using the ISCO dependence on the spin of the black hole, the latter parameter -- which is crucial to understanding the accretion history of supermassive black holes -- may be determined. 
%\subsubsection{Binary black holes and transient phenomena}

%Moreover, the excellent energy resolution of Athena  will enable emission line components originating in more distant matter (ubiquitously present in the X-ray spectra of AGN) to be separated from broad emission features originating in the innermost disk.

\subsubsection{Accreting black hole physics close to the event horizon}

The gas in the accretion disk may lose up to almost half of its energy within 1000 gravitational radii, resulting in powerful UV and X-ray emission. The strong gravity field implies that general and special relativity effects are detectable from the emitted radiation not only in the hard X-ray domain but in the optical and UV as well.  The close proximity to the event horizon is where differences in the spacetime  metric  due to the black hole rotation become appreciable. Lense-Tirring precession may be at the origin of warped disks in the case the angular momentum of the accreting material is misaligned with the spin angular momentum of the black hole \citep{bardeenpetterson75}. Effects on optical and UV emission line profiles are expected \citep{bachev99} although  their assessment  requires large samples of high-quality spectroscopical data. The DR 14 of the SDSS now lists over half a million quasars \citep{parisetal17} and millions of quasars are expected to be discovered and  observed by Euclid. The potential of large samples of spectroscopic data are largely unexploited. From single-epoch spectra a wealth of information can be retrieved considering the trends along the quasar ``main sequence'' \citep{sulenticetal00a}, including an estimate of the BLR size,  physical and dynamical conditions, and  chemical composition of the BLR gas \citep{negreteetal12,negreteetal13}. The methods are in place but still not widely applied also because of the lack of high S/N data that are however expected to grow in availability in the next decades. {We expect that the synergy between reverberation mapping and systematic inter-comparison of optical and UV emission lines coming from ionic species of low- and  high-ionization potential for large samples of individual spectra organized along the main sequence  yields contextualized inputs leading to  full physical and dynamical models of the broad line emitting region, as a function of black hole mass, Eddington ratio, spin, and environment \citep{marzianietal18}. This issue  -- conceptually analogous to the definition and interpretation of the stellar Hertzsprung-Russell diagram -- has been an unremitting problem of quasar research that  has remained basically unsolved since quasar  discovery \citep[see \citealt{sulenticetal12a} in ][]{donofrioetal12}.} 

 The ISCO of the accreting gas also depends on the black hole spin, being closest to the black hole for maximally rotating black holes.  In other words, the determination of the ISCO is a key endeavor because it is an indirect measurement of the black hole spin. This has important consequences for the radiation emitted by the accretion disk which is hotter in the case of high spin \citep{wangetal14}.  It is hard to conceive a black hole that is not spinning at all (everything in the Universe spins, even cometary nuclei or  asteroids!), but how many are spinning close to the maximally rotating case? How does the spin   evolves with cosmic time? We  think these questions are going to find an answer as the next generation of X-ray observatories will be launched. Much is needed even after  successful missions such as XMM. K$\alpha$\ line profile --  associated with reflection from the accretion disk --  remains to be observed in most quasars. While  ASCA detected the extended wing in the K$\alpha$\ profile associated with gravitational and transverse redshift \citep{tanakaetal95}, {XMM spectra are revealing additional complexity in the K$\alpha$\ profiles that may not be associated with relativistic effects.  Several methods have been proposed to estimate the ISCO; the one that has proved viable at least in a large fraction of cases is the K$\alpha$\ profile modelling, but other methods are possible as well: for example, if quasi periodic oscillations are associated with the ISCO angular frequency \citep{brenneman13}. }  X-ray observatories  that are expected be in service after 2025 may provide a  view of the spin and ISCO  distributions for supermassive black holes over a broad range of $z$.  
 
LISA design is suited  to detect the signal for coalescing   black holes with masses  $\lesssim 10^{6}$  M$_{\odot}$ in the source frame up to $z \approx 9$, and to enable the measurement of the dimensionless spin of the largest  \mbh\ with an absolute error better than 0.1. It would allow the detection of spin misalignment  with the orbital angular momentum with a precision better than 10 degrees \citep{danzmann17}. These multimessenger abilities should give an unprecedented view of the obscured populations well within the dark ages that preceded full reonization i.e., beyond redshift $\approx 6$. Within reach in the next 30-40 years (after LISA) is even the detection of SMBH merging with space-based interferometry (a feat that could be obtained even earlier with the pulsar timing arrays, \citealt{mingarellietal17}).

\subsubsection{Zooming closer to the Innermost Stable Circular Orbit (ISCO)}

In radio-quiet quasars, X-rays   are produced by Comptonization of thermal disk photons in a hot corona. Among radio-loud quasars, photons from the radio jet also contribute as seeds in the inverse Compton scattering process \citep{bottcheretal13,pianetal99,bottacinietal16}. Part of the resulting power-law continuum illuminates the disk, where it is reprocessed and reflected, both in the soft and hard X-ray domain, yielding a Compton hump peaking at  energy $\epsilon \approx 40 $ keV and Fe K$\alpha$ emission at 6.4 KeV. The disk may not always extend down to the ISCO; in the Magnetically Arrested Disk (MAD) configuration, the disk is  truncated and the truncation makes it possible to launch  relativistic jets \citep[][]{punslyetal09,rusineksikora17}. The next generation of X-ray observatories should enable measurements of  reflection features (Compton hump and K$\alpha$\ line), and  allow us to measure the ISCO  and spin  (from the ISCO) of black holes much beyond the local Universe. This feat should be achieved by measuring the general relativistic effects predicted on the K$\alpha$\ line profile, most notably the gravitational redshifting of the line base that is sensitive to the ISCO value. 

It is not clear what the hot corona might actually be: a compact ``sphere'' (as it is often modeled) or clumps above the disk that illuminate the disk? How does the corona depend on the accretion status of the SMBH?  The time lag between changes in primary radiation emission from the corona and the reprocessed emission from the disk provides a tool to measure the distance between corona and illuminated disk, as in the case of optical and UV reverberation mapping. { Results on the K$\alpha$\ response of several nearby AGN are already available \citep{karaetal16} and is presumable that they will be extended in the coming decades. } The determination of the disk-corona geometry via transfer function fitting  requires data of much better quality than presently available \citep{dovciaketal13}.  This is a task meant for X-ray observatories such as  Athena, with the possibility of long uninterrupted observations on its orbit at L2. Athena should maximize the synergies with the LISA gravitational wave observatory, notably in locating the mergers of massive and seed black holes expected to be detected by LISA. 

Observations and theoretical considerations suggest that the supermassive black hole, Sgr A*, in the center of our Milky Way is surrounded by a compact, foggy emission region radiating at and above 230 GHz. It has been predicted that the event horizon of Sgr A* should cast its shadow onto that emission region, which could be detectable with a global VLBI array of radio telescopes, the global Event Horizon Telescope. Ultimately, a space array at THz frequencies, the Event Horizon Imager, could produce much more detailed images of black holes that may allow for a measurement of the ISCO (and hence of the spin \citep{falcke17}. { There is no doubt about the large concentration of mass in a restricted volume of space \citep[e.g., ][]{petersonwandel99}, but the existence of an event horizon has been questioned also on theoretical grounds in the context of quantum gravity.  Gravitational wave detections from black hole merging events may finally distinguish between black holes and two main competitors: gravastar \citep{mazurmottola04} and dark-energy stars \citep{chapline05}, classes of compact objects that are innerly sustained by the negative pressure of dark energy, and that avoid some of the  paradoxes associated with the black hole singularity and event horizon. }

\subsubsection{What is the origin of VHE processes?} 
\label{sec:vhe}

Radio-loud AGN  are producing collimated relativistic outflows by a still poorly-understood process. Acceleration occurs extremely close to the SMBH (to explain remarkably short variability timescales), within a few tens gravitational radii, but what are the sufficient conditions for an efficient accelerations to ultra-relativistic speed (Lorentz factors $\gtrsim$ 10 -- 100)?    This may indicate  indicate  a Lorentz factor much larger than previosly thought, or hadronic acceleration. Very and ultra high energy (VHE and UHE) observations  are the best tool to probe the physics of  jet formation and the interaction of the black-hole magnetosphere with the accretion disk corona. 

The SED of bright blazars is well explained by leptonic emission scenarios, where the radiative output throughout the electromagnetic spectrum is assumed to be dominated by electrons and possibly positrons \citep{celottighisellini08}. Radio-astron observation of the jet base of powerful radio sources  \citep{kardashevetal15} with resolution of $20\mu$arcsec isolated the ``nozzle" at a brightness temperature $T\sim 10^{15}$ K, a very high temperature in excess to the Compton temperature. A very rapid variability of the high-energy flux (on timescales of a few minutes)  as well as the temperature revealed by RadioAstron require  extremely high Lorentz factors which are challenging for   leptonic models \citep{bottcheretal13}.  In hadronic models, both primary electrons and protons are accelerated and produce VHE photons in the $\gamma$ domain. Protons however exceed the threshold for p-$\gamma$ photo-pion production, leading to high-energy emission  due to several processes  involving  synchrotron and Compton emission decay products of pions which include the highest energy neutrinos \citep{kunetal18}, among other particles. The synergy between $\gamma$-ray such as CTA and neutrino observatories should yield insight on this issue, which is ultimately about the workings of the mechanism that is  extracting energy from the black hole spin \citep{blandfordznajek77}. {Alternatively, the energy could be extracted by the rotating accretion disk \citep{blandfordpayne82}. In both cases, it is not clear how an ordered magnetic field can be transported down to a few tens of gravitational radii. The increase in computing power as well a the improved constraints on jet acceleration (a proton component? is the jet heavy?) from $\gamma$-ray and neutrino observation may lead to a more detailed description of magnetic field transport over a broad range of spatial scale and of  the ultimate  energy source of relativistic ejecta.}  Understanding the momentum flow of relativistic ejecta  is related to the effects that they may have on  the host galaxy. And CTA may unveil that $\gamma$-ray sources are much more frequent than inferred from present detection rates \citep{costamanteetal18}. 

Radio-loud quasars are  one of the  likely sites of the acceleration of UHE CRs, with energies up to around 100 EeV.   $\gamma$-ray and neutrino observations also allow to search for   UHECRs.    $\gamma$-ray imaging observatories such as the CTA are expected  to explore with unprecedented sensitivity  the $\gamma$ rays in the energy range from 50 keV-2 MeV which are the best tracers of CRs: low event statistics and deviation of charged particles in extra-galactic and Galactic magnetic fields make it difficult to direct search for UHECR sources. The $\gamma$-ray   observations are expected to identify  beacons (the $\gamma$ RL quasars) that track the cosmological evolution of black holes down to the epochs of  galaxy formation. {Gravitational wave, $\gamma$ and hard X-ray observations could provide a solution of the long standing problem of the energy source of reionization, and of the role of accreting black holes in the formation of protogalaxies.}

\begin{table}[]
\footnotesize
    \centering
    \begin{tabular}{|c|c|c|c|c|}\hline
     Sources & Parameters & Basic relation & References & Notes \\ \hline
      extremely radiating   &  hard-X-ray slope & 
       
\multirow{2}{*}{
$d_\mathrm{L} \propto \left[ { l_0\left(1+a\ln \dot{m}\right) R_0 } \right]^{\frac{1-\alpha}{2}}   \frac{ \mathrm{FWHM}^{\frac{1}{1-\alpha}}}{ f_{5100}^{1/2}}$
}
     & \citet{wangetal13} & xA, V \\
     quasars (xA)     & velocity dispersion & & & \\ 
     \hline
     extremely radiating & virial velocity dispersion
        & \multirow{2}{*}{$L \propto {\rm FWHM}(\beta )^4$} & \citet{marzianisulentic14}& xA, V\\
     quasars (xA) &   FWHM(H$\beta$), $L/L_\mathrm{Edd} = const.$ &  & & \\ \hline
     general quasar & X-ray variability,   & \multirow{2}{*}{  $\log L + 4 \log FWHM = a \log \sigma^2 + b$} & \citet{lafrancaetal14} & V \\  
     population     & velocity dispersion & & & \\ \hline
     mainly quasars &    Reverberation mapping &  \multirow{2}{*}{  $d_\mathrm{L} \propto \frac{\tau}{\sqrt{f_{5100}}}$} & \citet{watsonetal11}& \\
     at $z \lesssim 1$ & time delay $\tau$     & &  \citet{czernyetal12}& \\ \hline
     general quasar &  non-linear      & \multirow{2}{*}{$\log d_\mathrm{L} \propto \log f_\mathrm{X} - \gamma \log f_\mathrm{UV} $} & \citet{risalitilusso15} & \\
     population     &  UV-X retation  &  & & \\
     \hline
    \end{tabular}
    \caption{{Table illustrating the foundations of several methods that have been proposed to exploit quasars as distance indicators. All methods have been developed in the last decade. $d_\mathrm{L}$\ indicates the luminosity distance, $f$ the fluxes observed at 5100, and in the UV and X ray ranges, $\dot{m}$\ the dimensionless accretion rate, $\sigma$ the excess variance in X ray fluxes. xA: extreme Population A sources following \citet{sulenticetal00a}, V: use of a virial broadening estimator (i.e., the FWHM of a suitable low-ionization line such as H$\beta$ or Pa$\beta$ of Hydrogen. } }
    \label{tab:cosmo}
\end{table}

%\begin{figure}[h!]
%\vspace{-2cm}
%	\begin{center}
%		\includegraphics[width=12cm, angle=90]{tablecosmo.pdf}						
%			\end{center}\vspace{-2cm}
%	\caption{  \label{tab:cosmo} }
%\end{figure}

\begin{figure}[h!]
\vspace{0cm}
	\begin{center}
		\includegraphics[width=12cm, angle=0]{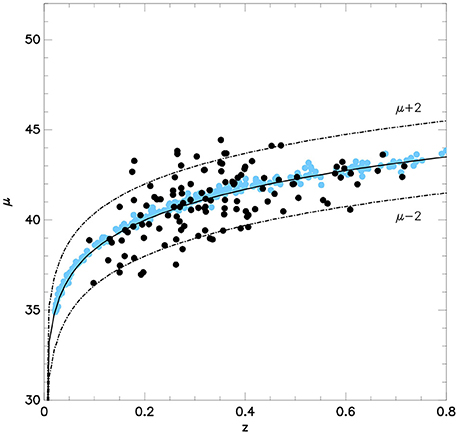}						\end{center} 
	\caption{{ Distance modulus of the ``extreme sample" of Negrete et al. (black dots) and the \citet{Kessleretal2009} SDSS-II SN sample (cyan dots) as a function of redshift. The solid curve is the expected distance modulus $\mu$ for  concordance cosmology. The dotted-dash  lines delimit the area with $|\delta \mu| \le 2$  magnitudes with respect to the concordance model. From \citet{negreteetal17}. } \label{fig:hd} }

\end{figure}

\subsubsection{Quasars and observational cosmology}
\label{obscosmo}

The quasars spatial distribution has been  used as a tracer of large scale structures and BAOs \citep[e.g.,][and references therein]{zarrouketal18}. 
 The array of optical/IR instruments described in Sect. \ref{sec:instr} should provide an enormous gain in statistics (at least two orders of magnitude) of optical SN Ia measurements,  allowing the extension of the SN Ia Hubble diagram, but only up to $z \approx 1.5$ \citep{hook13}. Also the DESI instruments may not go beyond 3 Gpc. Therefore, a large fraction of the early cosmic epochs will be left uncharted.  }  

%Intervening spectral absorption lines have been used to gain inferences on possible  variations of the fine structure constant \citep[e.g., ][]{songailacowie14}.

Quasars have a tremendous potential for cosmology, but their potential is as yet unexploited since they are not standard candles in conventional terms.  Early efforts to establish correlations between luminosity and one or more parameters {(for example, the equivalent width of high-ionization lines, the so-called Baldwin effects, \citealt{baldwinetal78})}    did not live up to cosmological expectations \citep[some early and some recent attempts  are reviewed in the Chapter by \citealt {bartelmannetal09} in ][]{donofrioburigana09}.   Nonetheless,  in the last few years several methods have been proposed for the use of quasars as  redshift-independent distance indicators, or as standard rulers. Four methods are summarized in Table \ref{tab:cosmo}, and some of them are widely discussed by \citet{czernyetal18}.  Their aims are to provide independent measures of the cosmic density of matter and dark energy $\Omega_\mathrm{M}$ and $\Omega_\Lambda$,  as well as to gain constraints on the equation of state of dark energy, and to test  {alternative cosmological scenarios}.  {   Perhaps an accelerating Universe is not anymore an issue, but the nature of dark energy yielding a non-zero $\Omega_\Lambda$\ is still mysterious.  We choose to focus on these methods since they represent novel lines of investigations which rely on an improved understanding of quasar inner structure and all have the potential to constrain the geometry of the Universe in the redshift range $2 \lesssim z \lesssim 4$, a still-uncharted territory. Needless to say, these methods still need years of testing before they may provide useful data, and not all of them may turn out to be feasible.  The main parameters, the basic relations and the sources for which the methods are applicable are summarized in Table \ref{tab:cosmo}. }  

  At very high accretion rate, the luminosity-to-black hole mass ratio ($L$/\mbh $\propto$ \lledd)  tends  toward a well-defined value (more precisely it grows with the logarithm of the mass accretion rate).  The resulting ``slim'' accretion disk is expected to emit a steep soft and hard X-ray spectrum,  with hard X-ray photon index (computed between 2 and 20 KeV) converging toward $\Gamma_\mathrm{hard} \approx 2.5$.  The steep slope of their hard X-ray spectrum  allows for the identification  of super-Eddington accretors \citep{wangetal13,wangetal14a}.  A  challenge in this case is the sample size, but the next generation of X-ray instruments could make possible to identify sizable sample of extremely accreting quasars.     
 
Highly accreting quasars can be considered as ``Eddington standard candles:''  \lledd$\propto L$/\mbh, so that, if \mbh\ can be retrieved under the virial assumption, an estimate of the luminosity becomes possible \citep[][Table \ref{tab:cosmo}]{marzianisulentic14,lafrancaetal14}. This approach  is conceptually analogous to the use of the link between the rotational velocity  of virialized systems such as the disks of spiral galaxies and their luminosity \citep{tullyfisher77}: in both cases, $L \propto \delta v^{4}$. In the case of quasars, $\delta v$\ is provided by the FWHM of H$\beta$\  or Pa-$\beta$.   An explorative analysis  confirms the conceptual validity of the quasar ``virial luminosity'' estimates, although the scatter on the distance modulus   is still too large to draw meaningful inferences for cosmology \citep{negreteetal17}. {Fig. \ref{fig:hd} shows that the scatter is no match for supernov\ae, being around 0.3 dex rms.}  A deeper understanding of the dynamics and physical conditions of the BLR of extremely accreting quasars may yield models of line broadening that may in turn improve the accuracy and precision for cosmology. 
 
The BLR size has been suggested as standard ruler, {in a way that is conceptually analogous to the BAOs}.   The cross-correlation function between the continuum and the emission line  light curve  measures a time lag $\tau$, meaning that the distance of the BLR from a central continuum source can be  written as: $ r_\mathrm{ BLR} = {c \tau}$. If $r_\mathrm{BLR}  = c \tau \propto \sqrt{L}$\ \citep[][and references thererin; see also \citealt{kaspietal05}]{bentzetal13},  the ratio $\tau/\sqrt{f_{5100}}$ \ (where $f_{5100}$\ is the continuum flux at 5100 \AA) is proportional to the luminosity distance that depends on the cosmological parameters  $  H_0, \Omega_\mathrm{M}, \Omega_\Lambda $\ \citep{watsonetal11,czernyetal12,melia15}. This method may yield important results in the next decades thanks to the SDSS-V and to other reverberation mapping campaigns that exploit the increasing multiplexing and wide field coverage of the forthcoming generation of spectrographs.  
 
The non-linear relation between soft X-ray and UV has also been used to build a Hubble diagram \citep{risalitilusso15}. Much is dependent on our understanding of high-energy continuum emission in quasars  that is, as outlined above, not yet satisfactorily modeled. However, the data coming from the next surveys and   advancements in our understanding in quasar structure may yield significant constraints for cosmology that are independent from all methods used to-date.

\section{Conclusion}

For the first time in human history, the next decades will see the ability to cover   the sky in a panchromatic fashion, with a resolution clearly variable    across the electromagnetic spectrum, but sufficient to resolve at least the brightest extragalactic sources from the low-frequency radio to the $\gamma$-ray domain. This ability will be enhanced by ``synoptical monitoring'' capabilities,  at least in the visual bands. Optical sky surveys will make data available for sources down to  $\approx 28 $ mag, and for brighter but still very faint sources ($i \lesssim 21$) in the X-ray domain. If the surveys that have been carried out in the 1990s and 2000s (or even earlier) are of any  help, astronomers will have a treasure trove that will require decades to be fully exploited (IRAS data collected in the 1980s are still used today!). {Multimessenger astronomy is literally still ``in the making," but its revolutionary potential may well go beyond our expectations, revealing a population of massive black holes that we are almost completely missing today.}

{We are confident that several of the main issues that are hotly debated and that need observational and computational improvement will become if not fully settled, at least better understoood: (1) the role of nuclear activity on the host galaxy evolution, over a broad range of redshift. This fundamental issue will benefit from the ability to trace nuclear activity phenomena in obscured source of radio, mm, X-ray and gravitational wave observatories; (2) the reionization main players at the redshift frontier; (3) the inner structure of quasars involving  the emission line region physics and dynamics, including the disk wind physics and modelization; (4) the real nature of the massive compact object in the nuclei of galaxies. We may obtain a final answer to the question: is it really a black hole? (5) The origin of the relativistic radio jets and the mysterious high-energy phenomena occurring in AGN should  become more constrained by the radio, X and $\gamma$\ ray observational developments. (6) Last, the possibility that quasars may be exploited as distance indicator will be certainly explored by several groups.   }
 
 {The same could be said for what concern our understanding of galaxies and clusters. Multi-messenger data will likely allow to map the star formation history of galaxies down to the first epochs, in close connection with the development of SMBHs. Most of the questions outlined before will find possible answers. Thanks to the enormous mass of data for billion of galaxies we will also have in a nearby future a much better understanding of the large scale structure of the Universe and in close connection with this, it will be likely possible to clarify which is the nature of DM and DE.}

We should not be oblivious that many of the advancements hypothesized in the previous sections  depend on the preservation of the social and economic conditions that should make possible for science to progress.  They shall also imply vast educational efforts. However, if the advancements progress as expected, the amount of data and model sophistication  may appear overwhelming.  {Large size elite collaboration may monopolize the frontiers fields of astronomical research, with costly dedicated instruments. Will there still be a place for the work of amateur astronomers and of citizen scientists?}  

We may even ask whether there will be anything left to discover? Will astronomers be reduced to priestesses  and priests of a static wisdom, just monitoring that nothing unexpected or unpredicted by models is happening? We believe there is a  chance that this might happen,  {although not in the next 30 years and perhaps not even before the end of the century. } As the boundary of humankind is going to spread beyond our home planet Earth in the next decades, we will be able to see further and further developments, and interferometers with longer and longer baselines, probing deeper and deeper into the dark ages at the cosmic frontiers.  There will be still fainter sources that will escape detection, and spatial details that we will not able to resolve. As well as intrinsically stochastic processes that will be impossible to predict or model.  And we may still face challenging aspects related to the inability to see beyond the cosmic horizon, if we attempt to analyze the global topology of the Universe \citep{luminet16}. 

%\textcolor{magenta}{Stay tuned!}

%\section*{Author Contributions}

%\section*{Funding}

%\section*{Acknowledgments}

%\section*{Supplemental Data}

\bibliographystyle{frontiersinSCNS_ENG_HUMS} % for Science, Engineering and Humanities and Social Sciences articles, for Humanities and Social Sciences articles please include page numbers in the in-text citations
%\bibliographystyle{frontiersinHLTH&FPHY} % for Health, Physics and Mathematics articles
%\bibliography{biblioletter2}

\begin{thebibliography}{209}
\providecommand{\natexlab}[1]{#1}
\expandafter\ifx\csname urlstyle\endcsname\relax
  \providecommand{\doi}[1]{doi:\discretionary{}{}{}#1}\else
  \providecommand{\doi}{doi:\discretionary{}{}{}\begingroup
  \urlstyle{rm}\Url}\fi
\providecommand{\selectlanguage}[1]{\relax}
\providecommand{\bibAnnoteFile}[1]{%
  \IfFileExists{#1}{\begin{quotation}\noindent\textsc{Key:} #1\\
  \textsc{Annotation:}\ \input{#1}\end{quotation}}{}}
\providecommand{\bibAnnote}[2]{%
  \begin{quotation}\noindent\textsc{Key:} #1\\
  \textsc{Annotation:}\ #2\end{quotation}}

\bibitem[{{Abbott} et~al.(2017){Abbott}, {Abbott}, {Abbott}, {Acernese},
  {Ackley}, {Adams} et~al.}]{abbottetal17}
{Abbott}, B.~P., {Abbott}, R., {Abbott}, T.~D., {Acernese}, F., {Ackley}, K.,
  {Adams}, C., et~al. (2017).
\newblock {Multi-messenger Observations of a Binary Neutron Star Merger}.
\newblock \emph{\apjl} 848, L12.
\newblock \doi{10.3847/2041-8213/aa91c9}
\bibAnnoteFile{abbottetal17}

\bibitem[{{Abdalla} et~al.(2015){Abdalla}, {Bull}, {Camera}, {Benoit-L{\'e}vy},
  {Joachimi}, {Kirk} et~al.}]{abdallaetal15}
{Abdalla}, F.~B., {Bull}, P., {Camera}, S., {Benoit-L{\'e}vy}, A., {Joachimi},
  B., {Kirk}, D., et~al. (2015).
\newblock {Cosmology from HI galaxy surveys with the SKA}.
\newblock \emph{Advancing Astrophysics with the Square Kilometre Array
  (AASKA14)} , 17
\bibAnnoteFile{abdallaetal15}

\bibitem[{{Alard}(2007)}]{Alard2007}
{Alard}, C. (2007).
\newblock {Gravitational arcs as a perturbation of the perfect ring}.
\newblock \emph{\mnras} 382, L58--L62.
\newblock \doi{10.1111/j.1745-3933.2007.00391.x}
\bibAnnoteFile{Alard2007}

\bibitem[{{Ananthakrishnan}(2005)}]{anathakrishnan05}
{Ananthakrishnan}, S. (2005).
\newblock {The Giant Metrewave Radio Telescope (GMRT): Salient features and
  recent results}.
\newblock \emph{International Cosmic Ray Conference} 10, 125
\bibAnnoteFile{anathakrishnan05}

\bibitem[{{Aniyan} and {Thorat}(2017)}]{aniyanthorat17}
{Aniyan}, A.~K. and {Thorat}, K. (2017).
\newblock {Classifying Radio Galaxies with the Convolutional Neural Network}.
\newblock \emph{\apjs} 230, 20.
\newblock \doi{10.3847/1538-4365/aa7333}
\bibAnnoteFile{aniyanthorat17}

\bibitem[{{Austin}(2016)}]{austin16}
{Austin}, R. (2016).
\newblock Megatelescope releases its first image.
\newblock \emph{Physics Today} 69, 42
\bibAnnoteFile{austin16}

\bibitem[{{Ba{\~n}ados} et~al.(2016){Ba{\~n}ados}, {Venemans}, {Decarli},
  {Farina}, {Mazzucchelli}, {Walter} et~al.}]{banadosetal16}
{Ba{\~n}ados}, E., {Venemans}, B.~P., {Decarli}, R., {Farina}, E.~P.,
  {Mazzucchelli}, C., {Walter}, F., et~al. (2016).
\newblock {The Pan-STARRS1 Distant z{\,}$>${\,}5.6 Quasar Survey: More than 100
  Quasars within the First Gyr of the Universe}.
\newblock \emph{\apjs} 227, 11.
\newblock \doi{10.3847/0067-0049/227/1/11}
\bibAnnoteFile{banadosetal16}

\bibitem[{{Bachev}(1999)}]{bachev99}
{Bachev}, R. (1999).
\newblock {Emission lines from illuminated warped accretion disks in AGN}.
\newblock \emph{\aap} 348, 71--76
\bibAnnoteFile{bachev99}

\bibitem[{{Baldwin} et~al.(1978){Baldwin}, {Burke}, {Gaskell}, and
  {Wampler}}]{baldwinetal78}
{Baldwin}, J.~A., {Burke}, W.~L., {Gaskell}, C.~M., and {Wampler}, E.~J.
  (1978).
\newblock {Relative quasar luminosities determined from emission line
  strengths}.
\newblock \emph{\nat} 273, 431--435.
\newblock \doi{10.1038/273431a0}
\bibAnnoteFile{baldwinetal78}

\bibitem[{{Barai} et~al.(2018){Barai}, {Gallerani}, {Pallottini}, {Ferrara},
  {Marconi}, {Cicone} et~al.}]{baraietal18}
{Barai}, P., {Gallerani}, S., {Pallottini}, A., {Ferrara}, A., {Marconi}, A.,
  {Cicone}, C., et~al. (2018).
\newblock {Quasar outflows at z {$\ge$} 6: the impact on the host galaxies}.
\newblock \emph{\mnras} 473, 4003--4020.
\newblock \doi{10.1093/mnras/stx2563}
\bibAnnoteFile{baraietal18}

\bibitem[{{Bardeen} and {Petterson}(1975)}]{bardeenpetterson75}
{Bardeen}, J.~M. and {Petterson}, J.~A. (1975).
\newblock {The Lense-Thirring Effect and Accretion Disks around Kerr Black
  Holes}.
\newblock \emph{\apjl} 195, L65.
\newblock \doi{10.1086/181711}
\bibAnnoteFile{bardeenpetterson75}

\bibitem[{{Bartelmann} et~al.(2009){Bartelmann}, {Bennett}, {Burigana},
  {Chiosi}, {D'Onofrio}, {Dressler} et~al.}]{bartelmannetal09}
{Bartelmann}, M., {Bennett}, C.~L., {Burigana}, C., {Chiosi}, C., {D'Onofrio},
  M., {Dressler}, A., et~al. (2009).
\newblock {Fundamental Cosmological Observations and Data Interpretation}.
\newblock In \emph{Questions of Modern Cosmology: Galileo's Legacy, by
  D'Onofrio, Mauro; Burigana, Carlo, ISBN 978-3-642-00791-0.~Berlin:
  Springer-Verlag Heidelberg, 2009, p.~7-202}, ed. {D'Onofrio, M.~\& Burigana,
  C.} (Springer Verlag, Berlin-Heidelberg), 7--202
\bibAnnoteFile{bartelmannetal09}

\bibitem[{{Bartelmann} et~al.(2002){Bartelmann}, {Perrotta}, and
  {Baccigalupi}}]{Bartelmannetal2002}
{Bartelmann}, M., {Perrotta}, F., and {Baccigalupi}, C. (2002).
\newblock {Halo concentrations and weak-lensing number counts in dark energy
  cosmologies}.
\newblock \emph{\aap} 396, 21--30.
\newblock \doi{10.1051/0004-6361:20021417}
\bibAnnoteFile{Bartelmannetal2002}

\bibitem[{Basart et~al.(1997)Basart, Burns, Dennison, Weiler, Kassim, Castillo
  et~al.}]{basartetal97}
Basart, J.~P., Burns, J.~O., Dennison, B.~K., Weiler, K.~W., Kassim, N.~E.,
  Castillo, S.~P., et~al. (1997).
\newblock Directions for space‐based low frequency radio astronomy: 1. system
  considerations.
\newblock \emph{Radio Science} 32, 251--263.
\newblock \doi{10.1029/96RS02407}
\bibAnnoteFile{basartetal97}

\bibitem[{{Becker} et~al.(2001){Becker}, {Fan}, {White}, {Strauss},
  {Narayanan}, {Lupton} et~al.}]{beckeretal01}
{Becker}, R.~H., {Fan}, X., {White}, R.~L., {Strauss}, M.~A., {Narayanan},
  V.~K., {Lupton}, R.~H., et~al. (2001).
\newblock {Evidence for Reionization at z\~{}6: Detection of a Gunn-Peterson
  Trough in a z=6.28 Quasar}.
\newblock \emph{\aj} 122, 2850--2857.
\newblock \doi{10.1086/324231}
\bibAnnoteFile{beckeretal01}

\bibitem[{{Becker} et~al.(1995){Becker}, {White}, and {Helfand}}]{beckeretal95}
{Becker}, R.~H., {White}, R.~L., and {Helfand}, D.~J. (1995).
\newblock {The FIRST Survey: Faint Images of the Radio Sky at Twenty
  Centimeters}.
\newblock \emph{\apj} 450, 559.
\newblock \doi{10.1086/176166}
\bibAnnoteFile{beckeretal95}

\bibitem[{{Bell} et~al.(2009){Bell}, {Davis}, {Dey}, {van Dokkum}, {Ellis},
  {Eisenstein} et~al.}]{belletal09}
{Bell}, E., {Davis}, M., {Dey}, A., {van Dokkum}, P., {Ellis}, R.,
  {Eisenstein}, D., et~al. (2009).
\newblock {Understanding the Astrophysics of Galaxy Evolution: the role of
  spectroscopic surveys in the next decade}.
\newblock In \emph{astro2010: The Astronomy and Astrophysics Decadal Survey}.
  vol. 2010 of \emph{Astronomy}
\bibAnnoteFile{belletal09}

\bibitem[{{Bellazzini}(2016)}]{bellazzini16}
{Bellazzini}, M. (2016).
\newblock \emph{MAORY for dummies}.
\newblock Technical note (TNO) E-MAO-000-INA-TNO-001, INAF
\bibAnnoteFile{bellazzini16}

\bibitem[{{Belli} et~al.(2017){Belli}, {Newman}, and {Ellis}}]{Bellietal2017}
{Belli}, S., {Newman}, A.~B., and {Ellis}, R.~S. (2017).
\newblock {MOSFIRE Spectroscopy of Quiescent Galaxies at 1.5 $\lt$ z $\lt$ 2.5.
  I. Evolution of Structural and Dynamical Properties}.
\newblock \emph{\apj} 834, 18.
\newblock \doi{10.3847/1538-4357/834/1/18}
\bibAnnoteFile{Bellietal2017}

\bibitem[{{Bentz} et~al.(2013){Bentz}, {Denney}, {Grier}, {Barth}, {Peterson},
  {Vestergaard} et~al.}]{bentzetal13}
{Bentz}, M.~C., {Denney}, K.~D., {Grier}, C.~J., {Barth}, A.~J., {Peterson},
  B.~M., {Vestergaard}, M., et~al. (2013).
\newblock {The Low-luminosity End of the Radius-Luminosity Relationship for
  Active Galactic Nuclei}.
\newblock \emph{\apj} 767, 149.
\newblock \doi{10.1088/0004-637X/767/2/149}
\bibAnnoteFile{bentzetal13}

\bibitem[{{Beringer} et~al.(2012){Beringer}, {Arguin}, {Barnett}, {Copic},
  {Dahl}, {Groom} et~al.}]{beringeretal12}
{Beringer}, J., {Arguin}, J.-F., {Barnett}, R.~M., {Copic}, K., {Dahl}, O.,
  {Groom}, D.~E., et~al. (2012).
\newblock {Review of Particle Physics}.
\newblock \emph{\prd} 86, 010001.
\newblock \doi{10.1103/PhysRevD.86.010001}
\bibAnnoteFile{beringeretal12}

\bibitem[{{Blandford} and {Narayan}(1992)}]{Blandford&Narayan1992}
{Blandford}, R.~D. and {Narayan}, R. (1992).
\newblock {Cosmological applications of gravitational lensing}.
\newblock \emph{\araa} 30, 311--358.
\newblock \doi{10.1146/annurev.astro.30.1.311}
\bibAnnoteFile{Blandford&Narayan1992}

\bibitem[{{Blandford} and {Payne}(1982)}]{blandfordpayne82}
{Blandford}, R.~D. and {Payne}, D.~G. (1982).
\newblock {Hydromagnetic flows from accretion discs and the production of radio
  jets}.
\newblock \emph{\mnras} 199, 883--903.
\newblock \doi{10.1093/mnras/199.4.883}
\bibAnnoteFile{blandfordpayne82}

\bibitem[{{Blandford} and {Znajek}(1977)}]{blandfordznajek77}
{Blandford}, R.~D. and {Znajek}, R.~L. (1977).
\newblock {Electromagnetic extraction of energy from Kerr black holes}.
\newblock \emph{\mnras} 179, 433--456
\bibAnnoteFile{blandfordznajek77}

\bibitem[{{Bom} et~al.(2017){Bom}, {Makler}, {Albuquerque}, and
  {Brandt}}]{Bometal2017}
{Bom}, C.~R., {Makler}, M., {Albuquerque}, M.~P., and {Brandt}, C.~H. (2017).
\newblock {A neural network gravitational arc finder based on the Mediatrix
  filamentation method}.
\newblock \emph{\aap} 597, A135.
\newblock \doi{10.1051/0004-6361/201629159}
\bibAnnoteFile{Bometal2017}

\bibitem[{{Bon} et~al.(2012){Bon}, {Jovanovi{\'c}}, {Marziani}, {Shapovalova},
  {Bon}, {Borka Jovanovi{\'c}} et~al.}]{bonetal12}
{Bon}, E., {Jovanovi{\'c}}, P., {Marziani}, P., {Shapovalova}, A.~I., {Bon},
  N., {Borka Jovanovi{\'c}}, V., et~al. (2012).
\newblock {The First Spectroscopically Resolved Sub-parsec Orbit of a
  Supermassive Binary Black Hole}.
\newblock \emph{\apj} 759, 118.
\newblock \doi{10.1088/0004-637X/759/2/118}
\bibAnnoteFile{bonetal12}

\bibitem[{{Bottacini} et~al.(2016){Bottacini}, {B{\"o}ttcher}, {Pian}, and
  {Collmar}}]{bottacinietal16}
{Bottacini}, E., {B{\"o}ttcher}, M., {Pian}, E., and {Collmar}, W. (2016).
\newblock {3C 279 in Outburst in 2015 June: A Broadband SED Study Based on the
  INTEGRAL Detection}.
\newblock \emph{\apj} 832, 17.
\newblock \doi{10.3847/0004-637X/832/1/17}
\bibAnnoteFile{bottacinietal16}

\bibitem[{{B{\"o}ttcher} et~al.(2013){B{\"o}ttcher}, {Reimer}, {Sweeney}, and
  {Prakash}}]{bottcheretal13}
{B{\"o}ttcher}, M., {Reimer}, A., {Sweeney}, K., and {Prakash}, A. (2013).
\newblock {Leptonic and Hadronic Modeling of Fermi-detected Blazars}.
\newblock \emph{\apj} 768, 54.
\newblock \doi{10.1088/0004-637X/768/1/54}
\bibAnnoteFile{bottcheretal13}

\bibitem[{{Brenneman}(2013)}]{brenneman13}
{Brenneman}, L. (2013).
\newblock \emph{{Measuring the Angular Momentum of Supermassive Black Holes}}.
\newblock \doi{10.1007/978-1-4614-7771-6}
\bibAnnoteFile{brenneman13}

\bibitem[{{Buta}(2013)}]{Buta2013}
{Buta}, R.~J. (2013).
\newblock \emph{{Galaxy Morphology}}.
\newblock 1.
\newblock \doi{10.1007/978-94-007-5609-0_1}
\bibAnnoteFile{Buta2013}

\bibitem[{{Cappellari} et~al.(2013){Cappellari}, {McDermid}, {Alatalo},
  {Blitz}, {Bois}, {Bournaud} et~al.}]{Cappellarietal2013}
{Cappellari}, M., {McDermid}, R.~M., {Alatalo}, K., {Blitz}, L., {Bois}, M.,
  {Bournaud}, F., et~al. (2013).
\newblock {The ATLAS$^{3D}$ project - XX. Mass-size and mass-{$\sigma$}
  distributions of early-type galaxies: bulge fraction drives kinematics,
  mass-to-light ratio, molecular gas fraction and stellar initial mass
  function}.
\newblock \emph{\mnras} 432, 1862--1893.
\newblock \doi{10.1093/mnras/stt644}
\bibAnnoteFile{Cappellarietal2013}

\bibitem[{{Carrami{\~n}ana} and {HAWC Collaboration}(2017)}]{carraminanaetal17}
{Carrami{\~n}ana}, A. and {HAWC Collaboration} (2017).
\newblock {Black hole astrophysics with HAWC, the High Altitude Water Cherenkov
  {$\gamma$}-ray observatory}.
\newblock In \emph{New Frontiers in Black Hole Astrophysics}, ed. A.~{Gomboc}.
  vol. 324 of \emph{IAU Symposium}, 309--316.
\newblock \doi{10.1017/S1743921317002289}
\bibAnnoteFile{carraminanaetal17}

\bibitem[{{Celotti} and {Ghisellini}(2008)}]{celottighisellini08}
{Celotti}, A. and {Ghisellini}, G. (2008).
\newblock {The power of blazar jets}.
\newblock \emph{\mnras} 385, 283--300.
\newblock \doi{10.1111/j.1365-2966.2007.12758.x}
\bibAnnoteFile{celottighisellini08}

\bibitem[{{Chapline}(2005)}]{chapline05}
{Chapline}, G. (2005).
\newblock {Dark Energy Stars}.
\newblock In \emph{22nd Texas Symposium on Relativistic Astrophysics}, eds.
  P.~{Chen}, E.~{Bloom}, G.~{Madejski}, and V.~{Patrosian}. 101--104
\bibAnnoteFile{chapline05}

\bibitem[{{Cherenkov Telescope Array Consortium} et~al.(2017){Cherenkov
  Telescope Array Consortium}, {:}, {Acharya}, {Agudo}, {Samarai}, {Alfaro}
  et~al.}]{CTAconsortium17}
{Cherenkov Telescope Array Consortium}, T., {:}, {Acharya}, B.~S., {Agudo}, I.,
  {Samarai}, I.~A., {Alfaro}, R., et~al. (2017).
\newblock {Science with the Cherenkov Telescope Array}.
\newblock \emph{ArXiv e-prints}
\bibAnnoteFile{CTAconsortium17}

\bibitem[{{Chiosi} et~al.(2017){Chiosi}, {Sciarratta}, {D{\rsquo}Onofrio},
  {Chiosi}, {Brotto}, {De Michele} et~al.}]{Choisetal2017}
{Chiosi}, C., {Sciarratta}, M., {D{\rsquo}Onofrio}, M., {Chiosi}, E., {Brotto},
  F., {De Michele}, R., et~al. (2017).
\newblock {Cosmic Star Formation: A Simple Model of the SFRD(z)}.
\newblock \emph{\apj} 851, 44.
\newblock \doi{10.3847/1538-4357/aa99d5}
\bibAnnoteFile{Choisetal2017}

\bibitem[{{Cicone} et~al.(2014){Cicone}, {Maiolino}, {Sturm},
  {Graci{\'a}-Carpio}, {Feruglio}, {Neri} et~al.}]{ciconeetal14}
{Cicone}, C., {Maiolino}, R., {Sturm}, E., {Graci{\'a}-Carpio}, J., {Feruglio},
  C., {Neri}, R., et~al. (2014).
\newblock {Massive molecular outflows and evidence for AGN feedback from CO
  observations}.
\newblock \emph{\aap} 562, A21.
\newblock \doi{10.1051/0004-6361/201322464}
\bibAnnoteFile{ciconeetal14}

\bibitem[{{Cohen} et~al.(2007){Cohen}, {Lane}, {Cotton}, {Kassim}, {Lazio},
  {Perley} et~al.}]{cohenetal07}
{Cohen}, A.~S., {Lane}, W.~M., {Cotton}, W.~D., {Kassim}, N.~E., {Lazio},
  T.~J.~W., {Perley}, R.~A., et~al. (2007).
\newblock {The VLA Low-Frequency Sky Survey}.
\newblock \emph{\aj} 134, 1245--1262.
\newblock \doi{10.1086/520719}
\bibAnnoteFile{cohenetal07}

\bibitem[{{Cole} et~al.(2005){Cole}, {Percival}, {Peacock}, {Norberg}, {Baugh},
  {Frenk} et~al.}]{Coleetal2005}
{Cole}, S., {Percival}, W.~J., {Peacock}, J.~A., {Norberg}, P., {Baugh}, C.~M.,
  {Frenk}, C.~S., et~al. (2005).
\newblock {The 2dF Galaxy Redshift Survey: power-spectrum analysis of the final
  data set and cosmological implications}.
\newblock \emph{\mnras} 362, 505--534.
\newblock \doi{10.1111/j.1365-2966.2005.09318.x}
\bibAnnoteFile{Coleetal2005}

\bibitem[{{Collett}(2015)}]{Collett2015}
{Collett}, T.~E. (2015).
\newblock {The Population of Galaxy-Galaxy Strong Lenses in Forthcoming Optical
  Imaging Surveys}.
\newblock \emph{\apj} 811, 20.
\newblock \doi{10.1088/0004-637X/811/1/20}
\bibAnnoteFile{Collett2015}

\bibitem[{{Comastri}(2004)}]{comastri04}
{Comastri}, A. (2004).
\newblock {Compton-Thick AGN: The Dark Side of the X-Ray Background}.
\newblock In \emph{Supermassive Black Holes in the Distant Universe}, ed. A.~J.
  {Barger}. vol. 308 of \emph{Astrophysics and Space Science Library}, 245.
\newblock \doi{10.1007/978-1-4020-2471-9_8}
\bibAnnoteFile{comastri04}

\bibitem[{{Costamante} et~al.(2018){Costamante}, {Cutini}, {Tosti}, {Antolini},
  and {Tramacere}}]{costamanteetal18}
{Costamante}, L., {Cutini}, S., {Tosti}, G., {Antolini}, E., and {Tramacere},
  A. (2018).
\newblock {On the origin of gamma-rays in Fermi blazars: beyondthe broad-line
  region}.
\newblock \emph{\mnras} 477, 4749--4767.
\newblock \doi{10.1093/mnras/sty887}
\bibAnnoteFile{costamanteetal18}

\bibitem[{{Cross}(2016)}]{cross16}
{Cross}, T. (2016).
\newblock r moore's law after moore's law -- double, double, toil and trouble.
\newblock \emph{The Economist}
\bibAnnoteFile{cross16}

\bibitem[{{Czerny} et~al.(2018){Czerny}, {Beaton}, {Bejger}, {Cackett},
  {Dall'Ora}, {Holanda} et~al.}]{czernyetal18}
{Czerny}, B., {Beaton}, R., {Bejger}, M., {Cackett}, E., {Dall'Ora}, M.,
  {Holanda}, R.~F.~L., et~al. (2018).
\newblock {Astronomical Distance Determination in the Space Age. Secondary
  Distance Indicators}.
\newblock \emph{Space Science Reviews} 214, \#32.
\newblock \doi{10.1007/s11214-018-0466-9}
\bibAnnoteFile{czernyetal18}

\bibitem[{{Czerny} et~al.(2012){Czerny}, {Hryniewicz}, {Maity},
  {Schwarzenberg-Czerny}, {Zycki}, and {Bilicki}}]{czernyetal12}
{Czerny}, B., {Hryniewicz}, K., {Maity}, I., {Schwarzenberg-Czerny}, A.,
  {Zycki}, P.~T., and {Bilicki}, M. (2012).
\newblock {Towards equation of state of dark energy from quasar monitoring:
  Reverberation strategy}.
\newblock \emph{ArXiv e-prints}
\bibAnnoteFile{czernyetal12}

\bibitem[{{Daddi} et~al.(2005){Daddi}, {Renzini}, {Pirzkal}, {Cimatti},
  {Malhotra}, {Stiavelli} et~al.}]{Daddietal2005}
{Daddi}, E., {Renzini}, A., {Pirzkal}, N., {Cimatti}, A., {Malhotra}, S.,
  {Stiavelli}, M., et~al. (2005).
\newblock {Passively Evolving Early-Type Galaxies at 1.4 $\lt$\~{} z $\lt$\~{}
  2.5 in the Hubble Ultra Deep Field}.
\newblock \emph{\apj} 626, 680--697.
\newblock \doi{10.1086/430104}
\bibAnnoteFile{Daddietal2005}

\bibitem[{{Dalcanton} et~al.(2015){Dalcanton}, {Seager}, {Aigrain}, {Battel},
  {Brandt}, {Conroy} et~al.}]{dalcantonetal15}
{Dalcanton}, J., {Seager}, S., {Aigrain}, S., {Battel}, S., {Brandt}, N.,
  {Conroy}, C., et~al. (2015).
\newblock {From Cosmic Birth to Living Earths: The Future of UVOIR Space
  Astronomy}.
\newblock \emph{ArXiv e-prints}
\bibAnnoteFile{dalcantonetal15}

\bibitem[{{Danzmann} and {et al.}(2017)}]{danzmann17}
{Danzmann}, K. and {et al.} (2017).
\newblock Lisa: laser interferometer space antenna
\bibAnnoteFile{danzmann17}

\bibitem[{{Dekel} et~al.(2009){Dekel}, {Sari}, and {Ceverino}}]{Dekeletal2009}
{Dekel}, A., {Sari}, R., and {Ceverino}, D. (2009).
\newblock {Formation of Massive Galaxies at High Redshift: Cold Streams, Clumpy
  Disks, and Compact Spheroids}.
\newblock \emph{\apj} 703, 785--801.
\newblock \doi{10.1088/0004-637X/703/1/785}
\bibAnnoteFile{Dekeletal2009}

\bibitem[{{Diamond}(2008)}]{diamond08}
{Diamond}, P. (2008).
\newblock {Present and future radio instrumentation}.
\newblock In \emph{Gas and Stars in Galaxies - A Multi-Wavelength 3D
  Perspective}. 72
\bibAnnoteFile{diamond08}

\bibitem[{{D'Onofrio} and {Burigana}(2009)}]{donofrioburigana09}
{D'Onofrio}, M. and {Burigana}, C. (2009).
\newblock \emph{{Questions of Modern Cosmology: Galileo's Legacy}} (Springer
  Verlag).
\newblock \doi{10.1007/978-3-642-00792-7}
\bibAnnoteFile{donofrioburigana09}

\bibitem[{{D'Onofrio} et~al.(2017){D'Onofrio}, {Cariddi}, {Chiosi}, {Chiosi},
  and {Marziani}}]{Donofrioetal2017}
{D'Onofrio}, M., {Cariddi}, S., {Chiosi}, C., {Chiosi}, E., and {Marziani}, P.
  (2017).
\newblock {On the Origin of the Fundamental Plane and Faber-Jackson Relations:
  Implications for the Star Formation Problem}.
\newblock \emph{\apj} 838, 163.
\newblock \doi{10.3847/1538-4357/aa6540}
\bibAnnoteFile{Donofrioetal2017}

\bibitem[{{D'Onofrio} et~al.(2012){D'Onofrio}, {Marziani}, and {
  Sulentic}}]{donofrioetal12}
{D'Onofrio}, M., {Marziani}, P., and { Sulentic}, J.~W. (eds.) (2012).
\newblock \emph{Fifty Years of Quasars From Early Observations and Ideas to
  Future Research}, vol. 386 of \emph{Astrophysics and Space Science Library}
  (Springer Verlag, Berlin-Heidelberg)
\bibAnnoteFile{donofrioetal12}

\bibitem[{{D'Onofrio} et~al.(2016){D'Onofrio}, {Rampazzo}, and
  {Zaggia}}]{donofrioetal16a}
{D'Onofrio}, M., {Rampazzo}, R., and {Zaggia}, S. (eds.) (2016).
\newblock \emph{{From the Realm of the Nebulae to Populations of Galaxies}},
  vol. 435 of \emph{Astrophysics and Space Science Library}.
\newblock \doi{10.1007/978-3-319-31006-0}
\bibAnnoteFile{donofrioetal16a}

\bibitem[{{Dovciak} et~al.(2013){Dovciak}, {Matt}, {Bianchi}, {Boller},
  {Brenneman}, {Bursa} et~al.}]{dovciaketal13}
{Dovciak}, M., {Matt}, G., {Bianchi}, S., {Boller}, T., {Brenneman}, L.,
  {Bursa}, M., et~al. (2013).
\newblock {The Hot and Energetic Universe: The close environments of
  supermassive black holes}.
\newblock \emph{ArXiv e-prints}
\bibAnnoteFile{dovciaketal13}

\bibitem[{{Eisenhauer} et~al.(2011){Eisenhauer}, {Perrin}, {Brandner},
  {Straubmeier}, {Perraut}, {Amorim} et~al.}]{eisenhaueretal11}
{Eisenhauer}, F., {Perrin}, G., {Brandner}, W., {Straubmeier}, C., {Perraut},
  K., {Amorim}, A., et~al. (2011).
\newblock {GRAVITY: Observing the Universe in Motion}.
\newblock \emph{The Messenger} 143, 16--24
\bibAnnoteFile{eisenhaueretal11}

\bibitem[{{Eisenstein} et~al.(2005){Eisenstein}, {Zehavi}, {Hogg},
  {Scoccimarro}, {Blanton}, {Nichol} et~al.}]{Eisensteinetal2005}
{Eisenstein}, D.~J., {Zehavi}, I., {Hogg}, D.~W., {Scoccimarro}, R., {Blanton},
  M.~R., {Nichol}, R.~C., et~al. (2005).
\newblock {Detection of the Baryon Acoustic Peak in the Large-Scale Correlation
  Function of SDSS Luminous Red Galaxies}.
\newblock \emph{\apj} 633, 560--574.
\newblock \doi{10.1086/466512}
\bibAnnoteFile{Eisensteinetal2005}

\bibitem[{{Elvis}(2000)}]{elvis00}
{Elvis}, M. (2000).
\newblock {A Structure for Quasars}.
\newblock \emph{\apj} 545, 63--76.
\newblock \doi{10.1086/317778}
\bibAnnoteFile{elvis00}

\bibitem[{ESA(2018)}]{euclid}
ESA (2018).
\newblock Euclid
\bibAnnoteFile{euclid}

\bibitem[{ESO(2006)}]{esoowl}
ESO (2006).
\newblock \emph{OWL Blue Book} (ESO)
\bibAnnoteFile{esoowl}

\bibitem[{ESO(2018)}]{esoelt}
ESO (2018).
\newblock The european extremely large telescope ("elt") project
\bibAnnoteFile{esoelt}

\bibitem[{{Esposito} et~al.(2016){Esposito}, {Agapito}, {Bonaglia}, {Busoni},
  {Fusco}, {Neichel} et~al.}]{espositoetal16}
{Esposito}, S., {Agapito}, G., {Bonaglia}, M., {Busoni}, L., {Fusco}, T.,
  {Neichel}, B., et~al. (2016).
\newblock {AOF upgrade for VLT UT4: an 8m class HST from ground}.
\newblock In \emph{Adaptive Optics Systems V}. vol. 9909 of \emph{Proceedings
  of SPIE}, 99093U.
\newblock \doi{10.1117/12.2234737}
\bibAnnoteFile{espositoetal16}

\bibitem[{{Evans}(2014)}]{evans14}
{Evans}, D. (2014).
\newblock Moore's law: how long will it last?
\bibAnnoteFile{evans14}

\bibitem[{{Fabian}(2012)}]{Fabian2012}
{Fabian}, A.~C. (2012).
\newblock {Observational Evidence of Active Galactic Nuclei Feedback}.
\newblock \emph{\araa} 50, 455--489.
\newblock \doi{10.1146/annurev-astro-081811-125521}
\bibAnnoteFile{Fabian2012}

\bibitem[{{Falcke}(2017)}]{falcke17}
{Falcke}, H. (2017).
\newblock {Imaging black holes: past, present and future}.
\newblock In \emph{Journal of Physics Conference Series}. vol. 942 of
  \emph{Journal of Physics Conference Series}, 012001.
\newblock \doi{10.1088/1742-6596/942/1/012001}
\bibAnnoteFile{falcke17}

\bibitem[{{Ferrarese} and {Merritt}(2000)}]{ferraresemerritt00}
{Ferrarese}, L. and {Merritt}, D. (2000).
\newblock {A Fundamental Relation between Supermassive Black Holes and Their
  Host Galaxies}.
\newblock \emph{\apjl} 539, L9--L12.
\newblock \doi{10.1086/312838}
\bibAnnoteFile{ferraresemerritt00}

\bibitem[{{Feruglio} et~al.(2015){Feruglio}, {Fiore}, {Carniani}, {Piconcelli},
  {Zappacosta}, {Bongiorno} et~al.}]{feruglioetal15}
{Feruglio}, C., {Fiore}, F., {Carniani}, S., {Piconcelli}, E., {Zappacosta},
  L., {Bongiorno}, A., et~al. (2015).
\newblock {The multi-phase winds of Markarian 231: from the hot, nuclear,
  ultra-fast wind to the galaxy-scale, molecular outflow}.
\newblock \emph{\aap} 583, A99.
\newblock \doi{10.1051/0004-6361/201526020}
\bibAnnoteFile{feruglioetal15}

\bibitem[{{Fiorentino} et~al.(2017){Fiorentino}, {Bellazzini}, {Ciliegi},
  {Chauvin}, {Dout{\'e}}, {D'Orazi} et~al.}]{Fiorentinoetal2017}
{Fiorentino}, G., {Bellazzini}, M., {Ciliegi}, P., {Chauvin}, G., {Dout{\'e}},
  S., {D'Orazi}, V., et~al. (2017).
\newblock {MAORY science cases white book}.
\newblock \emph{ArXiv e-prints}
\bibAnnoteFile{Fiorentinoetal2017}

\bibitem[{Fraix-Burnet et~al.(2017)Fraix-Burnet, D'Onofrio, and
  Marziani}]{fraix-burnetetal17b}
Fraix-Burnet, D., D'Onofrio, M., and Marziani, P. (2017).
\newblock Phylogenetic analyses of quasars and galaxies.
\newblock \emph{Frontiers in Astronomy and Space Sciences} 4, 20.
\newblock \doi{10.3389/fspas.2017.00020}
\bibAnnoteFile{fraix-burnetetal17b}

\bibitem[{Fraix-Burnet et~al.(2015)Fraix-Burnet, Thuillard, and
  Chattopadhyay}]{Fraix-Burnet2015}
Fraix-Burnet, D., Thuillard, M., and Chattopadhyay, A.~K. (2015).
\newblock Multivariate approaches to classification in extragalactic astronomy.
\newblock \emph{Frontiers in Astronomy and Space Sciences} 2.
\newblock \doi{10.3389/fspas.2015.00003}
\bibAnnoteFile{Fraix-Burnet2015}

\bibitem[{{France} et~al.(2017){France}, {Fleming}, {West}, {McCandliss},
  {Bolcar}, {Harris} et~al.}]{franceetal17}
{France}, K., {Fleming}, B., {West}, G., {McCandliss}, S.~R., {Bolcar}, M.~R.,
  {Harris}, W., et~al. (2017).
\newblock {The LUVOIR Ultraviolet Multi-Object Spectrograph (LUMOS): instrument
  definition and design}.
\newblock In \emph{Society of Photo-Optical Instrumentation Engineers (SPIE)
  Conference Series}. vol. 10397 of \emph{Society of Photo-Optical
  Instrumentation Engineers (SPIE) Conference Series}, 1039713.
\newblock \doi{10.1117/12.2272025}
\bibAnnoteFile{franceetal17}

\bibitem[{{Franx} et~al.(2003){Franx}, {Labb{\'e}}, {Rudnick}, {van Dokkum},
  {Daddi}, {F{\"o}rster Schreiber} et~al.}]{franxetal03}
{Franx}, M., {Labb{\'e}}, I., {Rudnick}, G., {van Dokkum}, P.~G., {Daddi}, E.,
  {F{\"o}rster Schreiber}, N.~M., et~al. (2003).
\newblock {A Significant Population of Red, Near-Infrared-selected
  High-Redshift Galaxies}.
\newblock \emph{\apjl} 587, L79--L82.
\newblock \doi{10.1086/375155}
\bibAnnoteFile{franxetal03}

\bibitem[{{Gardner} et~al.(2006){Gardner}, {Mather}, {Clampin}, {Doyon},
  {Greenhouse}, {Hammel} et~al.}]{gardneretal06}
{Gardner}, J.~P., {Mather}, J.~C., {Clampin}, M., {Doyon}, R., {Greenhouse},
  M.~A., {Hammel}, H.~B., et~al. (2006).
\newblock The james webb space telescope.
\newblock \emph{Space Science Reviews} 123, 485--606.
\newblock \doi{10.1007/s11214-006-8315-7}
\bibAnnoteFile{gardneretal06}

\bibitem[{{Gavazzi} et~al.(2014){Gavazzi}, {Marshall}, {Treu}, and
  {Sonnenfeld}}]{Gavazzietal2014}
{Gavazzi}, R., {Marshall}, P.~J., {Treu}, T., and {Sonnenfeld}, A. (2014).
\newblock {RINGFINDER: Automated Detection of Galaxy-scale Gravitational Lenses
  in Ground-based Multi-filter Imaging Data}.
\newblock \emph{\apj} 785, 144.
\newblock \doi{10.1088/0004-637X/785/2/144}
\bibAnnoteFile{Gavazzietal2014}

\bibitem[{{Gebhardt} et~al.(2000){Gebhardt}, {Bender}, {Bower}, {Dressler},
  {Faber}, {Filippenko} et~al.}]{gebhardtetal00}
{Gebhardt}, K., {Bender}, R., {Bower}, G., {Dressler}, A., {Faber}, S.~M.,
  {Filippenko}, A.~V., et~al. (2000).
\newblock {A Relationship between Nuclear Black Hole Mass and Galaxy Velocity
  Dispersion}.
\newblock \emph{\apjl} 539, L13--L16.
\newblock \doi{10.1086/312840}
\bibAnnoteFile{gebhardtetal00}

\bibitem[{{Georgakakis} et~al.(2013){Georgakakis}, {Carrera}, {Lanzuisi},
  {Brightman}, {Buchner}, {Aird} et~al.}]{georgakakisetal13}
{Georgakakis}, A., {Carrera}, F., {Lanzuisi}, G., {Brightman}, M., {Buchner},
  J., {Aird}, J., et~al. (2013).
\newblock {The Hot and Energetic Universe: Understanding the build-up of
  supermassive black holes and galaxies at the heyday of the Universe}.
\newblock \emph{ArXiv e-prints}
\bibAnnoteFile{georgakakisetal13}

\bibitem[{{Graham} et~al.(2015){Graham}, {Djorgovski}, {Stern}, {Drake},
  {Mahabal}, {Donalek} et~al.}]{grahametal15}
{Graham}, M.~J., {Djorgovski}, S.~G., {Stern}, D., {Drake}, A.~J., {Mahabal},
  A.~A., {Donalek}, C., et~al. (2015).
\newblock {A systematic search for close supermassive black hole binaries in
  the Catalina Real-time Transient Survey}.
\newblock \emph{\mnras} 453, 1562--1576.
\newblock \doi{10.1093/mnras/stv1726}
\bibAnnoteFile{grahametal15}

\bibitem[{{Guainazzi} and {Athena Study Team}(2017)}]{guainazzi17}
{Guainazzi}, M. and {Athena Study Team} (2017).
\newblock {Athena: mission concept, study status, and optics development}.
\newblock In \emph{The X-ray Universe 2017}, eds. J.-U. {Ness} and
  S.~{Migliari}. 15
\bibAnnoteFile{guainazzi17}

\bibitem[{{Gunn} and {Peterson}(1965)}]{gunnpeterson65}
{Gunn}, J.~E. and {Peterson}, B.~A. (1965).
\newblock {On the Density of Neutral Hydrogen in Intergalactic Space.}
\newblock \emph{\apj} 142, 1633--1641.
\newblock \doi{10.1086/148444}
\bibAnnoteFile{gunnpeterson65}

\bibitem[{{Halpern}(1984)}]{halpern84}
{Halpern}, J.~P. (1984).
\newblock {Variable X-ray absorption in the QSO MR 2251 - 178}.
\newblock \emph{\apj} 281, 90--94.
\newblock \doi{10.1086/162077}
\bibAnnoteFile{halpern84}

\bibitem[{{Harrison} et~al.(2013){Harrison}, {Craig}, {Christensen}, {Hailey},
  {Zhang}, {Boggs} et~al.}]{harrisonetal13}
{Harrison}, F.~A., {Craig}, W.~W., {Christensen}, F.~E., {Hailey}, C.~J.,
  {Zhang}, W.~W., {Boggs}, S.~E., et~al. (2013).
\newblock {The Nuclear Spectroscopic Telescope Array (NuSTAR) High-energy X-Ray
  Mission}.
\newblock \emph{\apj} 770, 103.
\newblock \doi{10.1088/0004-637X/770/2/103}
\bibAnnoteFile{harrisonetal13}

\bibitem[{{Hartwick}(2017)}]{Hartwick2017}
{Hartwick}, F.~D.~A. (2017).
\newblock {The Chemical Evolution of Galaxies: The Stellar Mass-Metallicity
  Relation and Cosmic Downsizing}.
\newblock \emph{Research Notes of the American Astronomical Society} 1, 8.
\newblock \doi{10.3847/2515-5172/aa974c}
\bibAnnoteFile{Hartwick2017}

\bibitem[{{Heckman} and {Best}(2014)}]{Heckman&Best2014}
{Heckman}, T.~M. and {Best}, P.~N. (2014).
\newblock {The Coevolution of Galaxies and Supermassive Black Holes: Insights
  from Surveys of the Contemporary Universe}.
\newblock \emph{\araa} 52, 589--660.
\newblock \doi{10.1146/annurev-astro-081913-035722}
\bibAnnoteFile{Heckman&Best2014}

\bibitem[{Hobbs et~al.(2010)Hobbs, Archibald, Arzoumanian, Backer, Bailes, Bhat
  et~al.}]{hobbsetal10}
Hobbs, G., Archibald, A., Arzoumanian, Z., Backer, D., Bailes, M., Bhat, N.
  D.~R., et~al. (2010).
\newblock The international pulsar timing array project: using pulsars as a
  gravitational wave detector.
\newblock \emph{Classical and Quantum Gravity} 27, 084013
\bibAnnoteFile{hobbsetal10}

\bibitem[{{Hook} et~al.(2009){Hook}, {Liske}, {Villegas}, and
  {Kissler-Patig}}]{hooketal09}
{Hook}, I., {Liske}, J., {Villegas}, D., and {Kissler-Patig}, M. (2009).
\newblock {Report on the ESO Workshop E-ELT Design Reference Mission and
  Science Plan}.
\newblock \emph{The Messenger} 137, 51--52
\bibAnnoteFile{hooketal09}

\bibitem[{{Hook}(2013)}]{hook13}
{Hook}, I.~M. (2013).
\newblock {Supernovae and cosmology with future European facilities}.
\newblock \emph{Royal Society of London Philosophical Transactions Series A}
  371, 20282.
\newblock \doi{10.1098/rsta.2012.0282}
\bibAnnoteFile{hook13}

\bibitem[{{Hough} et~al.(2005){Hough}, {Rowan}, and
  {Sathyaprakash}}]{houghrowan05}
{Hough}, J., {Rowan}, S., and {Sathyaprakash}, B.~S. (2005).
\newblock {The search for gravitational waves}.
\newblock \emph{Journal of Physics B Atomic Molecular Physics} 38, S497--S519.
\newblock \doi{10.1088/0953-4075/38/9/004}
\bibAnnoteFile{houghrowan05}

\bibitem[{{Icecube Collaboration}(2013)}]{icecubeetal13}
{Icecube Collaboration}, e.~a. (2013).
\newblock Evidence for high-energy extraterrestrial neutrinos at the icecube
  detector.
\newblock \emph{Science} 342.
\newblock \doi{10.1126/science.1242856}
\bibAnnoteFile{icecubeetal13}

\bibitem[{{Icecube Collaboration}(2018)}]{icecubeetal18}
{Icecube Collaboration}, e.~a. (2018).
\newblock Multimessenger observations of a flaring blazar coincident with
  high-energy neutrino icecube-170922a.
\newblock \emph{Science} 361.
\newblock \doi{10.1126/science.aat1378}
\bibAnnoteFile{icecubeetal18}

\bibitem[{{IceCube-Gen2 Collaboration} et~al.(2014){IceCube-Gen2
  Collaboration}, {:}, {Aartsen}, {Ackermann}, {Adams}, {Aguilar}
  et~al.}]{icecubegen214}
{IceCube-Gen2 Collaboration}, {:}, {Aartsen}, M.~G., {Ackermann}, M., {Adams},
  J., {Aguilar}, J.~A., et~al. (2014).
\newblock {IceCube-Gen2: A Vision for the Future of Neutrino Astronomy in
  Antarctica}.
\newblock \emph{ArXiv e-prints}
\bibAnnoteFile{icecubegen214}

\bibitem[{{Intema} et~al.(2017){Intema}, {Jagannathan}, {Mooley}, and
  {Frail}}]{intemaetal17}
{Intema}, H.~T., {Jagannathan}, P., {Mooley}, K.~P., and {Frail}, D.~A. (2017).
\newblock {The GMRT 150 MHz all-sky radio survey. First alternative data
  release TGSS ADR1}.
\newblock \emph{\aap} 598, A78.
\newblock \doi{10.1051/0004-6361/201628536}
\bibAnnoteFile{intemaetal17}

\bibitem[{{Ishibashi} and {Fabian}(2012)}]{ishibashifabian12}
{Ishibashi}, W. and {Fabian}, A.~C. (2012).
\newblock {Active galactic nucleus feedback and triggering of star formation in
  galaxies}.
\newblock \emph{\mnras} 427, 2998--3005.
\newblock \doi{10.1111/j.1365-2966.2012.22074.x}
\bibAnnoteFile{ishibashifabian12}

\bibitem[{{Ivezi{\'c}}(2017)}]{ivezic17}
{Ivezi{\'c}}, {\v Z}. (2017).
\newblock {LSST survey: millions and millions of quasars}.
\newblock In \emph{IAU Symposium}. vol. 324 of \emph{IAU Symposium}, 330--337.
\newblock \doi{10.1017/S1743921316012424}
\bibAnnoteFile{ivezic17}

\bibitem[{{Jiang} et~al.(2016){Jiang}, {McGreer}, {Fan}, {Strauss},
  {Ba{\~n}ados}, {Becker} et~al.}]{jiangetal16}
{Jiang}, L., {McGreer}, I.~D., {Fan}, X., {Strauss}, M.~A., {Ba{\~n}ados}, E.,
  {Becker}, R.~H., et~al. (2016).
\newblock {The Final SDSS High-redshift Quasar Sample of 52 Quasars at
  z$\gt$5.7}.
\newblock \emph{\apj} 833, 222.
\newblock \doi{10.3847/1538-4357/833/2/222}
\bibAnnoteFile{jiangetal16}

\bibitem[{{Joseph} et~al.(2014){Joseph}, {Courbin}, {Metcalf}, {Giocoli},
  {Hartley}, {Jackson} et~al.}]{Josephetal2014}
{Joseph}, R., {Courbin}, F., {Metcalf}, R.~B., {Giocoli}, C., {Hartley}, P.,
  {Jackson}, N., et~al. (2014).
\newblock {A PCA-based automated finder for galaxy-scale strong lenses}.
\newblock \emph{\aap} 566, A63.
\newblock \doi{10.1051/0004-6361/201423365}
\bibAnnoteFile{Josephetal2014}

\bibitem[{{Juarez} et~al.(2009){Juarez}, {Maiolino}, {Mujica}, {Pedani},
  {Marinoni}, {Nagao} et~al.}]{juarezetal09}
{Juarez}, Y., {Maiolino}, R., {Mujica}, R., {Pedani}, M., {Marinoni}, S.,
  {Nagao}, T., et~al. (2009).
\newblock {The metallicity of the most distant quasars}.
\newblock \emph{A\&Ap} 494, L25--L28.
\newblock \doi{10.1051/0004-6361:200811415}
\bibAnnoteFile{juarezetal09}

\bibitem[{{Kara} et~al.(2016){Kara}, {Alston}, {Fabian}, {Cackett}, {Uttley},
  {Reynolds} et~al.}]{karaetal16}
{Kara}, E., {Alston}, W.~N., {Fabian}, A.~C., {Cackett}, E.~M., {Uttley}, P.,
  {Reynolds}, C.~S., et~al. (2016).
\newblock {A global look at X-ray time lags in Seyfert galaxies}.
\newblock \emph{\mnras} 462, 511--531.
\newblock \doi{10.1093/mnras/stw1695}
\bibAnnoteFile{karaetal16}

\bibitem[{{Kardashev} et~al.(2015){Kardashev}, {Alakoz}, {Kovalev}, {Popov},
  {Sobolev}, and {Sokolovsky}}]{kardashevetal15}
{Kardashev}, N.~S., {Alakoz}, A.~V., {Kovalev}, Y.~Y., {Popov}, M.~V.,
  {Sobolev}, A.~M., and {Sokolovsky}, K.~V. (2015).
\newblock {Radioastron: Main results of the implementation of the early science
  program in studies of astronomical objects in the universe with ultra-high
  angular resolution}.
\newblock \emph{Solar System Research} 49, 573--579.
\newblock \doi{10.1134/S0038094615070096}
\bibAnnoteFile{kardashevetal15}

\bibitem[{{Kashikawa} et~al.(2011){Kashikawa}, {Shimasaku}, {Matsuda}, {Egami},
  {Jiang}, {Nagao} et~al.}]{kashikawaetal11}
{Kashikawa}, N., {Shimasaku}, K., {Matsuda}, Y., {Egami}, E., {Jiang}, L.,
  {Nagao}, T., et~al. (2011).
\newblock {Completing the Census of Ly{$\alpha$} Emitters at the Reionization
  Epoch}.
\newblock \emph{\apj} 734, 119.
\newblock \doi{10.1088/0004-637X/734/2/119}
\bibAnnoteFile{kashikawaetal11}

\bibitem[{{Kaspi} et~al.(2005){Kaspi}, {Maoz}, {Netzer}, {Peterson},
  {Vestergaard}, and {Jannuzi}}]{kaspietal05}
{Kaspi}, S., {Maoz}, D., {Netzer}, H., {Peterson}, B.~M., {Vestergaard}, M.,
  and {Jannuzi}, B.~T. (2005).
\newblock {The Relationship between Luminosity and Broad-Line Region Size in
  Active Galactic Nuclei}.
\newblock \emph{ApJ} 629, 61--71.
\newblock \doi{10.1086/431275}
\bibAnnoteFile{kaspietal05}

\bibitem[{{Katsianis} et~al.(2017){Katsianis}, {Tescari}, {Blanc}, and
  {Sargent}}]{Katsianisetal2017}
{Katsianis}, A., {Tescari}, E., {Blanc}, G., and {Sargent}, M. (2017).
\newblock {The evolution of the star formation rate function and cosmic star
  formation rate density of galaxies at z {\tilde} 1-4}.
\newblock \emph{\mnras} 464, 4977--4994.
\newblock \doi{10.1093/mnras/stw2680}
\bibAnnoteFile{Katsianisetal2017}

\bibitem[{{Kessler} et~al.(2009){Kessler}, {Becker}, {Cinabro}, {Vanderplas},
  {Frieman}, {Marriner} et~al.}]{Kessleretal2009}
{Kessler}, R., {Becker}, A.~C., {Cinabro}, D., {Vanderplas}, J., {Frieman},
  J.~A., {Marriner}, J., et~al. (2009).
\newblock {First-Year Sloan Digital Sky Survey-II Supernova Results: Hubble
  Diagram and Cosmological Parameters}.
\newblock \emph{\apjs} 185, 32--84.
\newblock \doi{10.1088/0067-0049/185/1/32}
\bibAnnoteFile{Kessleretal2009}

\bibitem[{{Kimball} et~al.(2015){Kimball}, {Lacy}, {Lonsdale}, and
  {Macquart}}]{kimballetal15}
{Kimball}, A.~E., {Lacy}, M., {Lonsdale}, C.~J., and {Macquart}, J.-P. (2015).
\newblock {ALMA detection of a disc-dominated [C II] emission line at z=4.6 in
  the luminous QSO J1554+1937}.
\newblock \emph{\mnras} 452, 88--98.
\newblock \doi{10.1093/mnras/stv1160}
\bibAnnoteFile{kimballetal15}

\bibitem[{{Kollmeier} et~al.(2017){Kollmeier}, {Zasowski}, {Rix}, {Johns},
  {Anderson}, {Drory} et~al.}]{kollmeieretal17}
{Kollmeier}, J.~A., {Zasowski}, G., {Rix}, H.-W., {Johns}, M., {Anderson},
  S.~F., {Drory}, N., et~al. (2017).
\newblock {SDSS-V: Pioneering Panoptic Spectroscopy}.
\newblock \emph{ArXiv e-prints}
\bibAnnoteFile{kollmeieretal17}

\bibitem[{{Koopmans} and {Treu}(2003)}]{Koopmans&Treu2003}
{Koopmans}, L.~V.~E. and {Treu}, T. (2003).
\newblock {The Structure and Dynamics of Luminous and Dark Matter in the
  Early-Type Lens Galaxy of 0047-281 at z = 0.485}.
\newblock \emph{\apj} 583, 606--615.
\newblock \doi{10.1086/345423}
\bibAnnoteFile{Koopmans&Treu2003}

\bibitem[{{Kormendy} and {Ho}(2013)}]{Kormendy&Ho2013}
{Kormendy}, J. and {Ho}, L.~C. (2013).
\newblock {Coevolution (Or Not) of Supermassive Black Holes and Host Galaxies}.
\newblock \emph{\araa} 51, 511--653.
\newblock \doi{10.1146/annurev-astro-082708-101811}
\bibAnnoteFile{Kormendy&Ho2013}

\bibitem[{{Kun} et~al.(2018){Kun}, {Biermann}, {Britzen}, and
  {Gergely}}]{kunetal18}
{Kun}, E., {Biermann}, P., {Britzen}, S., and {Gergely}, L. (2018).
\newblock {On the High-Energy Neutrino Emission from Active Galactic Nuclei}.
\newblock \emph{Universe} 4, 24.
\newblock \doi{10.3390/universe4020024}
\bibAnnoteFile{kunetal18}

\bibitem[{{La Franca} et~al.(2014){La Franca}, {Bianchi}, {Ponti}, {Branchini},
  and {Matt}}]{lafrancaetal14}
{La Franca}, F., {Bianchi}, S., {Ponti}, G., {Branchini}, E., and {Matt}, G.
  (2014).
\newblock {A New Cosmological Distance Measure Using Active Galactic Nucleus
  X-Ray Variability}.
\newblock \emph{\apjl} 787, L12.
\newblock \doi{10.1088/2041-8205/787/1/L12}
\bibAnnoteFile{lafrancaetal14}

\bibitem[{{LaMassa} et~al.(2015){LaMassa}, {Cales}, {Moran}, {Myers},
  {Richards}, {Eracleous} et~al.}]{lamassaetal15}
{LaMassa}, S.~M., {Cales}, S., {Moran}, E.~C., {Myers}, A.~D., {Richards},
  G.~T., {Eracleous}, M., et~al. (2015).
\newblock {The Discovery of the First ``Changing Look" Quasar: New Insights
  Into the Physics and Phenomenology of Active Galactic Nucleus}.
\newblock \emph{\apj} 800, 144.
\newblock \doi{10.1088/0004-637X/800/2/144}
\bibAnnoteFile{lamassaetal15}

\bibitem[{{Lane} et~al.(2012){Lane}, {Cotton}, {Helmboldt}, and
  {Kassim}}]{laneetal12}
{Lane}, W.~M., {Cotton}, W.~D., {Helmboldt}, J.~F., and {Kassim}, N.~E. (2012).
\newblock {VLSS redux: Software improvements applied to the Very Large Array
  Low-Frequency Sky Survey}.
\newblock \emph{Radio Science} 47, RS0K04.
\newblock \doi{10.1029/2011RS004941}
\bibAnnoteFile{laneetal12}

\bibitem[{{Lange} et~al.(2015){Lange}, {Driver}, {Robotham}, {Kelvin},
  {Graham}, {Alpaslan} et~al.}]{Langeetal2015}
{Lange}, R., {Driver}, S.~P., {Robotham}, A.~S.~G., {Kelvin}, L.~S., {Graham},
  A.~W., {Alpaslan}, M., et~al. (2015).
\newblock {Galaxy And Mass Assembly (GAMA): mass-size relations of z $\lt$ 0.1
  galaxies subdivided by S{\'e}rsic index, colour and morphology}.
\newblock \emph{\mnras} 447, 2603--2630.
\newblock \doi{10.1093/mnras/stu2467}
\bibAnnoteFile{Langeetal2015}

\bibitem[{Lawrence(2018)}]{lawrence18}
Lawrence, A. (2018).
\newblock Quasar viscosity crisis.
\newblock \emph{Nature Astronomy} 2, 102--103.
\newblock \doi{10.1038/s41550-017-0372-1}
\bibAnnoteFile{lawrence18}

\bibitem[{{Levi} et~al.(2013){Levi}, {Bebek}, {Beers}, {Blum}, {Cahn},
  {Eisenstein} et~al.}]{levietal13}
{Levi}, M., {Bebek}, C., {Beers}, T., {Blum}, R., {Cahn}, R., {Eisenstein}, D.,
  et~al. (2013).
\newblock {The DESI Experiment, a whitepaper for Snowmass 2013}.
\newblock \emph{ArXiv e-prints}
\bibAnnoteFile{levietal13}

\bibitem[{{Li} et~al.(2018){Li}, {Mao}, {Cappellari}, {Ge}, {Long}, {Li}
  et~al.}]{Lietal2018}
{Li}, H., {Mao}, S., {Cappellari}, M., {Ge}, J., {Long}, R.~J., {Li}, R.,
  et~al. (2018).
\newblock {SDSS-IV MaNGA: global stellar population and gradients for about
  2000 early-type and spiral galaxies on the mass-size plane}.
\newblock \emph{\mnras} 476, 1765--1775.
\newblock \doi{10.1093/mnras/sty334}
\bibAnnoteFile{Lietal2018}

\bibitem[{{Li} et~al.(2016){Li}, {Wang}, {Ho}, {Lu}, {Qiu}, {Du}
  et~al.}]{lietal16}
{Li}, Y.-R., {Wang}, J.-M., {Ho}, L.~C., {Lu}, K.-X., {Qiu}, J., {Du}, P.,
  et~al. (2016).
\newblock {Spectroscopic Indication of a Centi-parsec Supermassive Black Hole
  Binary in the Galactic Center of NGC 5548}.
\newblock \emph{\apj} 822, 4.
\newblock \doi{10.3847/0004-637X/822/1/4}
\bibAnnoteFile{lietal16}

\bibitem[{{Lian} et~al.(2018){Lian}, {Thomas}, {Maraston}, {Goddard},
  {Comparat}, {Gonzalez-Perez} et~al.}]{Lianetal2018}
{Lian}, J., {Thomas}, D., {Maraston}, C., {Goddard}, D., {Comparat}, J.,
  {Gonzalez-Perez}, V., et~al. (2018).
\newblock {The mass-metallicity relations for gas and stars in star-forming
  galaxies: strong outflow versus variable IMF}.
\newblock \emph{\mnras} 474, 1143--1164.
\newblock \doi{10.1093/mnras/stx2829}
\bibAnnoteFile{Lianetal2018}

\bibitem[{{Lilly} et~al.(2013){Lilly}, {Carollo}, {Pipino}, {Renzini}, and
  {Peng}}]{Lillyetal2013}
{Lilly}, S.~J., {Carollo}, C.~M., {Pipino}, A., {Renzini}, A., and {Peng}, Y.
  (2013).
\newblock {Gas Regulation of Galaxies: The Evolution of the Cosmic Specific
  Star Formation Rate, the Metallicity-Mass-Star-formation Rate Relation, and
  the Stellar Content of Halos}.
\newblock \emph{\apj} 772, 119.
\newblock \doi{10.1088/0004-637X/772/2/119}
\bibAnnoteFile{Lillyetal2013}

\bibitem[{{Liska} et~al.(2018){Liska}, {Hesp}, {Tchekhovskoy}, {Ingram}, {van
  der Klis}, and {Markoff}}]{liskaetal18}
{Liska}, M., {Hesp}, C., {Tchekhovskoy}, A., {Ingram}, A., {van der Klis}, M.,
  and {Markoff}, S. (2018).
\newblock {Formation of precessing jets by tilted black hole discs in 3D
  general relativistic MHD simulations}.
\newblock \emph{\mnras} 474, L81--L85.
\newblock \doi{10.1093/mnrasl/slx174}
\bibAnnoteFile{liskaetal18}

\bibitem[{{Lodato} and {Natarajan}(2007)}]{lodatonatarajan07}
{Lodato}, G. and {Natarajan}, P. (2007).
\newblock {The mass function of high-redshift seed black holes}.
\newblock \emph{\mnras} 377, L64--L68.
\newblock \doi{10.1111/j.1745-3933.2007.00304.x}
\bibAnnoteFile{lodatonatarajan07}

\bibitem[{{Longinotti} et~al.(2015){Longinotti}, {Krongold}, {Guainazzi},
  {Giroletti}, {Panessa}, {Costantini} et~al.}]{longinottietal15}
{Longinotti}, A.~L., {Krongold}, Y., {Guainazzi}, M., {Giroletti}, M.,
  {Panessa}, F., {Costantini}, E., et~al. (2015).
\newblock {X-ray high-resolution spectroscopy reveals feedback in a Seyfert
  galaxy from an ultra fast wind with complex ionization and velocity
  structure}.
\newblock \emph{ArXiv e-prints}
\bibAnnoteFile{longinottietal15}

\bibitem[{{Lopez} et~al.(2014){Lopez}, {Lagarde}, {Jaffe}, {Petrov},
  {Sch{\"o}ller}, {Antonelli} et~al.}]{lopezetal14}
{Lopez}, B., {Lagarde}, S., {Jaffe}, W., {Petrov}, R., {Sch{\"o}ller}, M.,
  {Antonelli}, P., et~al. (2014).
\newblock {An Overview of the MATISSE Instrument -- Science, Concept and
  Current Status}.
\newblock \emph{The Messenger} 157, 5--12
\bibAnnoteFile{lopezetal14}

\bibitem[{{Luminet}(2016)}]{luminet16}
{Luminet}, J.-P. (2016).
\newblock {The Status of Cosmic Topology after Planck Data}.
\newblock \emph{Universe} 2, 1.
\newblock \doi{10.3390/universe2010001}
\bibAnnoteFile{luminet16}

\bibitem[{{Macchetto} and {Penston}(1978)}]{macchettopenston78}
{Macchetto}, F. and {Penston}, M.~V. (1978).
\newblock {The International Ultraviolet Explorer}.
\newblock \emph{ESA Bulletin} 13, 9--14
\bibAnnoteFile{macchettopenston78}

\bibitem[{{Madau} and {Dickinson}(2014)}]{madaudickinson14}
{Madau}, P. and {Dickinson}, M. (2014).
\newblock {Cosmic Star-Formation History}.
\newblock \emph{\araa} 52, 415--486.
\newblock \doi{10.1146/annurev-astro-081811-125615}
\bibAnnoteFile{madaudickinson14}

\bibitem[{{Marziani} et~al.(2018){Marziani}, {Dultzin}, {Sulentic}, {Del Olmo},
  {Negrete}, {Martinez-Aldama} et~al.}]{marzianietal18}
{Marziani}, P., {Dultzin}, D., {Sulentic}, J.~W., {Del Olmo}, A., {Negrete},
  C.~A., {Martinez-Aldama}, M.~L., et~al. (2018).
\newblock {A main sequence for quasars}.
\newblock \emph{ArXiv e-prints} arXiv180205575M
\bibAnnoteFile{marzianietal18}

\bibitem[{{Marziani} and {Sulentic}(2014)}]{marzianisulentic14}
{Marziani}, P. and {Sulentic}, J.~W. (2014).
\newblock {Highly accreting quasars: sample definition and possible
  cosmological implications}.
\newblock \emph{\mnras} 442, 1211--1229.
\newblock \doi{10.1093/mnras/stu951}
\bibAnnoteFile{marzianisulentic14}

\bibitem[{{Mathur}(2000)}]{mathur00}
{Mathur}, S. (2000).
\newblock {Narrow-line Seyfert 1 galaxies and the evolution of galaxies and
  active galaxies}.
\newblock \emph{\mnras} 314, L17--L20.
\newblock \doi{10.1046/j.1365-8711.2000.03530.x}
\bibAnnoteFile{mathur00}

\bibitem[{{Matsuoka} et~al.(2018){Matsuoka}, {Iwasawa}, {Onoue}, {Kashikawa},
  {Strauss}, {Lee} et~al.}]{matsuokaetal18}
{Matsuoka}, Y., {Iwasawa}, K., {Onoue}, M., {Kashikawa}, N., {Strauss}, M.~A.,
  {Lee}, C.-H., et~al. (2018).
\newblock {Subaru High-z Exploration of Low-Luminosity Quasars (SHELLQs). IV.
  Discovery of 41 Quasars and Luminous Galaxies at 5.7 $<$ z $<$ 6.9}.
\newblock \emph{ArXiv e-prints}
\bibAnnoteFile{matsuokaetal18}

\bibitem[{{Maturi} et~al.(2011){Maturi}, {Fedeli}, and
  {Moscardini}}]{Maturietal2011}
{Maturi}, M., {Fedeli}, C., and {Moscardini}, L. (2011).
\newblock {Imprints of primordial non-Gaussianity on the number counts of
  cosmic shear peaks}.
\newblock \emph{\mnras} 416, 2527--2538.
\newblock \doi{10.1111/j.1365-2966.2011.18958.x}
\bibAnnoteFile{Maturietal2011}

\bibitem[{{Mazur} and {Mottola}(2004)}]{mazurmottola04}
{Mazur}, P.~O. and {Mottola}, E. (2004).
\newblock {Gravitational vacuum condensate stars}.
\newblock \emph{Proceedings of the National Academy of Science} 101,
  9545--9550.
\newblock \doi{10.1073/pnas.0402717101}
\bibAnnoteFile{mazurmottola04}

\bibitem[{{McCarthy}(2006)}]{mccarthy06}
{McCarthy}, P.~J. (2006).
\newblock {The Giant Magellan Telescope project}.
\newblock In \emph{The Scientific Requirements for Extremely Large Telescopes},
  eds. P.~{Whitelock}, M.~{Dennefeld}, and B.~{Leibundgut}. vol. 232 of
  \emph{IAU Symposium}, 420--428.
\newblock \doi{10.1017/S1743921306001050}
\bibAnnoteFile{mccarthy06}

\bibitem[{McQuinn(2016)}]{mcquinn16}
McQuinn, M. (2016).
\newblock The evolution of the intergalactic medium.
\newblock \emph{Annual Review of Astronomy and Astrophysics} 54, 313--362.
\newblock \doi{10.1146/annurev-astro-082214-122355}
\bibAnnoteFile{mcquinn16}

\bibitem[{{Melia}(2015)}]{melia15}
{Melia}, F. (2015).
\newblock {The AGN Hubble Diagram and its implications for cosmology}.
\newblock \emph{\apss} 359, 34.
\newblock \doi{10.1007/s10509-015-2483-4}
\bibAnnoteFile{melia15}

\bibitem[{Mingarelli et~al.(2017)Mingarelli, Lazio, Sesana, Greene, Ellis, Ma
  et~al.}]{mingarellietal17}
Mingarelli, C. M.~F., Lazio, T. J.~W., Sesana, A., Greene, J.~E., Ellis, J.~A.,
  Ma, C.-P., et~al. (2017).
\newblock The local nanohertz gravitational-wave landscape from supermassive
  black hole binaries.
\newblock \emph{Nature Astronomy} 1, 886--892.
\newblock \doi{10.1038/s41550-017-0299-6}
\bibAnnoteFile{mingarellietal17}

\bibitem[{{Morganti}(2017)}]{morganti17}
{Morganti}, R. (2017).
\newblock {The many routes to AGN feedback}.
\newblock \emph{Frontiers in Astronomy and Space Sciences} 4, 42.
\newblock \doi{10.3389/fspas.2017.00042}
\bibAnnoteFile{morganti17}

\bibitem[{{Negrete} et~al.(2012){Negrete}, {Dultzin}, {Marziani}, and
  {Sulentic}}]{negreteetal12}
{Negrete}, A., {Dultzin}, D., {Marziani}, P., and {Sulentic}, J. (2012).
\newblock {BLR Physical Conditions in Extreme Population A Quasars: a Method to
  Estimate Central Black Hole Mass at High Redshift}.
\newblock \emph{ApJ} 757, 62
\bibAnnoteFile{negreteetal12}

\bibitem[{{Negrete} et~al.(2013){Negrete}, {Dultzin}, {Marziani}, and
  {Sulentic}}]{negreteetal13}
{Negrete}, C.~A., {Dultzin}, D., {Marziani}, P., and {Sulentic}, J.~W. (2013).
\newblock {Reverberation and Photoionization Estimates of the Broad-line Region
  Radius in Low-z Quasars}.
\newblock \emph{\apj} 771, 31.
\newblock \doi{10.1088/0004-637X/771/1/31}
\bibAnnoteFile{negreteetal13}

\bibitem[{Negrete et~al.(2017)Negrete, Dultzin, Marziani, Sulentic,
  Esparza-Arredondo, Mart{\'\i}nez-Aldama et~al.}]{negreteetal17}
Negrete, C.~A., Dultzin, D., Marziani, P., Sulentic, J.~W., Esparza-Arredondo,
  D., Mart{\'\i}nez-Aldama, M.~L., et~al. (2017).
\newblock Quasars as cosmological standard candles.
\newblock \emph{Frontiers in Astronomy and Space Sciences} 4, 59.
\newblock \doi{10.3389/fspas.2017.00059}
\bibAnnoteFile{negreteetal17}

\bibitem[{{Newman} et~al.(2015){Newman}, {Belli}, and {Ellis}}]{Newmanetal2015}
{Newman}, A.~B., {Belli}, S., and {Ellis}, R.~S. (2015).
\newblock {Discovery of a Strongly Lensed Massive Quiescent Galaxy at z =
  2.636: Spatially Resolved Spectroscopy and Indications of Rotation}.
\newblock \emph{\apjl} 813, L7.
\newblock \doi{10.1088/2041-8205/813/1/L7}
\bibAnnoteFile{Newmanetal2015}

\bibitem[{{Nyland} et~al.(2018){Nyland}, {Harwood}, {Mukherjee}, {Jagannathan},
  {Rujopakarn}, {Emonts} et~al.}]{nylandetal18}
{Nyland}, K., {Harwood}, J.~J., {Mukherjee}, D., {Jagannathan}, P.,
  {Rujopakarn}, W., {Emonts}, B., et~al. (2018).
\newblock {Revolutionizing Our Understanding of AGN Feedback and its Importance
  to Galaxy Evolution in the Era of the Next Generation Very Large Array}.
\newblock \emph{ArXiv e-prints}
\bibAnnoteFile{nylandetal18}

\bibitem[{{Oguri} and {Marshall}(2010)}]{Oguri&Marshall2010}
{Oguri}, M. and {Marshall}, P.~J. (2010).
\newblock {Gravitationally lensed quasars and supernovae in future wide-field
  optical imaging surveys}.
\newblock \emph{\mnras} 405, 2579--2593.
\newblock \doi{10.1111/j.1365-2966.2010.16639.x}
\bibAnnoteFile{Oguri&Marshall2010}

\bibitem[{{Padovani}(2017)}]{padovani17}
{Padovani}, P. (2017).
\newblock {Active Galactic Nuclei at All Wavelengths and from All Angles}.
\newblock \emph{Frontiers in Astronomy and Space Sciences} 4, 35.
\newblock \doi{10.3389/fspas.2017.00035}
\bibAnnoteFile{padovani17}

\bibitem[{{Padovani} et~al.(2016){Padovani}, {Resconi}, {Giommi}, {Arsioli},
  and {Chang}}]{padovanietal16}
{Padovani}, P., {Resconi}, E., {Giommi}, P., {Arsioli}, B., and {Chang}, Y.~L.
  (2016).
\newblock {Extreme blazars as counterparts of IceCube astrophysical neutrinos}.
\newblock \emph{\mnras} 457, 3582--3592.
\newblock \doi{10.1093/mnras/stw228}
\bibAnnoteFile{padovanietal16}

\bibitem[{{P{\^a}ris} et~al.(2017){P{\^a}ris}, {Petitjean}, {Ross}, {Myers},
  {Aubourg}, {Streblyanska} et~al.}]{parisetal17}
{P{\^a}ris}, I., {Petitjean}, P., {Ross}, N.~P., {Myers}, A.~D., {Aubourg},
  {\'E}., {Streblyanska}, A., et~al. (2017).
\newblock {The Sloan Digital Sky Survey Quasar Catalog: Twelfth data release}.
\newblock \emph{\aap} 597, A79.
\newblock \doi{10.1051/0004-6361/201527999}
\bibAnnoteFile{parisetal17}

\bibitem[{{Peterson} and {Wandel}(1999)}]{petersonwandel99}
{Peterson}, B.~M. and {Wandel}, A. (1999).
\newblock {Keplerian Motion of Broad-Line Region Gas as Evidence for
  Supermassive Black Holes in Active Galactic Nuclei}.
\newblock \emph{\apjl} 521, L95--L98.
\newblock \doi{10.1086/312190}
\bibAnnoteFile{petersonwandel99}

\bibitem[{{Pian} et~al.(1999){Pian}, {Urry}, {Maraschi}, {Madejski}, {McHardy},
  {Koratkar} et~al.}]{pianetal99}
{Pian}, E., {Urry}, C.~M., {Maraschi}, L., {Madejski}, G., {McHardy}, I.~M.,
  {Koratkar}, A., et~al. (1999).
\newblock {Ultraviolet and Multiwavelength Variability of the Blazar 3C 279:
  Evidence for Thermal Emission}.
\newblock \emph{\apj} 521, 112--120.
\newblock \doi{10.1086/307548}
\bibAnnoteFile{pianetal99}

\bibitem[{{Piconcelli} et~al.(2005){Piconcelli}, {Jimenez-Bail{\'o}n},
  {Guainazzi}, {Schartel}, {Rodr{\'{\i}}guez-Pascual}, and
  {Santos-Lle{\'o}}}]{piconcellietal05}
{Piconcelli}, E., {Jimenez-Bail{\'o}n}, E., {Guainazzi}, M., {Schartel}, N.,
  {Rodr{\'{\i}}guez-Pascual}, P.~M., and {Santos-Lle{\'o}}, M. (2005).
\newblock {The XMM-Newton view of PG quasars. I. X-ray continuum and
  absorption}.
\newblock \emph{\aap} 432, 15--30.
\newblock \doi{10.1051/0004-6361:20041621}
\bibAnnoteFile{piconcellietal05}

\bibitem[{{Predehl} et~al.(2010){Predehl}, {Andritschke}, {B{\"o}hringer},
  {Bornemann}, {Br{\"a}uninger}, {Brunner} et~al.}]{predheletal10}
{Predehl}, P., {Andritschke}, R., {B{\"o}hringer}, H., {Bornemann}, W.,
  {Br{\"a}uninger}, H., {Brunner}, H., et~al. (2010).
\newblock {eROSITA on SRG}.
\newblock In \emph{Space Telescopes and Instrumentation 2010: Ultraviolet to
  Gamma Ray}. vol. 7732 of \emph{Proceedings of SPIE}, 77320U.
\newblock \doi{10.1117/12.856577}
\bibAnnoteFile{predheletal10}

\bibitem[{{Proga} and {Kallman}(2004)}]{progakallman04}
{Proga}, D. and {Kallman}, T.~R. (2004).
\newblock {Dynamics of Line-driven Disk Winds in Active Galactic Nuclei. II.
  Effects of Disk Radiation}.
\newblock \emph{\apj} 616, 688--695.
\newblock \doi{10.1086/425117}
\bibAnnoteFile{progakallman04}

\bibitem[{Punsly et~al.(2009)Punsly, Igumenshchev, and Hirose}]{punslyetal09}
Punsly, B., Igumenshchev, I.~V., and Hirose, S. (2009).
\newblock Three-dimensional simulations of vertical magnetic flux in the
  immediate vicinity of black holes.
\newblock \emph{The Astrophysical Journal} 704, 1065
\bibAnnoteFile{punslyetal09}

\bibitem[{{Rau} et~al.(2017){Rau}, {Nandra}, {Meidinger}, and
  {Plattner}}]{rauetal17}
{Rau}, A., {Nandra}, K., {Meidinger}, N., and {Plattner}, M. (2017).
\newblock {The Wide Field Imager for Athena}.
\newblock In \emph{The X-ray Universe 2017}, eds. J.-U. {Ness} and
  S.~{Migliari}. 24
\bibAnnoteFile{rauetal17}

\bibitem[{{Read} et~al.(2016){Read}, {Iorio}, {Agertz}, and
  {Fraternali}}]{Readetal2016}
{Read}, J.~I., {Iorio}, G., {Agertz}, O., and {Fraternali}, F. (2016).
\newblock {Understanding the shape and diversity of dwarf galaxy rotation
  curves in {$\Lambda$}CDM}.
\newblock \emph{\mnras} 462, 3628--3645.
\newblock \doi{10.1093/mnras/stw1876}
\bibAnnoteFile{Readetal2016}

\bibitem[{{Reynolds}(1997)}]{reynolds97}
{Reynolds}, C.~S. (1997).
\newblock {An X-ray spectral study of 24 type 1 active galactic nuclei}.
\newblock \emph{\mnras} 286, 513--537.
\newblock \doi{10.1093/mnras/286.3.513}
\bibAnnoteFile{reynolds97}

\bibitem[{{Richards} et~al.(2011){Richards}, {Kruczek}, {Gallagher}, {Hall},
  {Hewett}, {Leighly} et~al.}]{richardsetal11}
{Richards}, G.~T., {Kruczek}, N.~E., {Gallagher}, S.~C., {Hall}, P.~B.,
  {Hewett}, P.~C., {Leighly}, K.~M., et~al. (2011).
\newblock {Unification of Luminous Type 1 Quasars through C IV Emission}.
\newblock \emph{\aj} 141, 167--+.
\newblock \doi{10.1088/0004-6256/141/5/167}
\bibAnnoteFile{richardsetal11}

\bibitem[{{Risaliti} and {Lusso}(2015)}]{risalitilusso15}
{Risaliti}, G. and {Lusso}, E. (2015).
\newblock {A Hubble Diagram for Quasars}.
\newblock \emph{\apj} 815, 33.
\newblock \doi{10.1088/0004-637X/815/1/33}
\bibAnnoteFile{risalitilusso15}

\bibitem[{Rusinek and Sikora(2017)}]{rusineksikora17}
Rusinek, K. and Sikora, M. (2017).
\newblock Confrontation of the magnetically arrested disc scenario with
  observations of fr ii sources.
\newblock \emph{Frontiers in Astronomy and Space Sciences} 4, 22.
\newblock \doi{10.3389/fspas.2017.00022}
\bibAnnoteFile{rusineksikora17}

\bibitem[{{S{\c a}dowski} et~al.(2014){S{\c a}dowski}, {Narayan}, {McKinney},
  and {Tchekhovskoy}}]{sadowskietal14}
{S{\c a}dowski}, A., {Narayan}, R., {McKinney}, J.~C., and {Tchekhovskoy}, A.
  (2014).
\newblock {Numerical simulations of super-critical black hole accretion flows
  in general relativity}.
\newblock \emph{\mnras} 439, 503--520.
\newblock \doi{10.1093/mnras/stt2479}
\bibAnnoteFile{sadowskietal14}

\bibitem[{{Scaringi} et~al.(2009){Scaringi}, {Cottis}, {Knigge}, and
  {Goad}}]{scaringietal09}
{Scaringi}, S., {Cottis}, C.~E., {Knigge}, C., and {Goad}, M.~R. (2009).
\newblock {Classifying broad absorption line quasars: metrics, issues and a new
  catalogue constructed from SDSS DR5}.
\newblock \emph{\mnras} 399, 2231--2238.
\newblock \doi{10.1111/j.1365-2966.2009.15426.x}
\bibAnnoteFile{scaringietal09}

\bibitem[{{Seidel} and {Bartelmann}(2007)}]{Seidel&Bartelmann2007}
{Seidel}, G. and {Bartelmann}, M. (2007).
\newblock {Arcfinder: an algorithm for the automatic detection of gravitational
  arcs}.
\newblock \emph{\aap} 472, 341--352.
\newblock \doi{10.1051/0004-6361:20066097}
\bibAnnoteFile{Seidel&Bartelmann2007}

\bibitem[{{Shen} et~al.(2003){Shen}, {Mo}, {White}, {Blanton}, {Kauffmann},
  {Voges} et~al.}]{Shenetal2003}
{Shen}, S., {Mo}, H.~J., {White}, S.~D.~M., {Blanton}, M.~R., {Kauffmann}, G.,
  {Voges}, W., et~al. (2003).
\newblock {The size distribution of galaxies in the Sloan Digital Sky Survey}.
\newblock \emph{\mnras} 343, 978--994.
\newblock \doi{10.1046/j.1365-8711.2003.06740.x}
\bibAnnoteFile{Shenetal2003}

\bibitem[{{Sijacki} et~al.(2015){Sijacki}, {Vogelsberger}, {Genel}, {Springel},
  {Torrey}, {Snyder} et~al.}]{Sijackietal2015}
{Sijacki}, D., {Vogelsberger}, M., {Genel}, S., {Springel}, V., {Torrey}, P.,
  {Snyder}, G.~F., et~al. (2015).
\newblock {The Illustris simulation: the evolving population of black holes
  across cosmic time}.
\newblock \emph{\mnras} 452, 575--596.
\newblock \doi{10.1093/mnras/stv1340}
\bibAnnoteFile{Sijackietal2015}

\bibitem[{{Spinoglio}(2016)}]{spinoglio16}
{Spinoglio}, L. (2016).
\newblock {SPICA - a space infrared telescope for cosmology and astrophysics}.
\newblock In \emph{Frontier Research in Astrophysics II, Proceedings of a
  meeting held 23-28 May, 2016 in Mondello (Palermo), Italy (FRAPWS2016).} 86
\bibAnnoteFile{spinoglio16}

\bibitem[{{Springel}(2005)}]{springeletal05a}
{Springel}, V. (2005).
\newblock {The cosmological simulation code GADGET-2}.
\newblock \emph{\mnras} 364, 1105--1134.
\newblock \doi{10.1111/j.1365-2966.2005.09655.x}
\bibAnnoteFile{springeletal05a}

\bibitem[{{Springel} et~al.(2005){Springel}, {White}, {Jenkins}, {Frenk},
  {Yoshida}, {Gao} et~al.}]{springeletal05}
{Springel}, V., {White}, S.~D.~M., {Jenkins}, A., {Frenk}, C.~S., {Yoshida},
  N., {Gao}, L., et~al. (2005).
\newblock {Simulations of the formation, evolution and clustering of galaxies
  and quasars}.
\newblock \emph{\nat} 435, 629--636.
\newblock \doi{10.1038/nature03597}
\bibAnnoteFile{springeletal05}

\bibitem[{{Stark} et~al.(2011){Stark}, {Ellis}, and {Ouchi}}]{starketal11}
{Stark}, D.~P., {Ellis}, R.~S., and {Ouchi}, M. (2011).
\newblock {Keck Spectroscopy of Faint 3$\gt$z$\gt$7 Lyman Break Galaxies: A
  High Fraction of Line Emitters at Redshift Six}.
\newblock \emph{\apjl} 728, L2.
\newblock \doi{10.1088/2041-8205/728/1/L2}
\bibAnnoteFile{starketal11}

\bibitem[{{Steidel} et~al.(1999){Steidel}, {Adelberger}, {Giavalisco},
  {Dickinson}, and {Pettini}}]{steideletal99}
{Steidel}, C.~C., {Adelberger}, K.~L., {Giavalisco}, M., {Dickinson}, M., and
  {Pettini}, M. (1999).
\newblock {Lyman-Break Galaxies at z$>$\~{}4 and the Evolution of the
  Ultraviolet Luminosity Density at High Redshift}.
\newblock \emph{\apj} 519, 1--17.
\newblock \doi{10.1086/307363}
\bibAnnoteFile{steideletal99}

\bibitem[{{Straatman} et~al.(2015){Straatman}, {Labb{\'e}}, {Spitler},
  {Glazebrook}, {Tomczak}, {Allen} et~al.}]{Straatmanetal2015}
{Straatman}, C.~M.~S., {Labb{\'e}}, I., {Spitler}, L.~R., {Glazebrook}, K.,
  {Tomczak}, A., {Allen}, R., et~al. (2015).
\newblock {The Sizes of Massive Quiescent and Star-forming Galaxies at z
  {\tilde} 4 with ZFOURGE and CANDELS}.
\newblock \emph{\apjl} 808, L29.
\newblock \doi{10.1088/2041-8205/808/1/L29}
\bibAnnoteFile{Straatmanetal2015}

\bibitem[{{Sulentic} et~al.(2007){Sulentic}, {Bachev}, {Marziani}, {Negrete},
  and {Dultzin}}]{sulenticetal07}
{Sulentic}, J.~W., {Bachev}, R., {Marziani}, P., {Negrete}, C.~A., and
  {Dultzin}, D. (2007).
\newblock {C IV {$\lambda$}1549 as an Eigenvector 1 Parameter for Active
  Galactic Nuclei}.
\newblock \emph{ApJ} 666, 757--777.
\newblock \doi{10.10\-86\-/\-51\-99\-16}
\bibAnnoteFile{sulenticetal07}

\bibitem[{{Sulentic} et~al.(2006){Sulentic}, {Dultzin-Hacyan}, {Marziani},
  {Bongardo}, {Braito}, {Calvani} et~al.}]{sulenticetal06a}
{Sulentic}, J.~W., {Dultzin-Hacyan}, D., {Marziani}, P., {Bongardo}, C.,
  {Braito}, V., {Calvani}, M., et~al. (2006).
\newblock {Low Redshift BAL QSOs in the Eigenvector 1 Context}.
\newblock \emph{Revista Mexicana de Astronomia y Astrofisica} 42, 23--39
\bibAnnoteFile{sulenticetal06a}

\bibitem[{{Sulentic} et~al.(2014){Sulentic}, {Marziani}, {del Olmo}, {Dultzin},
  {Perea}, and {Alenka Negrete}}]{sulenticetal14}
{Sulentic}, J.~W., {Marziani}, P., {del Olmo}, A., {Dultzin}, D., {Perea}, J.,
  and {Alenka Negrete}, C. (2014).
\newblock {GTC spectra of z {$\approx$} 2.3 quasars: comparison with local
  luminosity analogs}.
\newblock \emph{\aap} 570, A96.
\newblock \doi{10.1051/0004-6361/201423975}
\bibAnnoteFile{sulenticetal14}

\bibitem[{{Sulentic} et~al.(2000){Sulentic}, {Marziani}, and
  {Dultzin-Hacyan}}]{sulenticetal00a}
{Sulentic}, J.~W., {Marziani}, P., and {Dultzin-Hacyan}, D. (2000).
\newblock {Phenomenology of Broad Emission Lines in Active Galactic Nuclei}.
\newblock \emph{ARA\&A} 38, 521--571.
\newblock \doi{10.1146/annurev.astro.38.1.521}
\bibAnnoteFile{sulenticetal00a}

\bibitem[{{Sulentic} et~al.(2012){Sulentic}, {Marziani}, and
  M.}]{sulenticetal12a}
{Sulentic}, J.~W., {Marziani}, P., and M., D. (2012).
\newblock {Fifty Years of Quasars: Current Impressions and Future
  Perspectives}.
\newblock In \emph{{Fifty Years of Quasars: From Early Observations and Ideas
  to Future Research}}, eds. M.~{D'Onofrio}, P.~{Marziani}, and J.~W.
  {Sulentic} (Springer Verlag, Berlin-Heidelberg), vol. 386 of
  \emph{Astrophysics and Space Science Library}. 549.
\newblock \doi{10.1007/978-3-642-27564-7}
\bibAnnoteFile{sulenticetal12a}

\bibitem[{{Suyu} et~al.(2013){Suyu}, {Auger}, {Hilbert}, {Marshall}, {Tewes},
  {Treu} et~al.}]{Suyuetal2013}
{Suyu}, S.~H., {Auger}, M.~W., {Hilbert}, S., {Marshall}, P.~J., {Tewes}, M.,
  {Treu}, T., et~al. (2013).
\newblock {Two Accurate Time-delay Distances from Strong Lensing: Implications
  for Cosmology}.
\newblock \emph{\apj} 766, 70.
\newblock \doi{10.1088/0004-637X/766/2/70}
\bibAnnoteFile{Suyuetal2013}

\bibitem[{{Sweet} et~al.(2017){Sweet}, {Sharp}, {Glazebrook}, {Rigaut},
  {Carrasco}, {Brodwin} et~al.}]{Sweetetal2017}
{Sweet}, S.~M., {Sharp}, R., {Glazebrook}, K., {Rigaut}, F., {Carrasco}, E.~R.,
  {Brodwin}, M., et~al. (2017).
\newblock {The stellar mass-size relation for cluster galaxies at z = 1 with
  high angular resolution from the Gemini/GeMS multiconjugate adaptive optics
  system}.
\newblock \emph{\mnras} 464, 2910--2929.
\newblock \doi{10.1093/mnras/stw2411}
\bibAnnoteFile{Sweetetal2017}

\bibitem[{{Tanaka} et~al.(1995){Tanaka}, {Nandra}, {Fabian}, {Inoue}, {Otani},
  {Dotani} et~al.}]{tanakaetal95}
{Tanaka}, Y., {Nandra}, K., {Fabian}, A.~C., {Inoue}, H., {Otani}, C.,
  {Dotani}, T., et~al. (1995).
\newblock {Gravitationally redshifted emission implying an accretion disk and
  massive black hole in the active galaxy MCG-6-30-15}.
\newblock \emph{\nat} 375, 659--661.
\newblock \doi{10.1038/375659a0}
\bibAnnoteFile{tanakaetal95}

\bibitem[{{Tescari} et~al.(2014){Tescari}, {Katsianis}, {Wyithe}, {Dolag},
  {Tornatore}, {Barai} et~al.}]{Tescarietal2014}
{Tescari}, E., {Katsianis}, A., {Wyithe}, J.~S.~B., {Dolag}, K., {Tornatore},
  L., {Barai}, P., et~al. (2014).
\newblock {Simulated star formation rate functions at z {\tilde} 4-7, and the
  role of feedback in high-z galaxies}.
\newblock \emph{\mnras} 438, 3490--3506.
\newblock \doi{10.1093/mnras/stt2461}
\bibAnnoteFile{Tescarietal2014}

\bibitem[{{Tessore} et~al.(2016){Tessore}, {Bellagamba}, and
  {Metcalf}}]{Tessoreetal2016}
{Tessore}, N., {Bellagamba}, F., and {Metcalf}, R.~B. (2016).
\newblock {LENSED: a code for the forward reconstruction of lenses and sources
  from strong lensing observations}.
\newblock \emph{\mnras} 463, 3115--3128.
\newblock \doi{10.1093/mnras/stw2212}
\bibAnnoteFile{Tessoreetal2016}

\bibitem[{{Testi} and {Walsh}(2013)}]{testiwalsh13}
{Testi}, L. and {Walsh}, J. (2013).
\newblock {The Inauguration of the Atacama Large Millimeter/submillimeter
  Array}.
\newblock \emph{The Messenger} 152, 2--6
\bibAnnoteFile{testiwalsh13}

\bibitem[{{The Theia Collaboration} et~al.(2017){The Theia Collaboration},
  {Boehm}, {Krone-Martins}, {Amorim}, {Anglada-Escude}, {Brandeker}
  et~al.}]{Theia2017}
{The Theia Collaboration}, {Boehm}, C., {Krone-Martins}, A., {Amorim}, A.,
  {Anglada-Escude}, G., {Brandeker}, A., et~al. (2017).
\newblock {Theia: Faint objects in motion or the new astrometry frontier}.
\newblock \emph{ArXiv e-prints}
\bibAnnoteFile{Theia2017}

\bibitem[{{Tombesi} et~al.(2010){Tombesi}, {Sambruna}, {Reeves}, {Braito},
  {Ballo}, {Gofford} et~al.}]{tombesietal10}
{Tombesi}, F., {Sambruna}, R.~M., {Reeves}, J.~N., {Braito}, V., {Ballo}, L.,
  {Gofford}, J., et~al. (2010).
\newblock {Discovery of Ultra-fast Outflows in a Sample of Broad-line Radio
  Galaxies Observed with Suzaku}.
\newblock \emph{\apj} 719, 700--715.
\newblock \doi{10.1088/0004-637X/719/1/700}
\bibAnnoteFile{tombesietal10}

\bibitem[{{Torrey} et~al.(2017){Torrey}, {Vogelsberger}, {Marinacci}, {Pakmor},
  {Springel}, {Nelson} et~al.}]{Torreyetal2017}
{Torrey}, P., {Vogelsberger}, M., {Marinacci}, F., {Pakmor}, R., {Springel},
  V., {Nelson}, D., et~al. (2017).
\newblock {The evolution of the mass-metallicity relation in IllustrisTNG}.
\newblock \emph{ArXiv e-prints}
\bibAnnoteFile{Torreyetal2017}

\bibitem[{{Treu} and {Marshall}(2016)}]{Treu&Marshall2016}
{Treu}, T. and {Marshall}, P.~J. (2016).
\newblock {Time delay cosmography}.
\newblock \emph{\aapr} 24, 11.
\newblock \doi{10.1007/s00159-016-0096-8}
\bibAnnoteFile{Treu&Marshall2016}

\bibitem[{{Trujillo} et~al.(2006){Trujillo}, {F{\"o}rster Schreiber},
  {Rudnick}, {Barden}, {Franx}, {Rix} et~al.}]{Trujilloetal2006}
{Trujillo}, I., {F{\"o}rster Schreiber}, N.~M., {Rudnick}, G., {Barden}, M.,
  {Franx}, M., {Rix}, H.-W., et~al. (2006).
\newblock {The Size Evolution of Galaxies since z\~{}3: Combining SDSS, GEMS,
  and FIRES}.
\newblock \emph{\apj} 650, 18--41.
\newblock \doi{10.1086/506464}
\bibAnnoteFile{Trujilloetal2006}

\bibitem[{{Trump} et~al.(2006){Trump}, {Hall}, {Reichard}, {Richards},
  {Schneider}, {Vanden Berk} et~al.}]{trumpetal06}
{Trump}, J.~R., {Hall}, P.~B., {Reichard}, T.~A., {Richards}, G.~T.,
  {Schneider}, D.~P., {Vanden Berk}, D.~E., et~al. (2006).
\newblock {A Catalog of Broad Absorption Line Quasars from the Sloan Digital
  Sky Survey Third Data Release}.
\newblock \emph{\apjs} 165, 1--18.
\newblock \doi{10.1086/503834}
\bibAnnoteFile{trumpetal06}

\bibitem[{{Truong} et~al.(2018){Truong}, {Rasia}, {Mazzotta}, {Planelles},
  {Biffi}, {Fabjan} et~al.}]{Truongetal2018}
{Truong}, N., {Rasia}, E., {Mazzotta}, P., {Planelles}, S., {Biffi}, V.,
  {Fabjan}, D., et~al. (2018).
\newblock {Cosmological hydrodynamical simulations of galaxy clusters: X-ray
  scaling relations and their evolution}.
\newblock \emph{\mnras} 474, 4089--4111.
\newblock \doi{10.1093/mnras/stx2927}
\bibAnnoteFile{Truongetal2018}

\bibitem[{{Tuccillo} et~al.(2018){Tuccillo}, {Huertas-Company},
  {Decenci{\`e}re}, {Velasco-Forero}, {Dom{\'{\i}}nguez S{\'a}nchez}, and
  {Dimauro}}]{tuccilloetal18}
{Tuccillo}, D., {Huertas-Company}, M., {Decenci{\`e}re}, E., {Velasco-Forero},
  S., {Dom{\'{\i}}nguez S{\'a}nchez}, H., and {Dimauro}, P. (2018).
\newblock {Deep learning for galaxy surface brightness profile fitting}.
\newblock \emph{\mnras} 475, 894--909.
\newblock \doi{10.1093/mnras/stx3186}
\bibAnnoteFile{tuccilloetal18}

\bibitem[{{Tully} and {Fisher}(1977)}]{tullyfisher77}
{Tully}, R.~B. and {Fisher}, J.~R. (1977).
\newblock {A new method of determining distances to galaxies}.
\newblock \emph{\aap} 54, 661--673
\bibAnnoteFile{tullyfisher77}

\bibitem[{{Valentinuzzi} et~al.(2010){Valentinuzzi}, {Fritz}, {Poggianti},
  {Cava}, {Bettoni}, {Fasano} et~al.}]{Valentinuzzietal2010}
{Valentinuzzi}, T., {Fritz}, J., {Poggianti}, B.~M., {Cava}, A., {Bettoni}, D.,
  {Fasano}, G., et~al. (2010).
\newblock {Superdense Massive Galaxies in Wings Local Clusters}.
\newblock \emph{\apj} 712, 226--237.
\newblock \doi{10.1088/0004-637X/712/1/226}
\bibAnnoteFile{Valentinuzzietal2010}

\bibitem[{Vallenari(2018)}]{vallenari18}
Vallenari, A. (2018).
\newblock The future of astrometry in space.
\newblock \emph{Frontiers in Astronomy and Space Sciences} 5, 11.
\newblock \doi{10.3389/fspas.2018.00011}
\bibAnnoteFile{vallenari18}

\bibitem[{{van Dokkum} et~al.(2018){van Dokkum}, {Danieli}, {Cohen}, {Merritt},
  {Romanowsky}, {Abraham} et~al.}]{vandokkumetal18}
{van Dokkum}, P., {Danieli}, S., {Cohen}, Y., {Merritt}, A., {Romanowsky},
  A.~J., {Abraham}, R., et~al. (2018).
\newblock {A galaxy lacking dark matter}.
\newblock \emph{\nat} 555, 629--632.
\newblock \doi{10.1038/nature25767}
\bibAnnoteFile{vandokkumetal18}

\bibitem[{{van Dokkum} et~al.(2008){van Dokkum}, {Franx}, {Kriek}, {Holden},
  {Illingworth}, {Magee} et~al.}]{vanDokkumetal2008}
{van Dokkum}, P.~G., {Franx}, M., {Kriek}, M., {Holden}, B., {Illingworth},
  G.~D., {Magee}, D., et~al. (2008).
\newblock {Confirmation of the Remarkable Compactness of Massive Quiescent
  Galaxies at z \~{} 2.3: Early-Type Galaxies Did not Form in a Simple
  Monolithic Collapse}.
\newblock \emph{\apjl} 677, L5.
\newblock \doi{10.1086/587874}
\bibAnnoteFile{vanDokkumetal2008}

\bibitem[{{van Haarlem} et~al.(2013){van Haarlem}, {Wise}, {Gunst}, {Heald},
  {McKean}, {Hessels} et~al.}]{vanhaarlemetal13}
{van Haarlem}, M.~P., {Wise}, M.~W., {Gunst}, A.~W., {Heald}, G., {McKean},
  J.~P., {Hessels}, J.~W.~T., et~al. (2013).
\newblock {LOFAR: The LOw-Frequency ARray}.
\newblock \emph{\aap} 556, A2.
\newblock \doi{10.1051/0004-6361/201220873}
\bibAnnoteFile{vanhaarlemetal13}

\bibitem[{{Vaughan} et~al.(2016){Vaughan}, {Uttley}, {Markowitz},
  {Huppenkothen}, {Middleton}, {Alston} et~al.}]{vaughanetal16}
{Vaughan}, S., {Uttley}, P., {Markowitz}, A.~G., {Huppenkothen}, D.,
  {Middleton}, M.~J., {Alston}, W.~N., et~al. (2016).
\newblock {False periodicities in quasar time-domain surveys}.
\newblock \emph{\mnras} 461, 3145--3152.
\newblock \doi{10.1093/mnras/stw1412}
\bibAnnoteFile{vaughanetal16}

\bibitem[{{Vegetti} and {Koopmans}(2009)}]{Vegetti&Koopmans2009}
{Vegetti}, S. and {Koopmans}, L.~V.~E. (2009).
\newblock {Bayesian strong gravitational-lens modelling on adaptive grids:
  objective detection of mass substructure in Galaxies}.
\newblock \emph{\mnras} 392, 945--963.
\newblock \doi{10.1111/j.1365-2966.2008.14005.x}
\bibAnnoteFile{Vegetti&Koopmans2009}

\bibitem[{{Vogelsberger} et~al.(2014){Vogelsberger}, {Genel}, {Springel},
  {Torrey}, {Sijacki}, {Xu} et~al.}]{Vogelsbergeretal2014}
{Vogelsberger}, M., {Genel}, S., {Springel}, V., {Torrey}, P., {Sijacki}, D.,
  {Xu}, D., et~al. (2014).
\newblock {Introducing the Illustris Project: simulating the coevolution of
  dark and visible matter in the Universe}.
\newblock \emph{\mnras} 444, 1518--1547.
\newblock \doi{10.1093/mnras/stu1536}
\bibAnnoteFile{Vogelsbergeretal2014}

\bibitem[{{Wang} et~al.(2014{\natexlab{a}}){Wang}, {Du}, {Li}, {Ho}, {Hu}, and
  {Bai}}]{wangetal14}
{Wang}, J.-M., {Du}, P., {Li}, Y.-R., {Ho}, L.~C., {Hu}, C., and {Bai}, J.-M.
  (2014{\natexlab{a}}).
\newblock {A New Approach to Constrain Black Hole Spins in Active Galaxies
  Using Optical Reverberation Mapping}.
\newblock \emph{\apjl} 792, L13.
\newblock \doi{10.1088/2041-8205/792/1/L13}
\bibAnnoteFile{wangetal14}

\bibitem[{{Wang} et~al.(2013){Wang}, {Du}, {Valls-Gabaud}, {Hu}, and
  {Netzer}}]{wangetal13}
{Wang}, J.-M., {Du}, P., {Valls-Gabaud}, D., {Hu}, C., and {Netzer}, H. (2013).
\newblock {Super-Eddington Accreting Massive Black Holes as Long-Lived
  Cosmological Standards}.
\newblock \emph{Physical Review Letters} 110, 081301.
\newblock \doi{10.1103/PhysRevLett.110.081301}
\bibAnnoteFile{wangetal13}

\bibitem[{{Wang} et~al.(2014{\natexlab{b}}){Wang}, {Qiu}, {Du}, and
  {Ho}}]{wangetal14a}
{Wang}, J.-M., {Qiu}, J., {Du}, P., and {Ho}, L.~C. (2014{\natexlab{b}}).
\newblock {Self-shadowing Effects of Slim Accretion Disks in Active Galactic
  Nuclei: The Diverse Appearance of the Broad-line Region}.
\newblock \emph{\apj} 797, 65.
\newblock \doi{10.1088/0004-637X/797/1/65}
\bibAnnoteFile{wangetal14a}

\bibitem[{{Watson} et~al.(2011){Watson}, {Denney}, {Vestergaard}, and
  {Davis}}]{watsonetal11}
{Watson}, D., {Denney}, K.~D., {Vestergaard}, M., and {Davis}, T.~M. (2011).
\newblock {A New Cosmological Distance Measure Using Active Galactic Nuclei}.
\newblock \emph{\apjl} 740, L49.
\newblock \doi{10.1088/2041-8205/740/2/L49}
\bibAnnoteFile{watsonetal11}

\bibitem[{{Wayth} et~al.(2005){Wayth}, {Warren}, {Lewis}, and
  {Hewett}}]{Waythetal2005}
{Wayth}, R.~B., {Warren}, S.~J., {Lewis}, G.~F., and {Hewett}, P.~C. (2005).
\newblock {The lens and source of the optical Einstein ring gravitational lens
  ER 0047-2808}.
\newblock \emph{\mnras} 360, 1333--1344.
\newblock \doi{10.1111/j.1365-2966.2005.09118.x}
\bibAnnoteFile{Waythetal2005}

\bibitem[{{Wen} et~al.(2011){Wen}, {Jenet}, {Yardley}, {Hobbs}, and
  {Manchester}}]{wenetal11}
{Wen}, Z.~L., {Jenet}, F.~A., {Yardley}, D., {Hobbs}, G.~B., and {Manchester},
  R.~N. (2011).
\newblock {Constraining the Coalescence Rate of Supermassive Black-hole
  Binaries Using Pulsar Timing}.
\newblock \emph{\apj} 730, 29.
\newblock \doi{10.1088/0004-637X/730/1/29}
\bibAnnoteFile{wenetal11}

\bibitem[{{Witt} et~al.(2000){Witt}, {Mao}, and {Keeton}}]{Wittetal2000}
{Witt}, H.~J., {Mao}, S., and {Keeton}, C.~R. (2000).
\newblock {Analytic Time Delays and H$_{0}$ Estimates for Gravitational
  Lenses}.
\newblock \emph{\apj} 544, 98--103.
\newblock \doi{10.1086/317201}
\bibAnnoteFile{Wittetal2000}

\bibitem[{{WSO-UV Consortium}(2018)}]{wsouv}
{WSO-UV Consortium} (2018).
\newblock World space observatory - ultraviolet
\bibAnnoteFile{wsouv}

\bibitem[{{Wuyts} et~al.(2011){Wuyts}, {F{\"o}rster Schreiber}, {van der Wel},
  {Magnelli}, {Guo}, {Genzel} et~al.}]{Wuytsetal2011}
{Wuyts}, S., {F{\"o}rster Schreiber}, N.~M., {van der Wel}, A., {Magnelli}, B.,
  {Guo}, Y., {Genzel}, R., et~al. (2011).
\newblock {Galaxy Structure and Mode of Star Formation in the SFR-Mass Plane
  from z \~{} 2.5 to z \~{} 0.1}.
\newblock \emph{\apj} 742, 96.
\newblock \doi{10.1088/0004-637X/742/2/96}
\bibAnnoteFile{Wuytsetal2011}

\bibitem[{{Yang} et~al.(2017){Yang}, {Wu}, {Fan}, {Jiang}, {McGreer},
  {Shangguan} et~al.}]{yangetal17}
{Yang}, Q., {Wu}, X.-B., {Fan}, X., {Jiang}, L., {McGreer}, I., {Shangguan},
  J., et~al. (2017).
\newblock {Discovery of 21 New Changing-look AGNs in Northern Sky}.
\newblock \emph{ArXiv e-prints}
\bibAnnoteFile{yangetal17}

\bibitem[{{Yoshiura} et~al.(2017){Yoshiura}, {Hasegawa}, {Ichiki}, {Tashiro},
  {Shimabukuro}, and {Takahashi}}]{yoshiuraetal17}
{Yoshiura}, S., {Hasegawa}, K., {Ichiki}, K., {Tashiro}, H., {Shimabukuro}, H.,
  and {Takahashi}, K. (2017).
\newblock {Constraining the contribution of galaxies and active galactic nuclei
  to cosmic reionization}.
\newblock \emph{\mnras} 471, 3713--3726.
\newblock \doi{10.1093/mnras/stx1754}
\bibAnnoteFile{yoshiuraetal17}

\bibitem[{{Zanella} et~al.(2016){Zanella}, {Scarlata}, {Corsini}, {Bedregal},
  {Dalla Bont{\`a}}, {Atek} et~al.}]{Zanellaetal2016}
{Zanella}, A., {Scarlata}, C., {Corsini}, E.~M., {Bedregal}, A.~G., {Dalla
  Bont{\`a}}, E., {Atek}, H., et~al. (2016).
\newblock {The Role of Quenching Time in the Evolution of the Mass-size
  Relation of Passive Galaxies from the Wisp Survey}.
\newblock \emph{\apj} 824, 68.
\newblock \doi{10.3847/0004-637X/824/2/68}
\bibAnnoteFile{Zanellaetal2016}

\bibitem[{{Zarrouk} et~al.(2018){Zarrouk}, {Burtin}, {Gil-Mar{\'{\i}}n},
  {Ross}, {Tojeiro}, {P{\^a}ris} et~al.}]{zarrouketal18}
{Zarrouk}, P., {Burtin}, E., {Gil-Mar{\'{\i}}n}, H., {Ross}, A.~J., {Tojeiro},
  R., {P{\^a}ris}, I., et~al. (2018).
\newblock {The clustering of the SDSS-IV extended Baryon Oscillation
  Spectroscopic Survey DR14 quasar sample: measurement of the growth rate of
  structure from the anisotropic correlation function between redshift 0.8 and
  2.2}.
\newblock \emph{\mnras} \doi{10.1093/mnras/sty506}
\bibAnnoteFile{zarrouketal18}

\bibitem[{{Zschaechner} et~al.(2016){Zschaechner}, {Walter}, {Bolatto},
  {Farina}, {Kruijssen}, {Leroy} et~al.}]{zschaechneretal16}
{Zschaechner}, L.~K., {Walter}, F., {Bolatto}, A., {Farina}, E.~P.,
  {Kruijssen}, J.~M.~D., {Leroy}, A., et~al. (2016).
\newblock {The Molecular Wind in the Nearest Seyfert Galaxy Circinus Revealed
  by ALMA}.
\newblock \emph{\apj} 832, 142.
\newblock \doi{10.3847/0004-637X/832/2/142}
\bibAnnoteFile{zschaechneretal16}

\end{thebibliography}

%%% Make sure to upload the bib file along with the tex file and PDF
%%% Please see the test.bib file for some examples of references

%\section*{Figure captions}

%%% Use this if adding the figures directly in the mansucript, if so, please remember to also upload the files when submitting your article
%%% There is no need for adding the file termination, as long as you indicate where the file is saved. In the examples below the files (logo1.jpg and logo2.eps) are in the Frontiers LaTeX folder
%%% If using *.tif files convert them to .jpg or .png
%%% If using panelled/compiled figures with subcaptions, use the subcaption package as in the second example.  N.B. This package is incompatible with the subfigure and subfig packages
%%%  NB logo1.jpg is required in the path in order to correctly compile front page header %%%

%%% If you don't add the figures in the LaTeX files, please upload them when submitting the article.

%%% Frontiers will add the figures at the end of the provisional pdf automatically %%%

%%% The use of LaTeX coding to draw Diagrams/Figures/Structures should be avoided. They should be external callouts including graphics.

\end{document}